\documentclass[usegraphicx,usenatbib,useAMS]{mn2e}

\usepackage{times} 
\usepackage{aas_macros}     
\usepackage{amssymb}        

\newcommand{\gsim}{\gtrsim} 
\newcommand{\lsim}{\lesssim} 

\newcommand{\Msol}{M_\odot}
\newcommand{\hMsol}{h^{-1}M_\odot}
\newcommand{\Lsol}{L_\odot}
\newcommand{\hLsol}{h^{-2}L_\odot}

\newcommand{\m}{{\rm m}}
\newcommand{\mum}{\mu{\rm m}}

\newcommand{\Mpc}{{\rm Mpc}}
\newcommand{\hMpc}{h^{-1}{\rm Mpc}}
\newcommand{\kms}{{\rm km\,s^{-1}}}
\newcommand{\yr}{{\rm yr}}

\newcommand{\mJy}{{\rm mJy}}
\newcommand{\muJy}{\mu{\rm Jy}}
\newcommand{\GHz}{{\rm GHz}}
\newcommand{\GALFORM}{{\tt GALFORM}} 
\newcommand{\GRASIL}{{\tt GRASIL}}
\newcommand{\SPITZER}{{\it Spitzer}}
\newcommand{\IRAS}{{\it IRAS}}
\newcommand{\COBE}{{\it COBE}}
\newcommand{\ISO}{{\it ISO}}
\newcommand{\SCUBA}{{\it SCUBA}}
\newcommand{\JCMT}{{\it JCMT}}
\newcommand{\AKARI}{{\it AKARI}}
\newcommand{\HERSCHEL}{{\it Herschel}}
\newcommand{\GALEX}{{\it GALEX}}

\newcommand{\tesc}{t_{\rm esc}}

\title[Predictions for Herschel]
{Predictions for Herschel from $\Lambda$CDM: unveiling the cosmic star
formation history}

\author[Lacey  et al.]{C. G. Lacey
\thanks{E-mail: Cedric.Lacey@durham.ac.uk (CGL)},$^1$
C. M. Baugh,$^1$ C.S. Frenk,$^1$ A.J. Benson,$^2$ A. Orsi,$^1$
\newauthor{L. Silva,$^3$ G.L. Granato,$^3$  and A. Bressan,$^4$}\\
$^{1}$Institute for Computational Cosmology, Department of Physics,
University of Durham, South Road, Durham, DH1 3LE, UK\\
$^2$MC350-17, California Institute of Technology,
Pasadena, CA 91125, USA\\
$^3$INAF, Osservatorio Astronomico di Trieste, Via Tiepolo 11, I-34131
Trieste, Italy\\
$^4$INAF, Osservatorio Astronomico di Padova, Vicolo dell'Osservatorio
2, I-35122 Padova, Italy.
}

\begin{document}


\maketitle

\begin{abstract}
We use a model for the evolution of galaxies in the far-IR based on
the $\Lambda$CDM cosmology to make detailed predictions for upcoming
cosmological surveys with the \HERSCHEL\ Space Observatory. We use the
combined \GALFORM\ semi-analytical galaxy formation model and \GRASIL\
spectrophotometric code to compute galaxy SEDs including the
reprocessing of radiation by dust. The model, which is the same as
that in \cite{Baugh05}, assumes two different IMFs: a normal solar
neighbourhood IMF for quiescent star formation in disks, and a very
top-heavy IMF in starbursts triggered by galaxy mergers. We have shown
previously that the top-heavy IMF appears necessary to explain the
number counts and redshifts of faint sub-mm galaxies. In this paper,
we present predictions for galaxy luminosity functions, number counts
and redshift distributions in the \HERSCHEL\ imaging bands. We find
that source confusion will be a serious problem in the deepest planned
surveys. We also show predictions for physical properties such as star
formation rates and stellar, gas and halo masses, together with fluxes
at other wavelengths (from the far-UV to the radio) relevant for
multi-wavelength follow-up observations. We investigate what fraction
of the total IR emission from dust and of the high-mass star formation
over the history of the Universe should be resolved by planned surveys
with \HERSCHEL, and find a fraction $\sim 30-50\%$, depending on
confusion. Finally, we show that galaxies in \HERSCHEL\ surveys should
be significantly clustered.
\end{abstract}


\section{Introduction}

The \HERSCHEL\ Space Observatory was launched on 14~May 2009 and will
begin science observations 
towards the end of the year. It will observe the
Universe at far-infrared (IR) wavelengths, from 60 to 670$\mum$, and
will be far more sensitive at these wavelengths than any previous
telescope. One of its primary goals will be to probe the dust-obscured
part of the cosmic history of star formation. In this paper we
use a state-of-the-art theoretical model of galaxy formation based on
structure formation in the $\Lambda$CDM model, combined with a
detailed treatment of the reprocessing of stellar radiation to far-IR
wavelengths, to make predictions for what should be seen in
cosmological surveys with \HERSCHEL, and for how well \HERSCHEL\
should be able to achieve its goal of unveiling the cosmic star
formation history.

The \HERSCHEL\ satellite follows in a line of previous space-based IR
telescopes, starting in the 1980s and 1990s with \IRAS\ and \ISO\, and
including most recently \SPITZER\ and \AKARI. These have gradually
revealed the nature and evolution of the galaxy population at mid- and
far-IR wavelengths which are inaccessible from the ground, where
galaxy luminosities are dominated by emission from interstellar dust
grains ß\citep[see reviews by][]{Soifer87,Elbaz05,Soifer08}. More
recently, these observations from space have been complemented by
surveys at longer, sub-mm wavelengths from the ground, starting with
surveys at 850$\mum$ using the \SCUBA\ instrument on the \JCMT\
\citep{Smail97,Hughes98}. However, due to their poor sensitivities and
angular resolutions at these wavelengths, these earlier space missions
provided only very limited direct information on the evolution of
galaxies in the rest-frame far-IR range which contains most of the
energy re-radiated from dust grains heated by starlight. The
ground-based sub-mm surveys have also been limited to seeing only the
very highest luminosity galaxies at high redshifts, and have been
hampered by the fact that they only observe dust emission longwards of
the peak in the spectral energy distribution (SED). \HERSCHEL\ will,
for the first time, allow observations of galaxy SEDs around the
far-IR peak in dust emission out to high redshifts, and thus back to
when the Universe was only a fraction of its current age.

Although observations with \IRAS\ had already shown that certain types
of nearby star-forming galaxies (the Ultra-Luminous IR Galaxies, or
ULIRGs) emit most of their luminosity through dust emission in the
far-IR \citep[see the review by][]{Soifer87}, a landmark was achieved
with the discovery by \COBE\ of the cosmic far-IR background
\citep{Puget96,Hauser98,Fixsen98}. This far-IR background was found to
have an energy density comparable to the ultraviolet (UV)/optical
background from stars. The far-IR background is most naturally
interpreted as the emission from dust in galaxies heated by starlight,
integrated over the history of the Universe. (There is also a
contribution to the far-IR background from dust heated by radiation
from AGN, but, based on multi-wavelength, especially X-ray, studies,
the AGN contribution appears to be small, only $\sim$10\% overall
\citep[e.g.][]{Almaini99,Fardal07}. Since most of the heating of dust
in galaxies is due to radiation from young stars, the far-IR
background provides powerful evidence that the bulk of star formation
over the history of the Universe has been obscured by dust, with most
of the radiation from young stars having been reprocessed from  UV
to  far-IR wavelengths. It is hoped that observations with
\HERSCHEL\ will 
allow most of the far-IR
background to be directly resolved into individual
sources. Preliminary constraints on the far-IR source counts have
recently been obtained with the balloon-born BLAST telescope
\citep{Devlin09}, but surveys with \HERSCHEL\ will have both better
angular resolution and much better sensitivity.

Measuring the cosmic star formation history, and understanding it in
terms of physical models for galaxy formation and evolution, are among
the main goals of modern cosmology. Observational studies of the
cosmic SFR history began with optically-selected samples of galaxies
at different redshifts, which typically used the rest-frame UV
emission from galaxies as the SFR indicator
\citep{Lilly96,Madau96,Steidel99}. The earliest studies derived cosmic SFR
histories ignoring the effects of dust extinction altogether, but it
soon became apparent that this approach was inadequate, since studies
of local star-forming galaxies had previously shown that these
galaxies usually had appreciable dust extinctions in the rest-frame UV
\citep[e.g.][]{Meurer95}. Applying locally-derived relations
between UV extinction and UV spectral slope
\citep{Meurer95,Calzetti95} to high-redshift optically-selected
galaxy samples, in particular the Lyman-break galaxies (LBGs), implied
that these galaxies should have large UV dust extinctions, and thus
also large dust corrections to the SFRs inferred from their
rest-frame UV luminosities \citep{Steidel99,Meurer99}. However, in the
absence of rest-frame far-IR data on these high-redshift galaxies,
which would directly measure the amount of energy absorbed and
re-radiated by dust, these corrections for dust extinction remain very
uncertain. Correspondingly the cosmic SFR histories derived from
dust-corrected rest-frame UV data also have large uncertainties. 

There have also been attempts to estimate the cosmic SFR history from
sub-mm surveys \citep[e.g.][]{Hughes98,Chapman05}, but here the
problems have been that these surveys have detected only small numbers
of the most luminous galaxies at high redshift, that the redshifts of
many of these are uncertain, and that an  extrapolation of
the SED must be made from the sub-mm to the far-IR in order to derive
the total IR luminosity from dust, from which the SFR is
calculated. (In a highly dust-obscured star-forming galaxy, almost all
of the UV light from young stars is reprocessed into IR emission by
dust, and so the total dust luminosity provides a good measure of the
underlying UV luminosity, and thus of the SFR.) More recently,
observations of the mid-IR emission from dust in galaxies have been
used to try to infer the cosmic SFR history
\citep[e.g.][]{LeFloch05,Perez05}, but this method has the drawback
that the SED must be extrapolated from the mid-IR to the far-IR in
order to estimate the total dust luminosity. The mid-IR/far-IR ratio
is known from nearby examples to show large variations between galaxies, so this
extrapolation is likewise uncertain when applied to high-redshift
galaxies where there is no direct measurement of the
far-IR. Measurements of the cosmic SFR history using \HERSCHEL\ will
avoid all of the problems associated with these other SFR tracers by
measuring the far-IR emission directly.

An important issue which we have not yet discussed is that of the stellar
Initial Mass Function (IMF). All of the methods used to
estimate observationally the SFRs of high-redshift galaxies (whether
from rest-frame UV or IR emission, or emission line or radio
luminosities) actually only directly trace {\em massive} young stars
(typically $\gsim 5\Msol$), and so really provide measures of the SFR
for high-mass stars only. Deriving the total SFR for the whole range
of stellar mass ($\sim 0.1 - 100 \Msol$) requires an extrapolation
using an assumed IMF. It is conventional in observational estimates of
the cosmic SFR history to assume a universal IMF, usually taken to be
similar to that in the solar neighbourhood. However, this assumption
of a universal IMF has no direct observational basis, especially at
higher redshifts, and indeed already appears to lead to
inconsistencies between estimated cosmic SFR histories and independent
estimates of the evolution of the stellar mass density
\citep[e.g.][]{Perez08}. We return to this issue later in the
paper.

As already stated, the aim of this paper is to make predictions for
what cosmological surveys with \HERSCHEL\ should see, based on the
best current {\em ab initio} models both for how galaxies form and
evolve, and for how the radiation they emit is distributed over UV,
optical, IR and sub-mm wavelengths. Before we discuss our own approach
in more detail, we briefly review the different modelling strategies
for galaxy evolution in the mid- and far-IR. The models can be broadly
divided into two classes, {\em phenomenological models} and {\em
hierarchical galaxy formation models}. In {\em phenomenological
models}, the galaxy luminosity function (LF) in the IR and its
evolution are given by some phenomenological fit, and the IR SEDs are
likewise empirical fits, with the parameters in the evolving LF
adjusted to match various observational data
\citep[e.g.][]{Rowan-Robinson01,Lagache03,Rowan-Robinson09}. These
models are generally restricted to modelling the dust emission from
galaxies, and do not include the UV/optical/near-IR emission from
stars. They have been used by e.g. \citet{Fernandez08} to make
predictions for \HERSCHEL. In {\em hierarchical galaxy formation
models}, the mass assembly of galaxies is related to structure
formation in the dark matter, and star formation and galaxy merger
histories are calculated based on physical prescriptions for star
formation, supernova feedback, dynamical friction etc. These models
typically use theoretical stellar population synthesis models to
compute galaxy stellar luminosities, combined with some physical model
for the dust extinction, but then can be distinguished according to
how they compute the SED of dust emission. Some employ empirical or
phenomelogical SEDs for the dust \citep[e.g.][]{Devriendt00}; others
employ fully theoretical dust SEDs based on radiative transfer and
detailed modelling of heating and cooling of dust grains.  The
modelling approach we use here, which has been set out in
\citet{Granato00}, \citet{Baugh05} and \citet{Lacey08}, is of the
latter type, with detailed physical modelling both of galaxy formation
and of the galaxy SEDs. Other models of the same type include
\citet{Fontanot07}. We note here some of the advantages of
hierarchical galaxy formation models over phenomenological models for
galaxy evolution in the IR: they provide a direct link to theoretical
models of structure formation; typically they predict galaxy SEDs over
a wider wavelength range than the IR, including at least the UV and
optical; generally they predict a wide range of other galaxy
properties, such as masses, sizes, gas fractions and morphologies; and
finally, they allow direct predictions of galaxy clustering, without
any further assumptions.

This present paper is the fourth in a series, where we combine the
\GALFORM\ semi-analytical model of galaxy formation \citep{Cole00}
with the \GRASIL\ model for stellar and dust emission from galaxies
\citep{Silva98}. The \GALFORM\ model computes the evolution of
galaxies in the framework of the $\Lambda$CDM model of structure
formation, based on physical treatments of the assembly of dark matter
halos by merging, gas cooling in halos, star formation and supernova
feedback, galaxy mergers, and chemical enrichment. The \GRASIL\ model
computes the SED of a model galaxy from the far-UV to the radio, based
on theoretical models of stellar evolution and stellar atmospheres,
radiative transfer through a two-phase dust medium to calculate both
the dust extinction and dust heating, and a distribution of dust
temperatures in each galaxy calculated from a detailed dust grain
model. 

In the first paper in the series \citep{Granato00}, we modelled
the IR properties of galaxies in the local Universe. While this model
was very successful in explaining observations of the local Universe,
we found subsequently that it failed when confronted with observations
of star-forming galaxies at high redshifts, predicting far too few
sub-mm galaxies (SMGs) at $z\sim 2$ and Lyman-break galaxies (LBGs) at
$z\sim 3$. Therefore, in the second paper \citep{Baugh05}, we proposed
a new version of the model which assumes a top-heavy IMF in starbursts
(with slope $x=0$, compared to Salpeter slope $x=1.35$), but a normal
solar neighbourhood IMF for quiescent star formation. In this new
model, the star formation parameters were also changed to force more
star formation to happen in bursts. This revised model agreed well
with both the number counts and redshift distributions of SMGs
detected at 850$\mum$, and with the rest-frame far-UV luminosity
function of LBGs at $z\sim 3$, while still maintaining consistency
with galaxy properties in the local Universe such as the optical,
near-IR and far-IR luminosity functions, and gas fractions,
metallicities, morphologies and sizes. In the third paper
\citep{Lacey08}, we compared predictions from the same \citet{Baugh05}
model (without changing any parameters) with observational data on
galaxy evolution in the mid-IR from \SPITZER\, and found generally
good agreement.

This same model was found by \citet{LeDelliou05,LeDelliou06} and
\citet{Orsi08} to provide a good match to the observed evolution of
the population of Ly$\alpha$-emitting galaxies, and of their
clustering, over the redshift range $z\sim 3-6$. Support for the
controversial assumption of a top-heavy IMF in bursts came from the
studies of chemical enrichment in this model by
\citet{Nagashima05a,Nagashima05b}, who found that the metallicities of
both the intergalactic gas in galaxy clusters and the stars in
elliptical galaxies were predicted to be significantly lower than
observed values if a normal IMF was assumed for all star formation,
but agreed much better if a top-heavy IMF in bursts was assumed, as in
\citeauthor{Baugh05}.

We also mention here an alternative semi-analytical approach using the
\GRASIL\ SED model which has been developed by \citet{Granato04} and
\citet{Silva05}. This approach differs from that in the present paper,
in that the treatment of mass assembly of galaxies is greatly
simplified, neglecting halo and galaxy mergers, and modelling the disk
population phenomenologically, but it does include a detailed
treatment of the relation between QSO and spheroid formation,
including feedback from the QSO phase on galaxy formation. Predictions
for \HERSCHEL\ from this model have been presented in
\citet{Negrello07}.

The outline of this paper is as follows: In \S\ref{sec:model}, we
review the physical ingredients in our model and the motivations for
some of the key parameter choices. In \S\ref{sec:LF-evoln} and
\S\ref{sec:counts_redshifts}, we present model predictions for the
evolution of the galaxy luminosity function in the \HERSCHEL\ bands,
and for the consequent number counts and redshift distributions. Next,
in \S\ref{sec:properties}, we show predictions for some other key
physical properties of \HERSCHEL-selected galaxies, while in
\S\ref{sec:multi-wavelength}, we show predictions for fluxes at other
wavelengths from the far-UV to the radio. In \S\ref{sec:sfrhist} we try
to answer the key cosmological question: what fraction of the
dust-obscured star formation over the history of the Universe should
the planned surveys with \HERSCHEL\ be able to uncover? In
\S\ref{sec:clustering}, we briefly discuss what the model predicts for
clustering of galaxies in the \HERSCHEL\ bands. Finally, we present
our conclusions in \S\ref{sec:conc}.


\section{Model}
\label{sec:model}

The model used for the predictions in this paper is identical to that
described in \citet{Baugh05} and \citet{Lacey08} (apart from minor
code updates), so we give only a very brief summary here. We use the
\GALFORM\ semi-analytical galaxy formation model to predict the
physical properties of the galaxy population at different redshifts,
and combine it with the \GRASIL\ spectrophotometric model to predict
the detailed SEDs of model galaxies (including emission from dust).

\subsection{\GALFORM\ galaxy formation model}
\label{ssec:GALFORM}

We compute the formation and evolution of galaxies within the
framework of the $\Lambda$CDM model of structure formation using the
semi-analytical galaxy formation model \GALFORM. The general
methodology and approximations behind the \GALFORM\ model are set out
in detail in \citet{Cole00} (see also the review by
\citealt{Baugh06}). In summary, the \GALFORM\ model follows the main
processes which shape the formation and evolution of galaxies. These
include: (i) the collapse and merging of dark matter halos; (ii) the
shock-heating and radiative cooling of gas inside dark halos, leading
to the formation of galaxy disks; (iii) quiescent star formation in
galaxy disks; (iv) feedback both from supernova explosions and from
photo-ionization of the IGM; (v) chemical enrichment of the stars and
gas; (vi) galaxy mergers driven by dynamical friction within common
dark matter halos, leading to the formation of stellar spheroids, and
also triggering bursts of star formation.  The end product of the
calculations is a prediction of the numbers and properties of galaxies
that reside within dark matter haloes of different masses. The model
predicts the stellar and cold gas masses of the galaxies, along with
their star formation and merger histories, their disk and bulge sizes,
and their metallicities.

The prescriptions and parameters for the different processes which we
use in this paper are identical to those adopted by \citet{Baugh05}
and \citet{Lacey08}. The background cosmology is a spatially flat CDM
universe with a cosmological constant, with ``concordance'' parameters
$\Omega_{m}=0.3$, $\Omega_{\Lambda}=0.7$, $\Omega_{b}=0.04$, and $h
\equiv H_0/(100\kms\Mpc^{-1})=0.7$. The amplitude of the initial
spectrum of density fluctuations is set by the r.m.s. linear
fluctuation in a sphere of radius 8$h^{-1}\Mpc$, $\sigma_8=0.93$. The
assembly and merger histories of dark matter halos are computed using
a Monte Carlo method based on the extended Press-Schechter model
\citep{Cole00}. When used in galaxy formation models, these Monte
Carlo merger histories have been found to give very similar results to
using halo merger histories extracted directly from N-body simulations
\citep[e.g.][]{Helly03}. The evolution of the baryons within the dark
halos is then calculated using analytical prescriptions for gas
cooling, star formation, supernova feedback etc. The parameters of the
\GALFORM\ model describing these baryonic processes were chosen to
reproduce a range of properties of present-day galaxies (optical,
near-IR and far-IR luminosity functions, fraction of spheroid- vs
disk-dominated galaxies, and galaxy disk sizes, gas fractions and
metallicities as a function of luminosity), as well as the number
counts and redshift distribution of sub-mm galaxies, and the
rest-frame far-UV luminosity function of Lyman-break galaxies at high
redshift \citep{Baugh05}.

An important feature of our model is the existence of two modes of
star formation, the ``quiescent'' mode in galaxy disks, and the
``burst'' mode triggered by galaxy mergers. In the current model,
starbursts are triggered by both major galaxy mergers (which transform
stellar disks to spheroids) and minor galaxy mergers (which leave the
stellar disk of the larger galaxy unchanged). The two modes of star
formation are assumed to have different stellar Initial Mass Functions
(IMFs). Both IMFs are taken to be piecewise power laws, with slopes
$x$ defined by ${\rm d}N/{\rm d}\ln m \propto m^{-x}$, where $N$ is
the number of stars and $m$ is the stellar mass (so the Salpeter slope
is $x=1.35$), and covering a stellar mass range $0.15 < m <
120\Msol$. Quiescent star formation in galaxy disks is assumed to have
a solar neighbourhood IMF, for which we use the \citet{Kennicutt83}
parametrization, with slope $x=0.4$ for $m< \Msol$ and $x=1.5$ for $m>
\Msol$. (The \citet{Kennicutt83} IMF is similar to other popular
parametrizations of the solar neighbourhood IMF, such as that of
\citet{Kroupa01}.)  Bursts of star formation triggered by galaxy
mergers are assumed to form stars with a top-heavy IMF with slope
$x=0$. The different IMFs result in different luminosities and SEDs
for a stellar population, as well as different overall rates of gas
return and metal ejection from dying stars. These effects are all
taken into account self-consistently, based on the predictions from
stellar evolution models.

As discussed in detail in \citet{Baugh05}, the top-heavy IMF in bursts
was found to be required in order to reproduce the observed number
counts and redshift distributions of the faint sub-mm galaxies. The
top-heavy IMF results both in higher bolometric luminosities for young
stellar populations, and greater production of heavy elements and
hence also dust, both effects being important for reproducing the
properties of SMGs in the model. Furthermore, as shown by
\citet{Nagashima05a,Nagashima05b}, the predicted chemical abundances
of the X-ray emitting gas in galaxy clusters and of the stars in
elliptical galaxies also agree better with observational data in a
model with the top-heavy IMF in bursts, rather than a universal solar
neighbourhood IMF. Subsequent work using the same model has also shown
that it predicts galaxy evolution in the mid-IR in good agreement with
observations by \SPITZER\ \citep{Lacey08}. A more detailed comparison
of the model with the properties of observed SMGs has been carried out
by \citet{Swinbank08}. As shown by \citet{LeDelliou05,LeDelliou06} and
\citet{Orsi08}, the same model also reproduces the observed evolution
of the luminosity function and clustering of Ly$\alpha$ emitting
galaxies at high redshift. Finally, \citet{Gonzalez09a} have made a
detailed comparison of the model with the luminosity function,
colours, morphologies and sizes of galaxies in the SDSS survey of the
local Universe.

A variety of other observational evidence has accumulated which
suggests that the IMF in some environments may be top-heavy compared
to the solar neighbourhood IMF (see \citet{Elmegreen09} for a recent
review). \citet{Rieke93} argued for a top-heavy IMF in the nearby
starburst M82, based on modelling its integrated properties, while
\citet{Parra07} found possible evidence for a top-heavy IMF in the
ultra-luminous starburst Arp220 from the relative numbers of
supernovae of different types observed at radio wavelengths.  Evidence
has been found for a top-heavy IMF in some star clusters in intensely
star-forming regions, both in M82 \citep[e.g.][]{McCrady03}, and in
our own Galaxy
\citep[e.g.][]{Figer99,Stolte05,Harayama08}. Observations of both the
old and young stellar populations in the central 1~pc of our Galaxy
also favour a top-heavy IMF \citep{Paumard06,Maness07}. In the local
Universe, \citet{Meurer09} find evidence for variations in the IMF
between galaxies from variations in the H$\alpha$/UV luminosity
ratio. \citet{Fardal07} found that reconciling measurements of the
optical and IR extragalactic background with measurements of the
cosmic star formation history also seemed to require an average IMF
that was somewhat top-heavy. \citet{Perez08} compared observational
constraints on the evolution of the star formation rate density and
stellar mass density over cosmic time, and found that reconciling
these two types of data also favours a more top-heavy IMF at higher
redshifts, as had been hinted at by earlier studies. Finally,
\citet{Dokkum08} found that reconciling the colour and luminosity
evolution of early-type galaxies in clusters also favoured a top-heavy
IMF. \citet{Larson98} summarized other evidence for a top-heavy IMF
during the earlier phases of galaxy evolution, and argued that this
could be a natural consequence of the temperature-dependence of the
Jeans mass for gravitational instability in gas
clouds. \citet{Larson05} extended this to argue that a top-heavy IMF
might also be expected in starburst regions, where there is strong
heating of the dust by the young stars.

In our model, the fraction of star formation occuring
in the burst mode increases with redshift \citep[see][]{Baugh05}, so
the average IMF with which stars are being formed shifts from being
close to a solar neighbourhood IMF at the present day to being very
top-heavy at high redshift.  In this model, 30\% of star formation
occured in the burst mode when integrated over the past history of the
Universe, but only 6\% of the current stellar mass was formed in
bursts, because of the much larger fraction of mass recycled by dying
stars for the top-heavy IMF. We note that our predictions for the IR
and sub-mm luminosities of starbursts are not sensitive to the precise
form of the top-heavy IMF, but simply require a larger fraction of $m
\sim 5-20 \Msol$ stars relative to a solar neighbourhood IMF.

We note that the galaxy formation model in this paper, unlike some
other recent semi-analytical models, does not include AGN
feedback. Instead, the role of AGN feedback in reducing the amount of
gas cooling to form massive galaxies is taken by superwinds driven by
supernova explosions. These superwinds eject gas from galaxy halos,
reducing the mass of hot gas and hence also the rate of gas cooling in
halos. The first semi-analytical model to include AGN feedback was
that of \citet{Granato04}, who introduced a detailed model of feedback
from QSO winds during the formation phase of supermassive black holes
(SMBHs), with the aim of explaining the co-evolution of the spheroidal
components of galaxies and their SMBHs. The predictions of the
\citeauthor{Granato04} model for number counts and redshift
distributions in the IR have been computed by \citet{Silva05} using
the \GRASIL\ spectrophotometric model, and compared to \ISO\ and
\SPITZER\ data. However, the \citet{Granato04} model has the
limitations that it does not include the merging of galaxies or of
dark halos, and does not treat the formation and evolution of galactic
disks. More recently, several semi-analytical models have been
published which propose that heating of halo gas by relativistic jets
from an AGN in an optically inconspicuous or ``radio'' mode can
balance radiative cooling of gas in high-mass halos, thus suppressing
{\em hot accretion} of gas onto galaxies \citep{Bower06, Croton06,
Cattaneo06, Monaco07,Lagos08}. However, these AGN feedback models differ in
detail, and all are fairly schematic. None of these models has been
shown to reproduce both the observed number counts and and the
observed redshifts of the faint sub-mm galaxies.

The effects of our superwind feedback are qualitatively quite similar
to those of the radio-mode AGN feedback. Both superwind and AGN
feedback models contain free parameters, which are adjusted in order
to make the model fit the bright end of the observed present-day
galaxy luminosity function at optical and near-IR
wavelengths. However, since the physical mechanisms are different,
they make different predictions for how the galaxy luminosity function
should evolve with redshift. Current models for the radio-mode AGN
feedback are very schematic, but they have the advantage over the
superwind model that the energetic constraints are greatly relaxed,
since accretion onto black holes can convert mass into energy with a
much higher efficiency than can supernova explosions. We will
investigate the predictions of models with AGN feedback for the IR and
sub-mm evolution of galaxies in a future paper.

\subsection{\GRASIL\ model for stellar and dust emission}
\label{ssec:GRASIL}

For each galaxy in our model, we compute the spectral energy
distribution using the spectrophotometric model \GRASIL\ 
\citep{Silva98,Granato00}. \GRASIL\ computes the emission
from the stellar population, the absorption and emission of radiation
by dust, and also radio emission (thermal and synchrotron) powered by
massive stars \citep{Bressan02}.

The main features of the \GRASIL\ model are as follows:\\
(i) The stars
are assumed to have an axisymmetric distribution in a disk and a
bulge. \\
(ii) The SEDs of the stellar populations are calculated separately for
the disk and the bulge, according to the distribution of stars in age
and metallicity that is obtained from the corresponding star
formation and chemical enrichment histories. We use the population
synthesis model described in \citet{Bressan98}, which is based on the
Padova stellar evolution tracks and Kurucz model atmospheres.  This
model is able to reproduce fairly well the integrated UV properties of
Globular Clusters \citep{Chavez09} and the observed \SPITZER\ IRS
spectra and mid-IR colours of ellipticals in the Virgo and Coma
clusters \citep{Bressan06,Clemens09}, i.e. old stellar populations
likely at the two extremes of the metallicity range of stellar
systems. At intermediate and young ages, it compares well with recent
observations of LMC star clusters \citep[see e.g.][]{Molla09}\\
(iii) The cold gas and dust in a galaxy are
assumed to be in a 2-phase medium, consisting of dense gas in giant
molecular clouds embedded in a lower-density diffuse component. In a
quiescent galaxy, the dust and gas are assumed to be confined to the
disk, while for a galaxy undergoing a burst, the dust and gas are
confined to the spheroidal burst component. \\
(iv) Stars are assumed to
be born inside molecular clouds, and then to leak out into the diffuse
medium on a timescale $\tesc$. As a result, the youngest and most
massive stars are concentrated in the dustiest regions, so they
experience larger dust extinctions than older, typically lower-mass
stars, and dust in the clouds is also much more strongly heated than
dust in the diffuse medium. \\
(v) The extinction of the starlight by
dust is computed using a radiative transfer code; this is used also to
compute the intensity of the stellar radiation field heating the dust
at each point in a galaxy. \\
(vi) The dust is modelled as a mixture of
graphite and silicate grains with a continuous distribution of grain
sizes (varying between 8\AA\ and 0.25 $\mum$), and
also Polycyclic Aromatic Hydrocarbon (PAH) molecules with a
distribution of sizes. The equilibrium temperature in the local
interstellar radiation field is calculated for each type and size of
grain, at each point in the galaxy, and this information is then used
to calculate the emission from each grain. In the case of very small
grains and PAH molecules, temperature fluctuations are important, and
the probability distribution of the temperature is calculated. The
detailed spectrum of the PAH emission is obtained using the PAH
cross-sections from \citet{Li01}, as described in \citet{vega05}. The
grain size distribution is chosen to match the mean dust extinction
curve and emissivity in the local ISM, and is not varied, except that
the PAH abundance in molecular clouds is assumed to be $10^{-3}$ of
that in the diffuse medium \citep{vega05}. \\
(vii) Radio emission from
ionized gas in HII regions and  synchrotron radiation from
relativistic electrons accelerated in supernova remnant shocks are
calculated as described in \citet{Bressan02}. 

\begin{figure}

\begin{center}

\includegraphics[width=7cm]{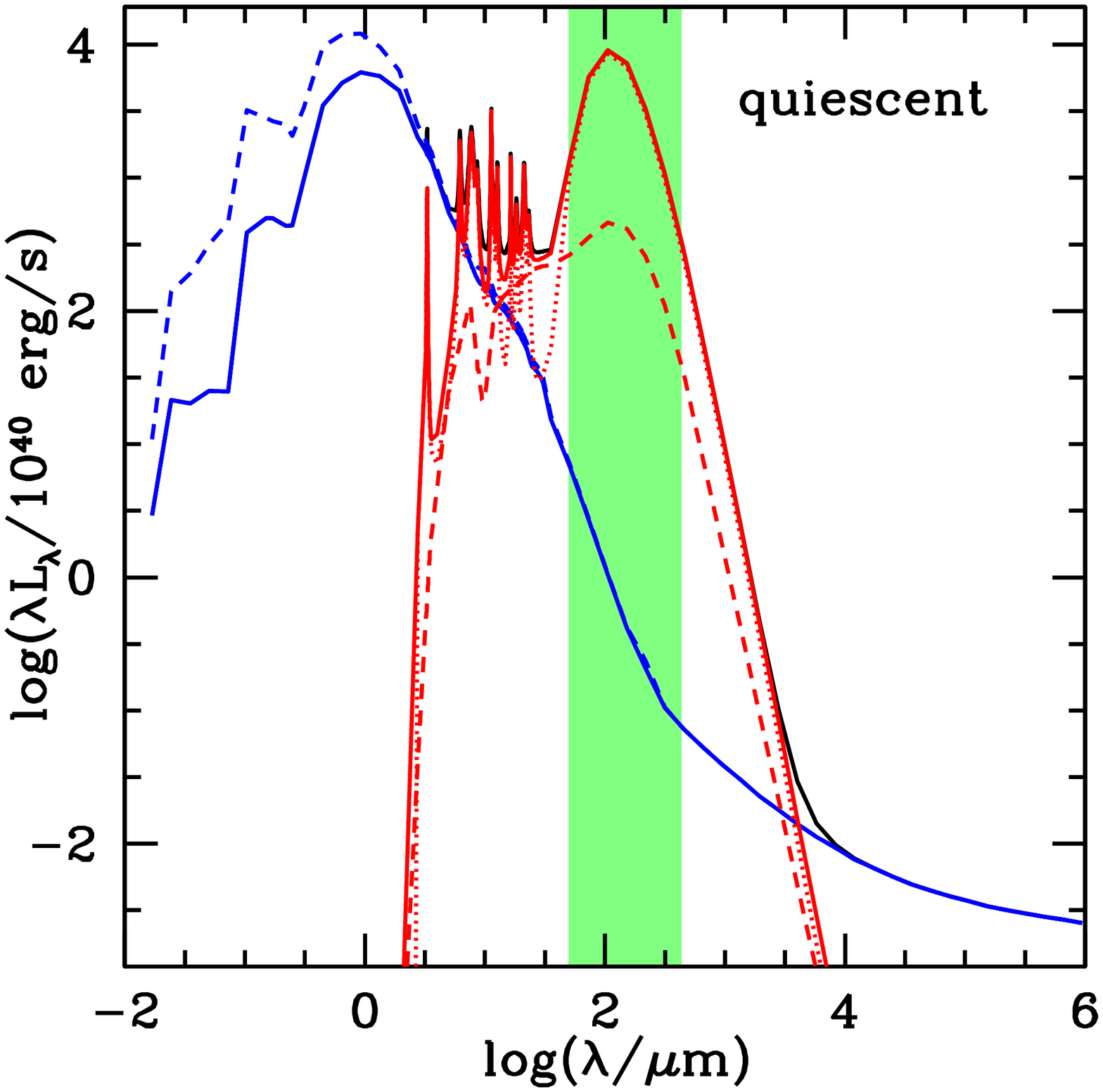}

\includegraphics[width=7cm]{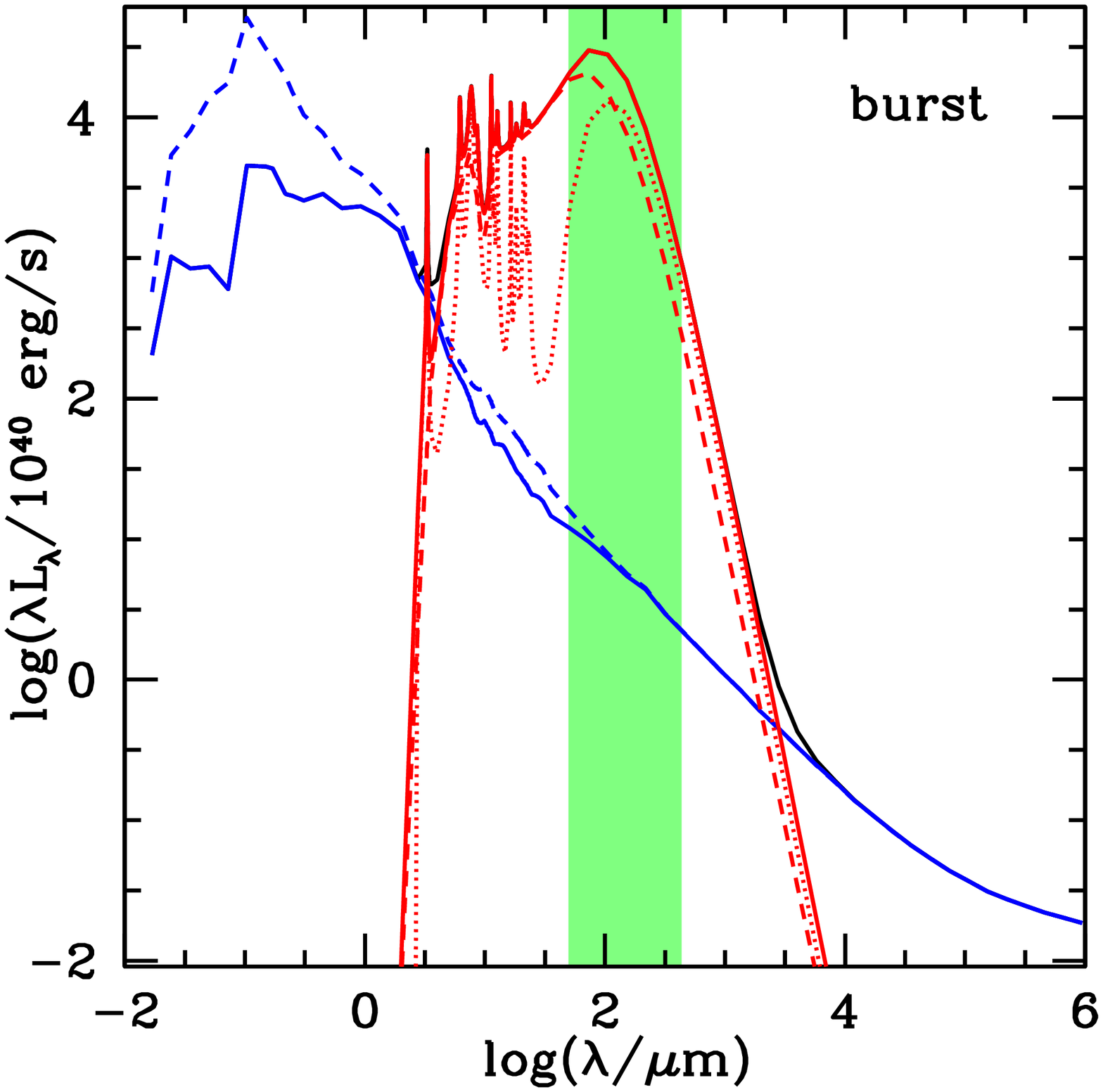}

\end{center}

\caption{Example SEDs from \GALFORM + \GRASIL. (a) Quiescently
  star-forming disk galaxy at $z=0$. (b) Starburst at $z=0$ seen one
  e-folding time after start of burst. The blue lines show the
  ``stellar'' emission (which includes emission from dust in AGB star
  envelopes, and thermal and synchrotron radio emission from the ISM),
  both with (solid lines) and without (dashed lines) dust
  extinction. The red lines show emission from interstellar dust, both
  the total (solid line) and the separate contributions from molecular
  clouds (dashed line) and the diffuse ISM (dotted line). The shaded
  green region shows the wavelength range 60-600$\mum$ covered by the
  \HERSCHEL\ imaging bands.
}

\label{fig:grasil-seds}
\end{figure}

The output from \GRASIL\ is then the complete SED of a galaxy from the
far-UV to the radio (wavelengths $100\AA \lsim \lambda \lsim
1\m$). The SED of the dust emission is computed as a sum over the
different types of grains, having different temperatures depending on
their size and their position in the galaxy. The dust SED is thus
intrinsically multi-temperature. \GRASIL\ has been shown to give an
excellent match to the observed SEDs of galaxies of all types, from
passive systems through to ULIRGs \citep{Silva98,Vega08}. We show some
example SEDs from the combined \GALFORM\ + \GRASIL\ code in
Fig.\ref{fig:grasil-seds}.

\begin{figure*}

\begin{center}

\begin{minipage}{7cm}
\includegraphics[width=7cm]{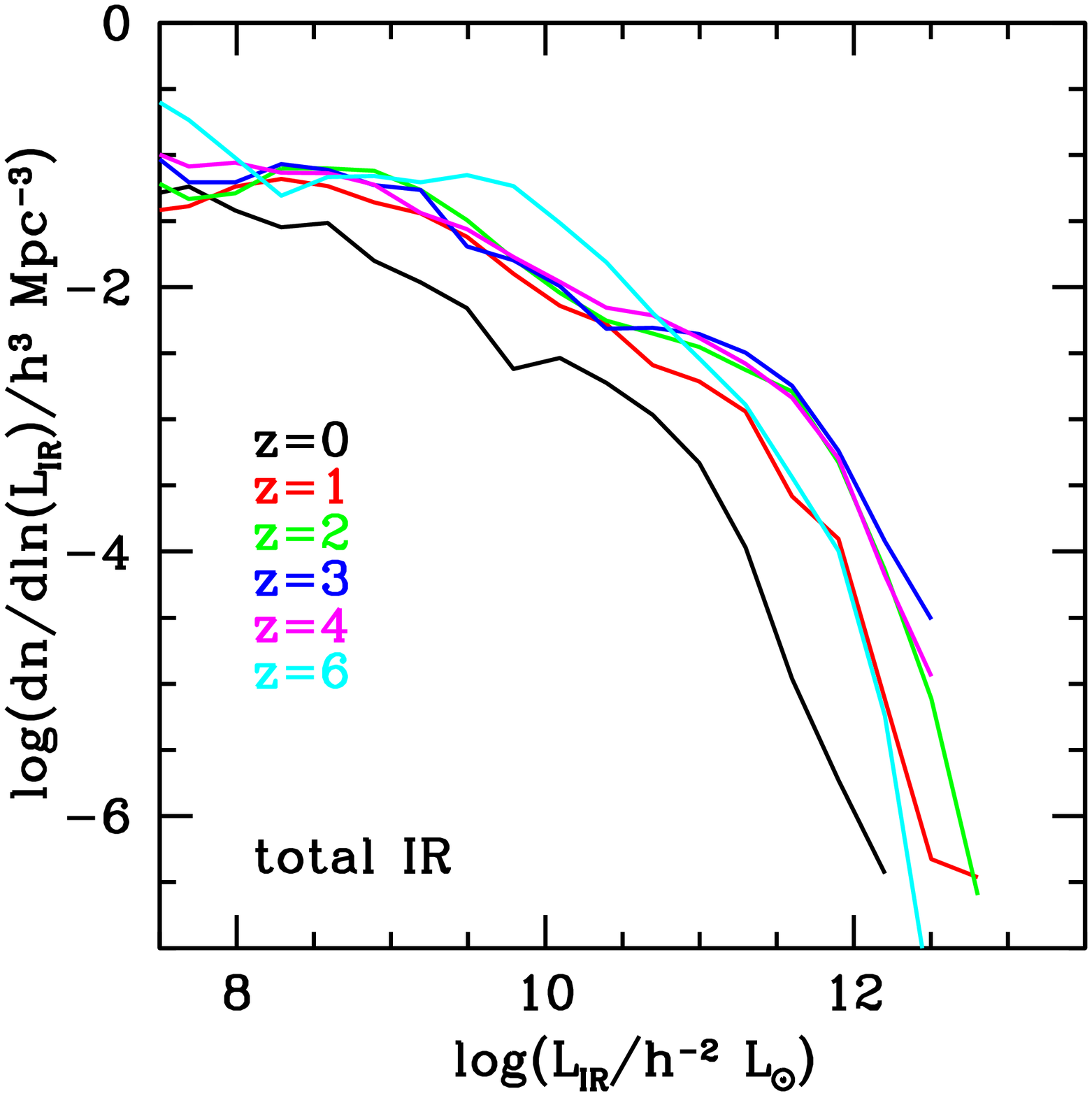}
\end{minipage}
\hspace{1cm}
\begin{minipage}{7cm}
\includegraphics[width=7cm]{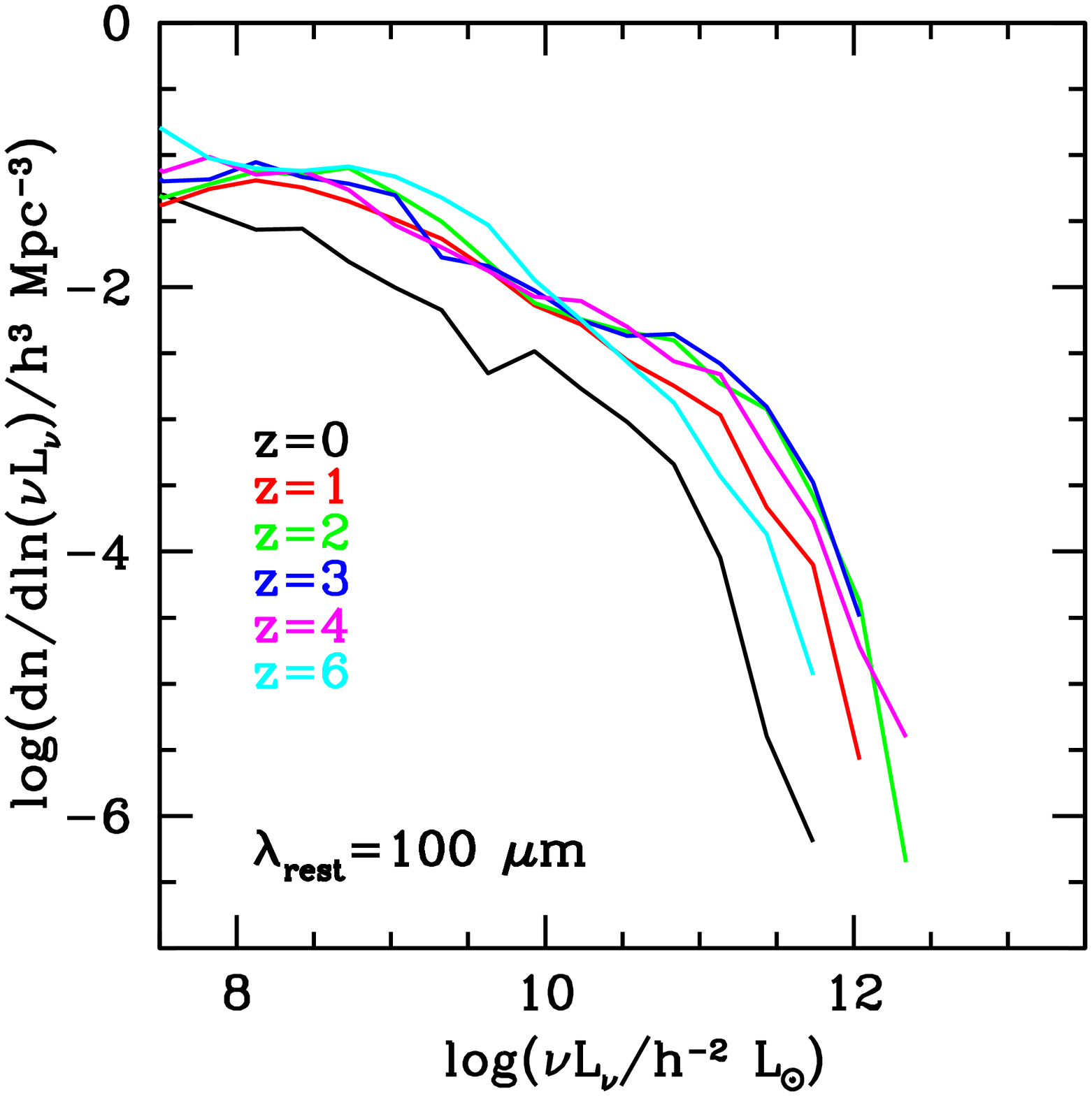}
\end{minipage}

\begin{minipage}{7cm}
\includegraphics[width=7cm]{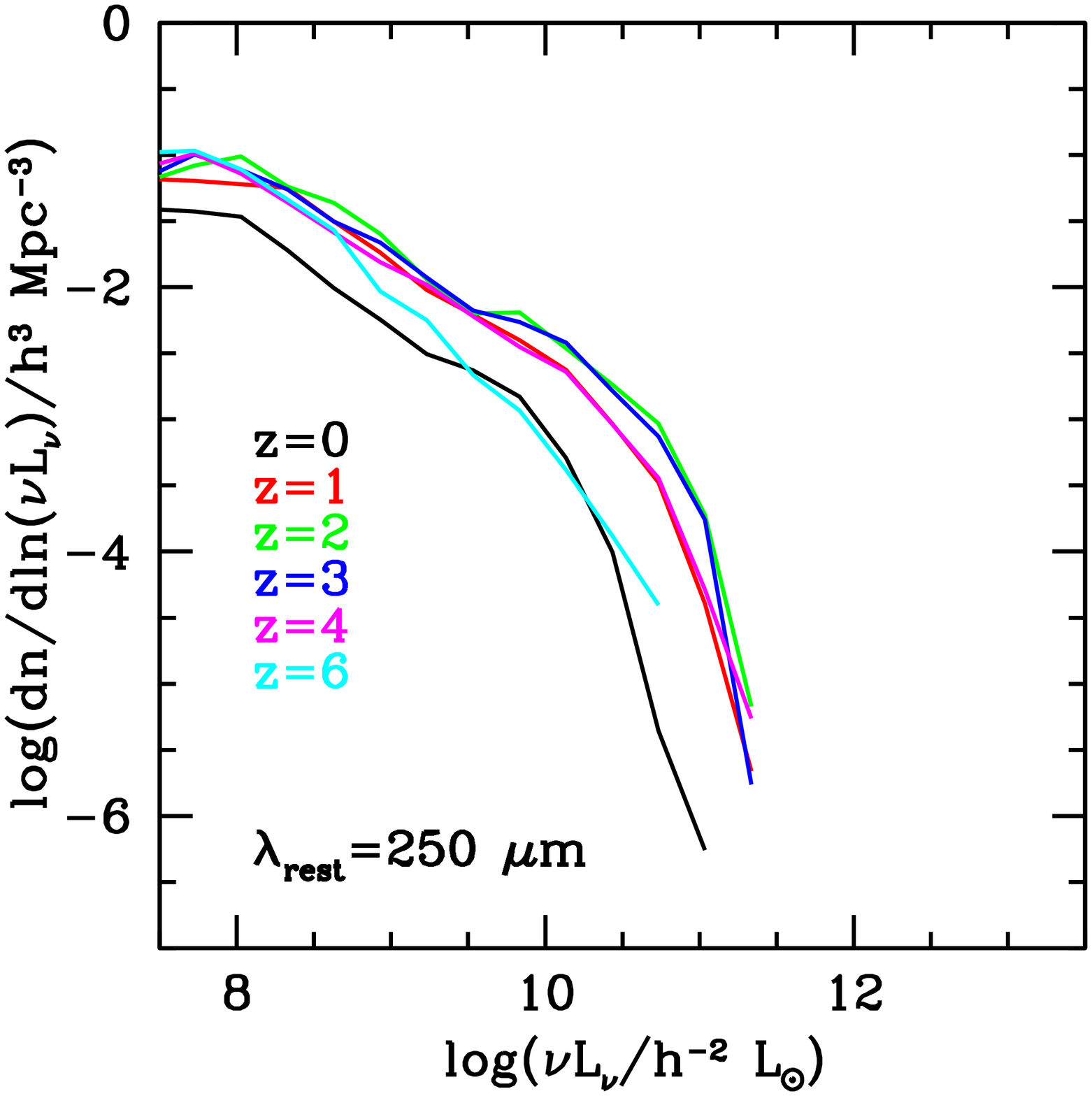}
\end{minipage}
\hspace{1cm}
\begin{minipage}{7cm}
\includegraphics[width=7cm]{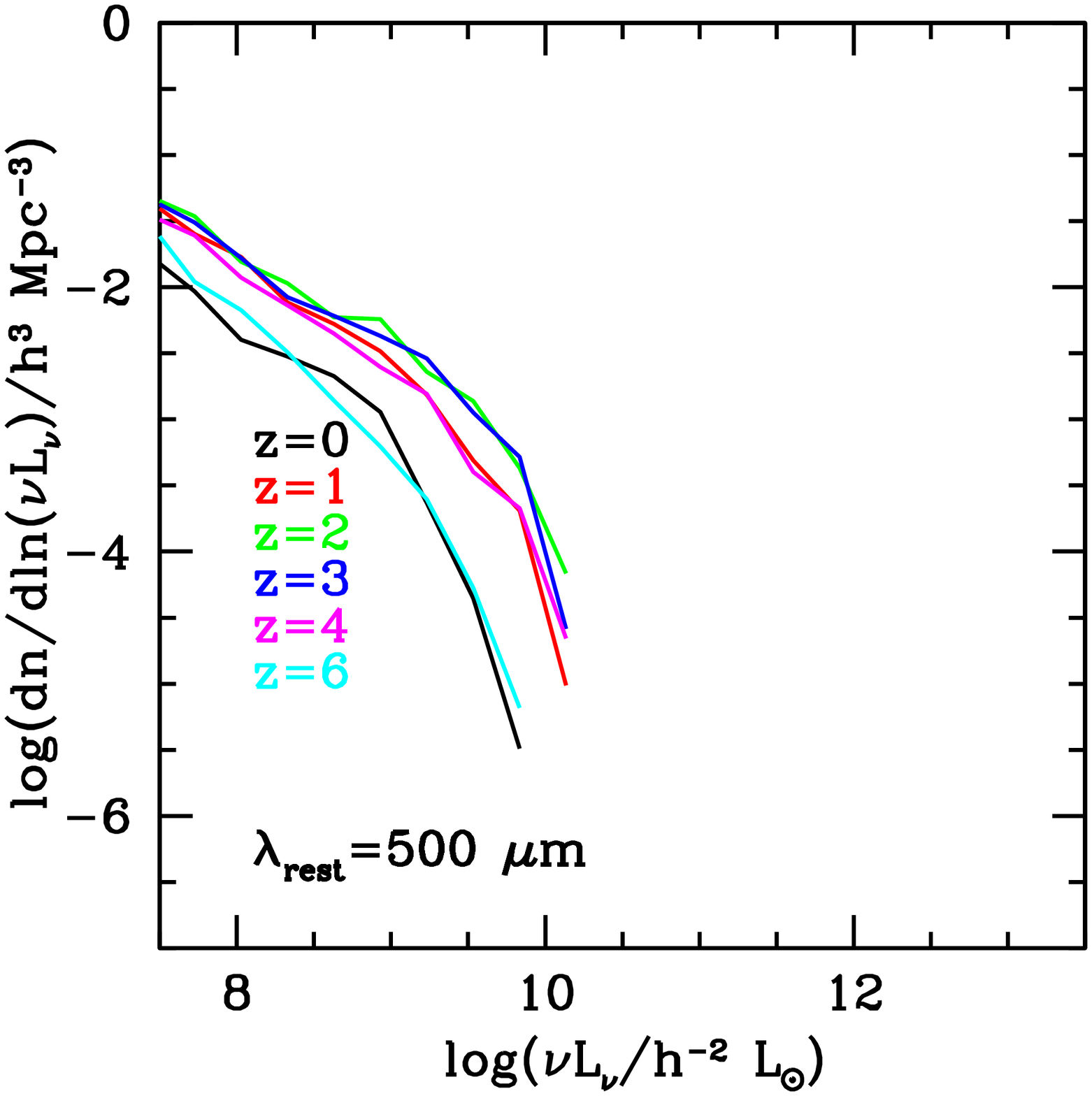}
\end{minipage}

\end{center}

\caption{Predicted evolution of the galaxy LF in the far-IR. The LFs
  are shown at redshifts $z=0,1,2,3,4,6$, with different redshifts
  shown by different colours, as indicated by the key. (a)
  Total IR (8-1000$\mum$) luminosity. (b) Rest-frame 100$\mum$. (c)
  Rest-frame 250$\mum$. (d) Rest-frame 500$\mum$. }

\label{fig:lf-evoln}
\end{figure*}

\section{Evolution of the galaxy luminosity function in the far-IR}
\label{sec:LF-evoln}

We begin by showing what the model predicts for the evolution of the
galaxy luminosity function (LF) at the far-IR wavelengths which will
be probed by \HERSCHEL, before moving on to predictions for more
immediately observable quantities in the next
section. Fig.~\ref{fig:lf-evoln} shows how the predicted luminosity
function evolves from $z=0$ to $z=6$. The top-left panel shows the LF
of the total mid+far-IR luminosity $L_{IR}$, defined as the integral
over the galaxy SED from 8 to 1000$\mum$. This range includes almost
all of the dust emission from a galaxy, but hardly any of the stellar
emission. The total IR LF evolves strongly with redshift, increasing
in characteristic luminosity and number density from $z=0$ to $z\sim
2$, having a plateau from $z\sim 2$ to $z\sim 4$, and then gradually
declining at higher redshifts. This evolution results from the
combined effects of the evolution of the galaxy SFR distribution and
the shift in the dominant mode of star formation from quiescent (with
a normal IMF) at low redshift to burst (with top-heavy IMF) at higher
redshift.  - this is discussed further in
\S\ref{sec:sfrhist}. Starting from very high redshift, the cosmic star
formation density increases to a peak at $z\sim 2-3$, driven by the
build-up of dark matter halos in which gas is able to cool and form
stars, and then declines down to $z=0$, driven by the declining
efficiency of gas cooling in halos as their masses become even larger.
Over the redshift range $z \sim 0-3$, the characteristic luminosity
increases by a factor $\sim 8$, while the number density of LIRGs
(Luminous Infrared Galaxies, defined to have $L_{IR} = 10^{11}\Lsol$)
increases by a factor $\sim 10$, and the number density of ULIRGs
(Ultra-luminous Infrared Galaxies, defined to have
$L_{IR}=10^{12}\Lsol$) increases by a factor $\sim 300$. The other
panels in Fig.~\ref{fig:lf-evoln} show the predicted evolution of the
LF at rest-frame wavelengths of 100, 250 and 500$\mum$ (calculated
through the PACS and SPIRE bandpasses). The evolution of the LF at
these wavelengths is qualitatively similar to that of the total IR
LF. The 100$\mum$ LF looks very similar to the total IR LF, both in
the form of the evolution and in normalization (when $\nu L_{\nu}$ is
used as the luminosity variable), reflecting the fact that the far-IR
SEDs peak around this wavelength. At longer wavelengths, the
characteristic luminosity is fainter (in terms of $\nu L_{\nu}$),
reflecting the decline in the SED longwards of the peak; the degree of
evolution is also somewhat less (a factor $\sim 5$ in characteristic
luminosity at 500$\mum$ over the redshift range $z \sim 0-3$),
reflecting a shift in the average far-IR SED shape to somewhat warmer
dust temperatures at higher redshifts.

\begin{figure*}

\begin{center}

\begin{minipage}{5.3cm}
\includegraphics[width=5.3cm]{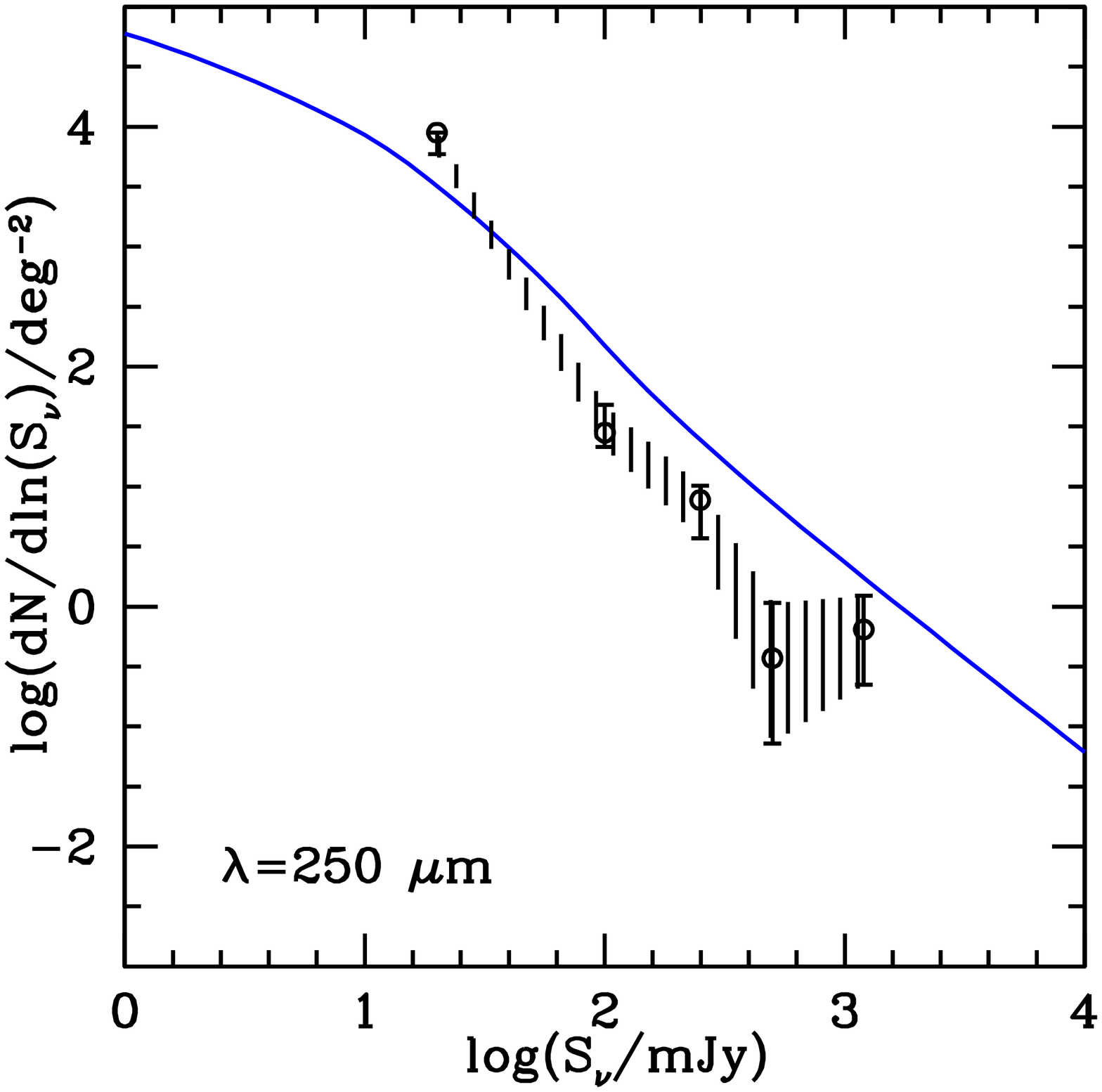}
\end{minipage}
\hspace{0.4cm}
\begin{minipage}{5.3cm}
\includegraphics[width=5.3cm]{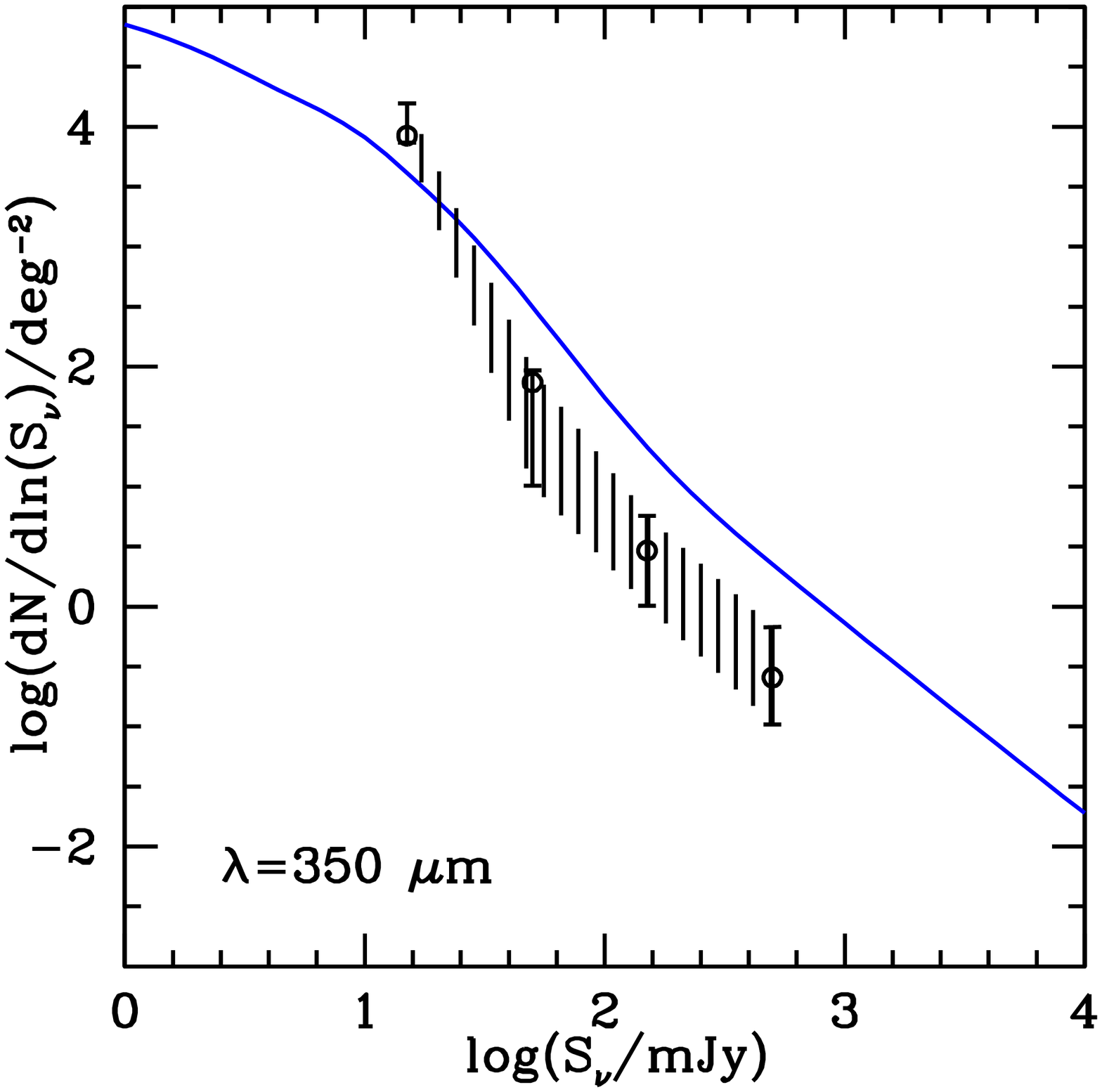}
\end{minipage}
\hspace{0.4cm}
\begin{minipage}{5.3cm}
\includegraphics[width=5.3cm]{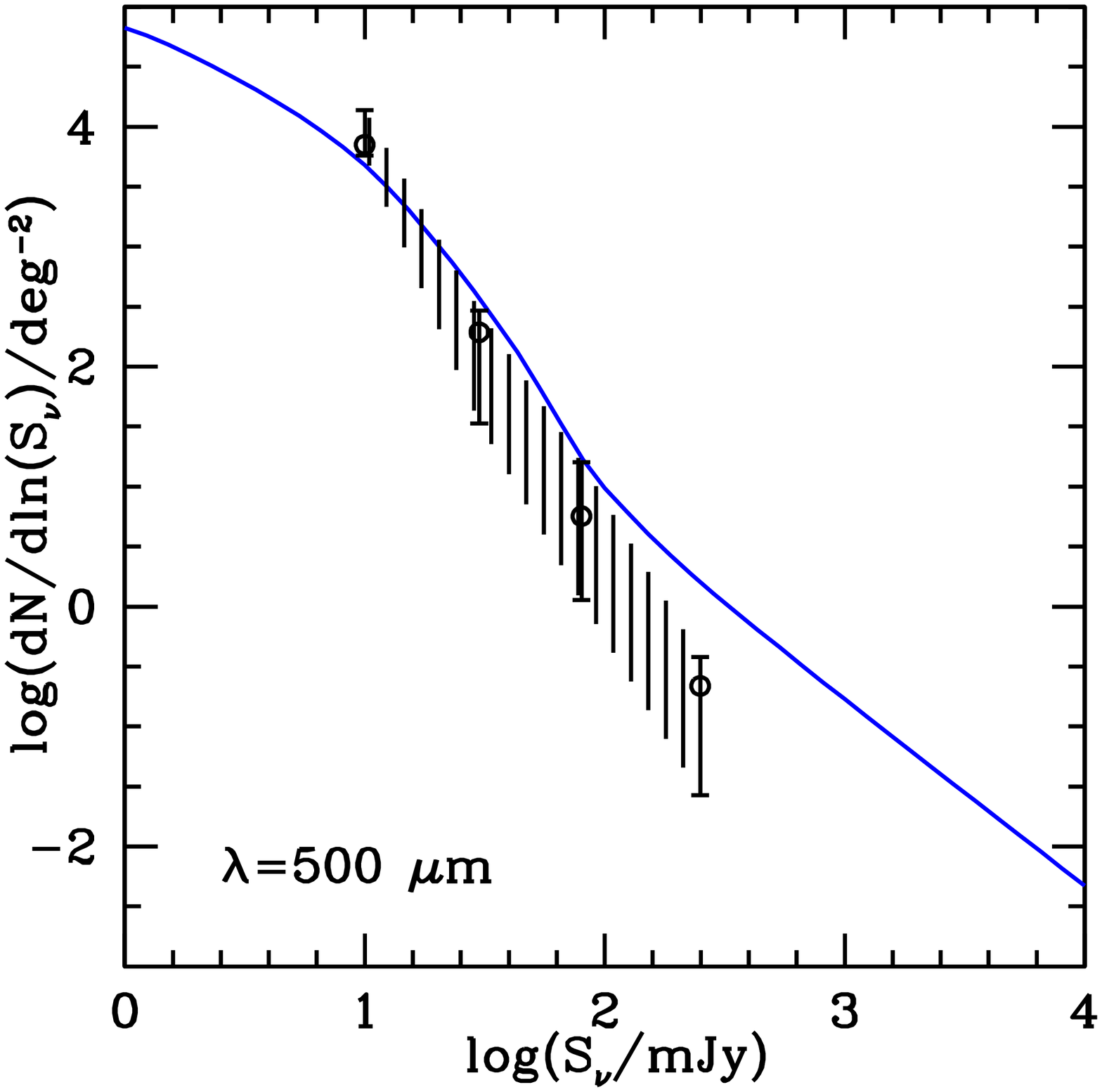}
\end{minipage}

\end{center}

\caption{ Predicted galaxy differential number counts in the SPIRE
  bands (blue curves) compared to observational estimates from BLAST
  \citep{Patanchon09}, shown in black. The observational estimates
  were obtained by modelling the P(D) distribution, rather than
  identifying individual sources. The points show the ``best
  estimates'', while the error bars and shaded regions show the 68\%
  confidence ranges. }

\label{fig:ncts_BLAST}
\end{figure*}

\begin{figure*}

\begin{center}

\begin{minipage}{7cm}
\includegraphics[width=7cm]{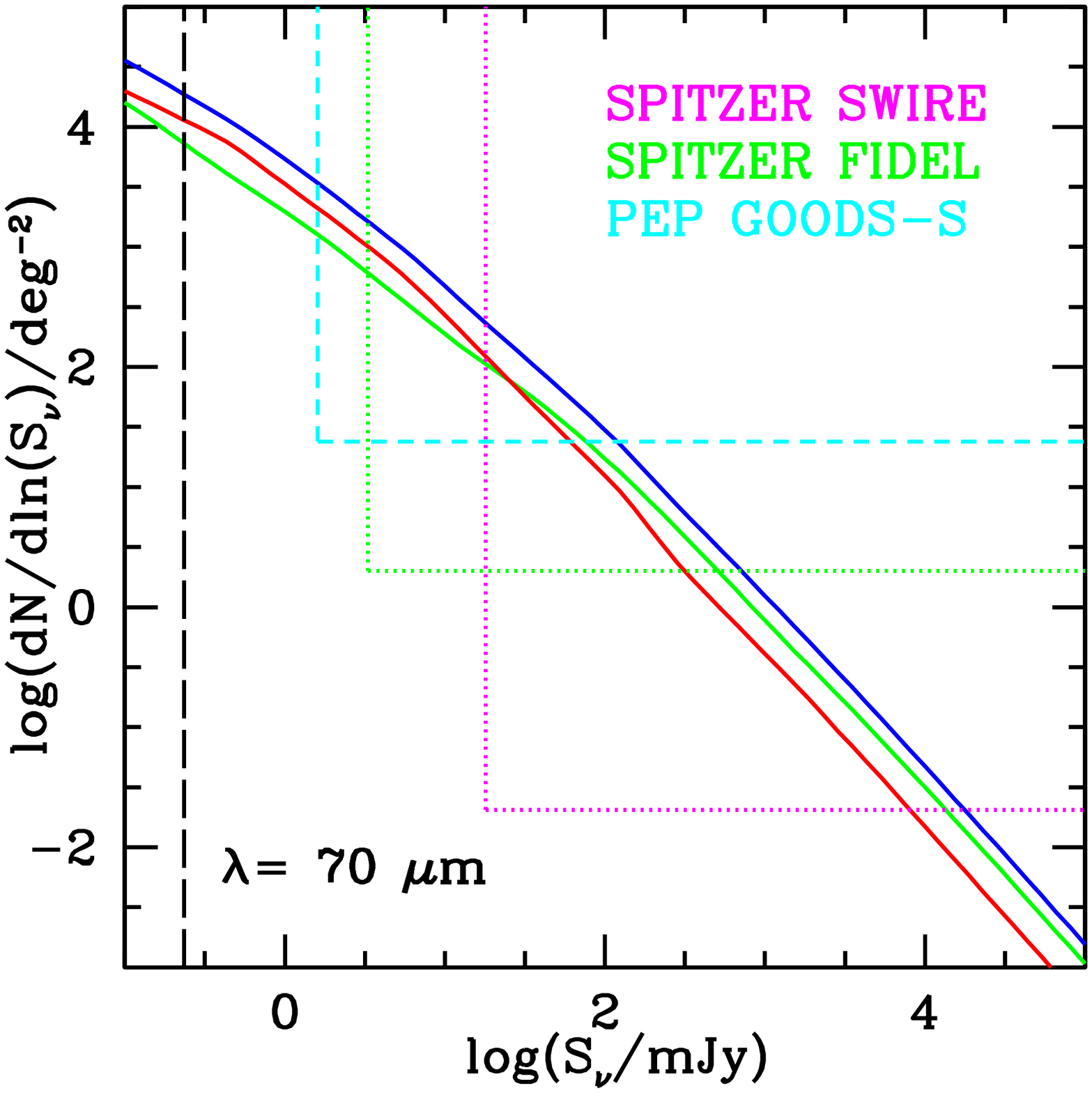}
\end{minipage}
\hspace{1cm}
\begin{minipage}{7cm}
\includegraphics[width=7cm]{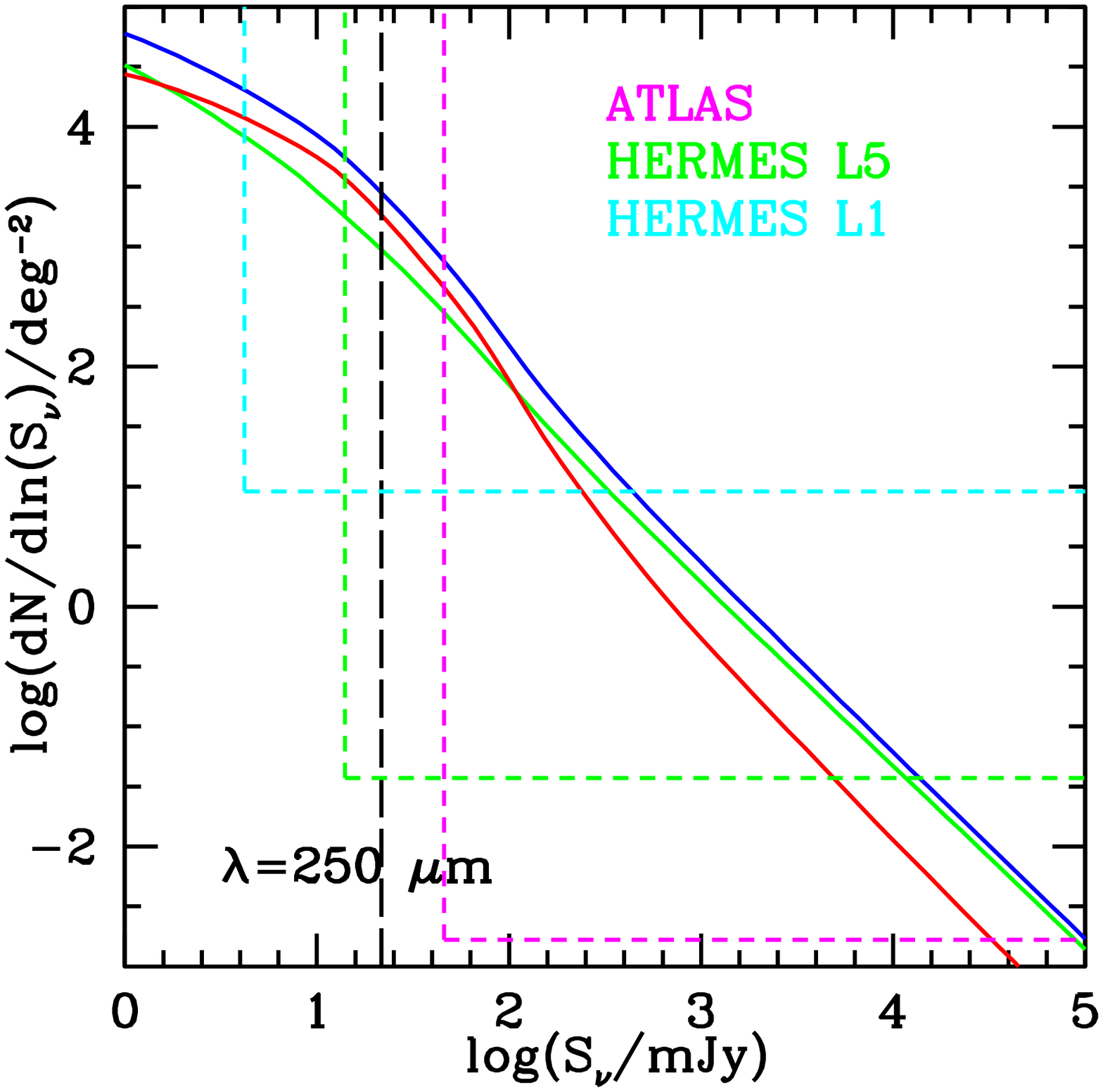}
\end{minipage}

\begin{minipage}{7cm}
\includegraphics[width=7cm]{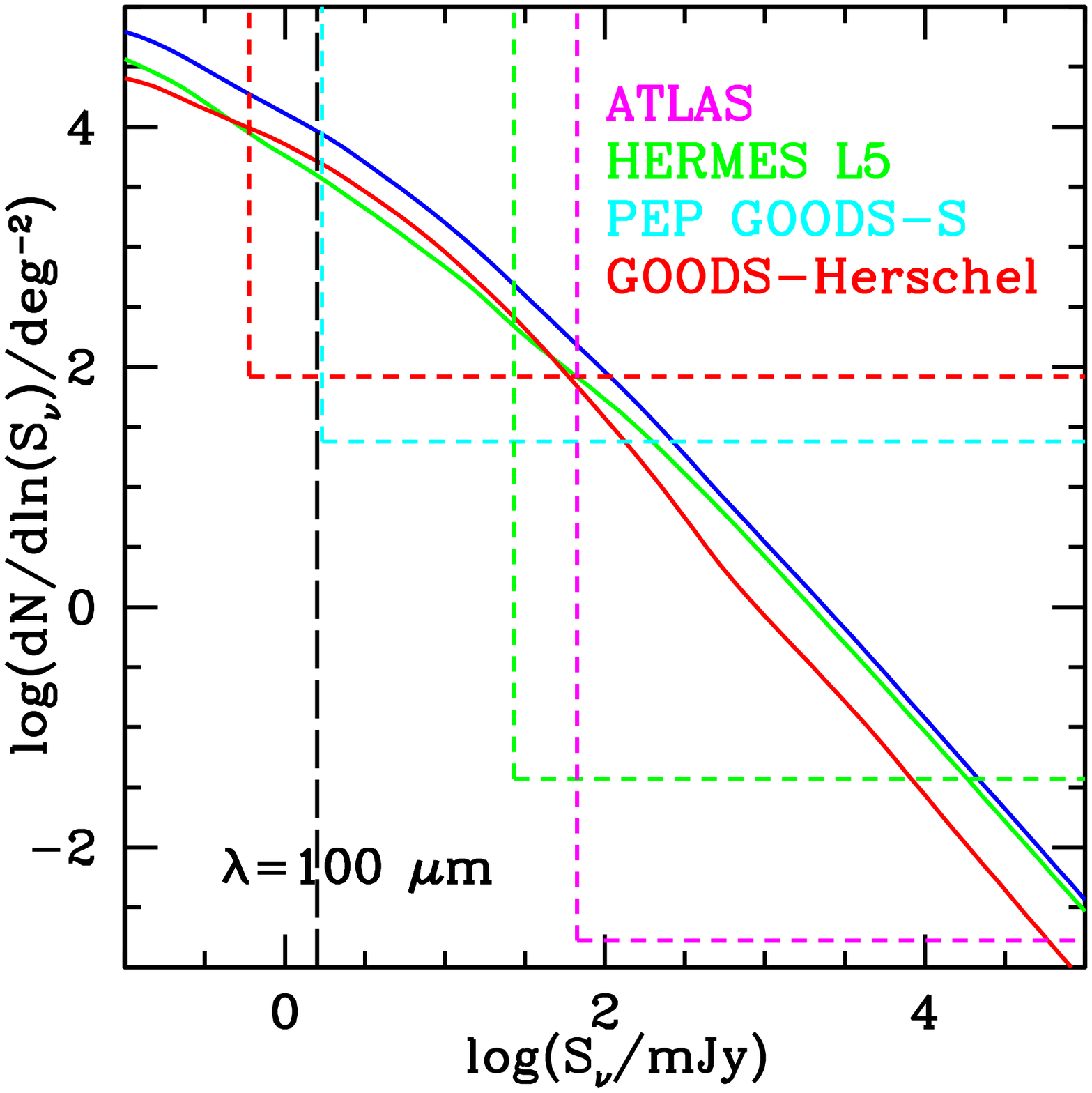}
\end{minipage}
\hspace{1cm}
\begin{minipage}{7cm}
\includegraphics[width=7cm]{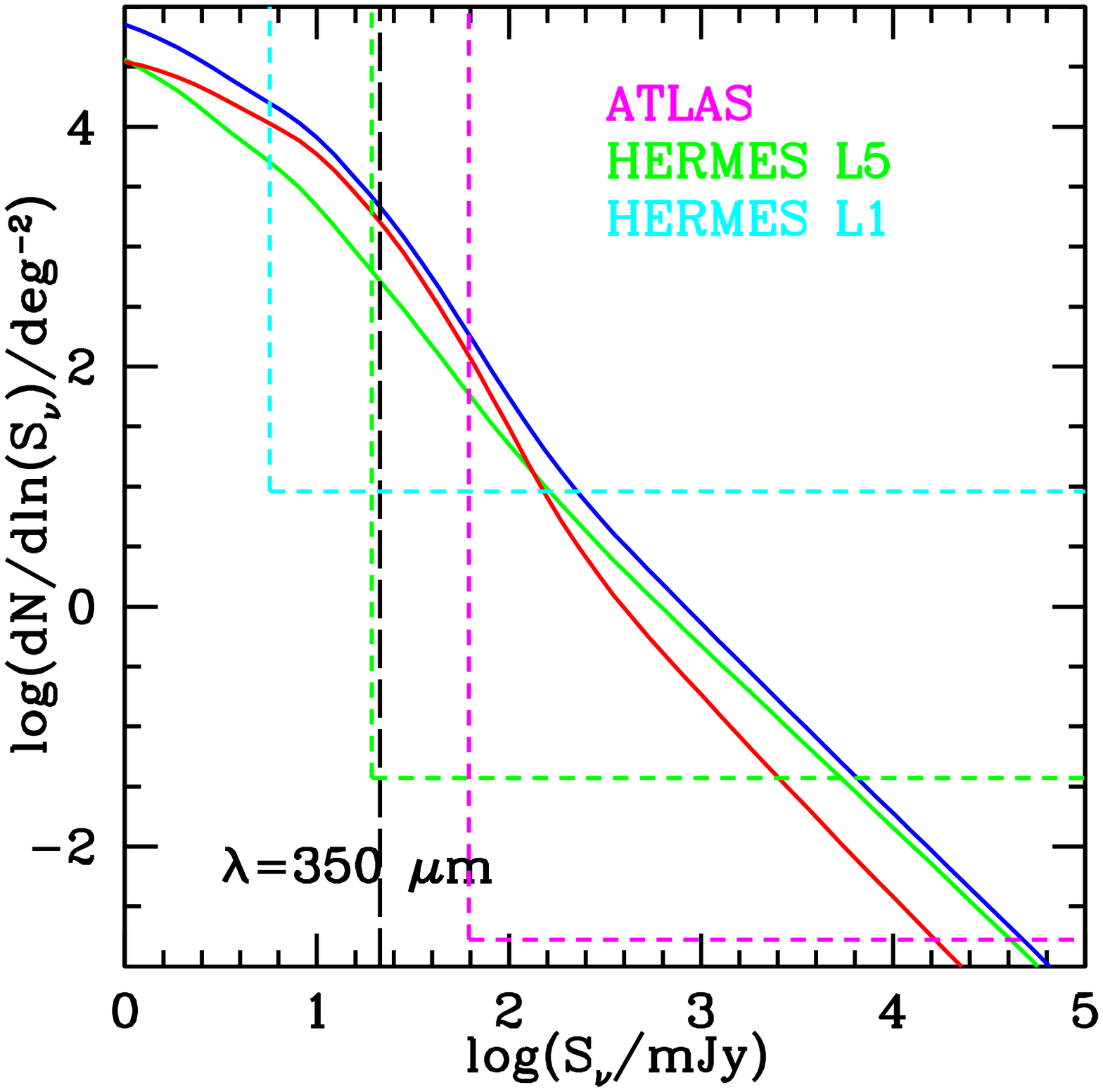}
\end{minipage}

\begin{minipage}{7cm}
\includegraphics[width=7cm]{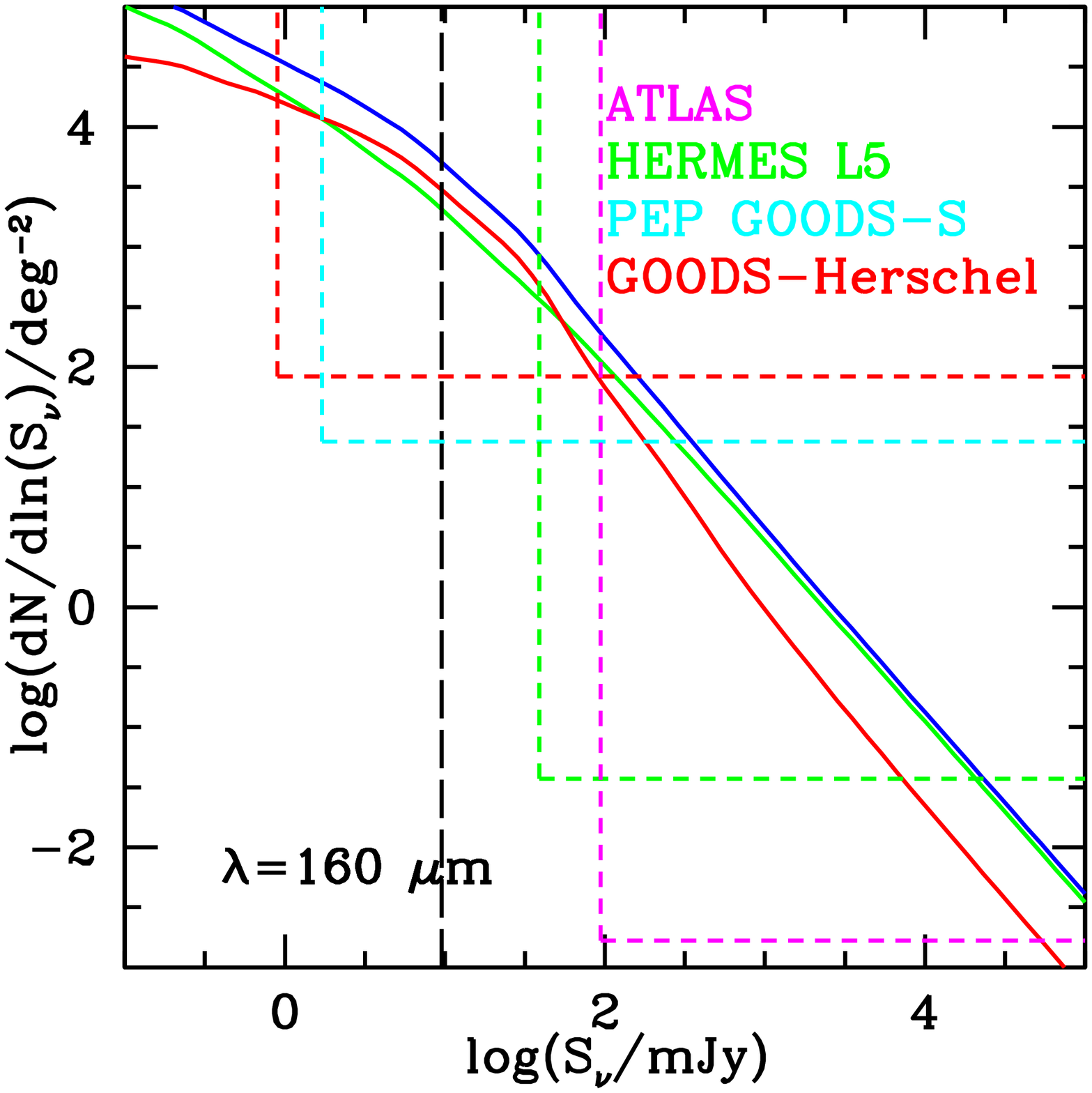}
\end{minipage}
\hspace{1cm}
\begin{minipage}{7cm}
\includegraphics[width=7cm]{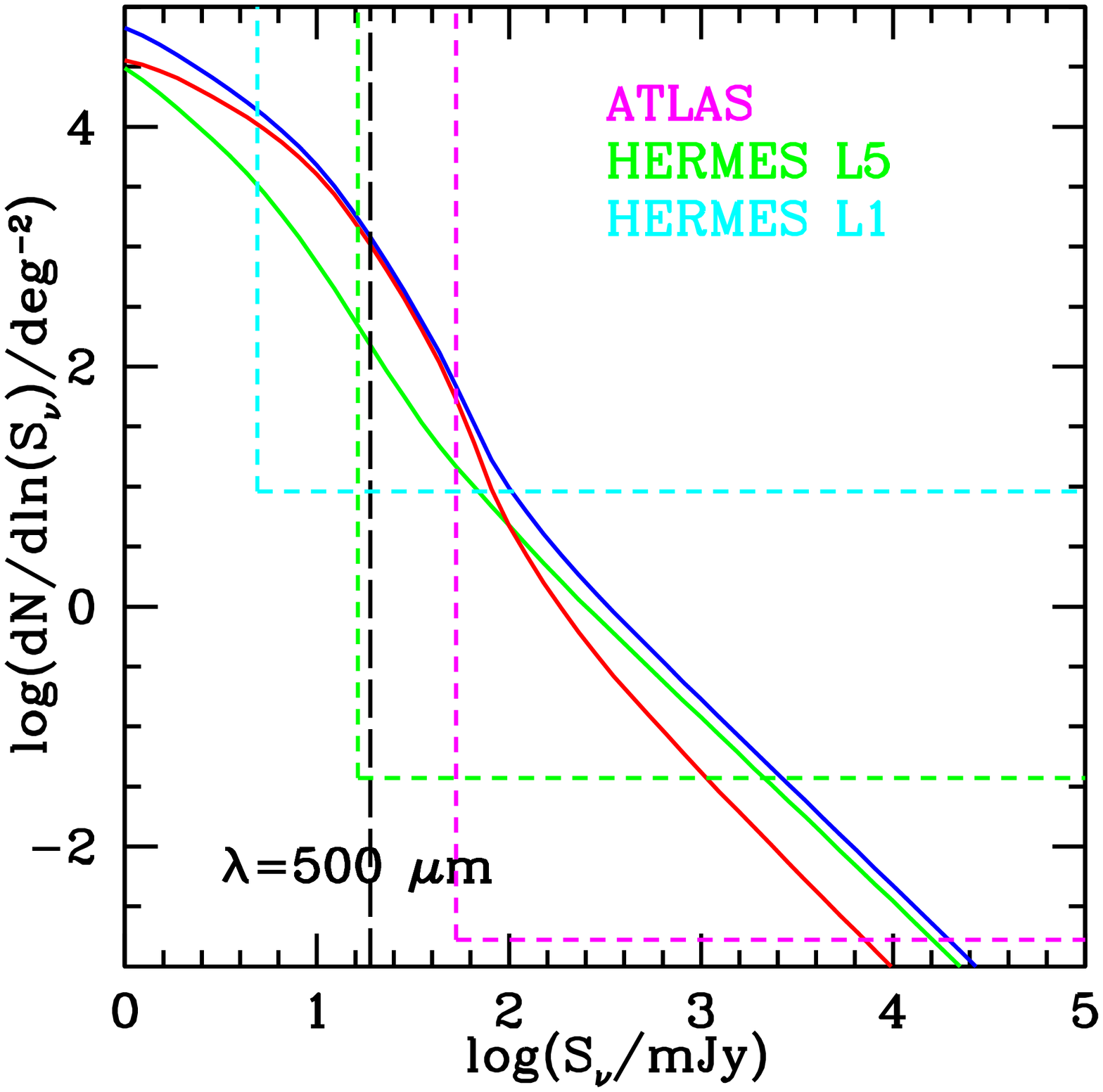}
\end{minipage}

\end{center}

\caption{Galaxy differential number counts in the six \HERSCHEL\
  bands. We show three different curves for our standard model - solid
  blue: total counts including dust extinction and emission; solid
  red: ongoing bursts; solid green: quiescent galaxies. The vertical
  dashed black lines show the estimated confusion limit for the model
  at each wavelength, calculated as described in
  \S\ref{ssec:counts}. The vertical and horizontal dashed coloured
  lines show the flux and area limits for some  planned Key
  Project surveys, as indicated in the key. }

\label{fig:ncts}
\end{figure*}

\section{Number counts and redshift distributions}
\label{sec:counts_redshifts}

\subsection{Number counts}
\label{ssec:counts}

We now move on to make predictions for the quantities which can be
measured most directly in cosmological galaxy surveys with
\HERSCHEL. The planned cosmological surveys will concentrate on
imaging in the three PACS bands (centred at 70, 100 and 160$\mum$) and
the three SPIRE bands (centred at 250, 350 and 500$\mum$); \HERSCHEL\
spectroscopy will be limited to fewer and brighter sources. The
simplest observable quantity is the number counts per solid angle of
galaxies as a function of flux, $S_\nu$, which we plot in the form
$dN/d\ln S_{\nu}$.

An observational estimate of the number counts in the SPIRE bands
has already been been made using the BLAST balloon-borne telescope
\citep{Devlin09,Patanchon09}, so we first compare our model
predictions with those data in Fig.~\ref{fig:ncts_BLAST}. The BLAST
results are for a single $9\deg^2$ field. The BLAST maps have two
times worse angular resolution than \HERSCHEL\, and are also quite
noisy, which leads to serious problems with confusion, incompleteness
and Eddington bias. As discussed by the BLAST authors, this means that
a direct determination of the counts by identifying individual sources
is not feasible. They therefore estimated the counts using a
``$P(D)$'' analysis, fitting a piecewise power-law model to the
distribution of pixel brightnesses in the maps at each wavelength. The
results of this are shown in the figure, with the points showing the
best estimate of the counts at each of the nodes of the model fit, and
the error bars and shaded region showing the 68\% confidence
region. Note that the BLAST $P(D)$ analysis assumed that the sources
were not clustered.  Including clustering could both change the best
estimate values, and increase the error bars on the estimated
counts. We see that the predicted counts agree well with the BLAST
estimates at faint fluxes, but are a little higher at bright fluxes,
especially at the shorter wavelengths. However, given the
aforementioned uncertainties in the BLAST results, and the limited sky
coverage, these differences should not be taken as conclusive at this
stage, and a definitive test must await the results from
\HERSCHEL.

Next, in Fig.~\ref{fig:ncts} we show the predicted differential number
counts in the PACS and SPIRE imaging bands, for the full range of
fluxes and wavelengths covered by \HERSCHEL, with each panel
corresponding to a different wavelength. In each panel, the blue curve
shows the predicted total counts, while the green and red curves show
the separate contributions to these from quiescent galaxies and
ongoing bursts. (A note about our terminology: by ``bursts'' we mean
any galaxy with ongoing star formation in the burst mode, while by
``quiescent'' we mean all other galaxies, whether undergoing star
formation only in the disk mode, or completely passive with no current
star formation. However, completely passive systems contribute very
little to the luminosity functions and number counts at far-IR
wavelengths.)  We see that quiescent galaxies dominate the counts at
brighter fluxes in all bands, while the bursts tend to dominate at
faint fluxes, reflecting the increasing domination of the starburst
mode of star formation at higher redshifts.

We can use our model to predict the flux levels at which sources
should become confused in the different \HERSCHEL\ bands. Source
confusion happens when multiple sources are separated in angle by less
than the angular resolution of the telescope, and so appear merged
together in images. Since the number of sources increases with
decreasing flux, confusion sets a lower limit to the flux at which one
can still identify separate sources in an image (regardless of the
integration time used). Confusion will be a serious problem in deep
cosmological surveys with \HERSCHEL\ due to the relatively poor
angular resolution of the telescope (compared to optical
telescopes). This confusion limit depends both on the angular
resolution of the telescope and on the actual surface density of
sources per solid angle as a function of flux. We estimate the
confusion limit using the {\em source density criterion}
\citep{Condon74,Vaisanen01}: if the telescope has a FWHM (full width
at half maximum) beamwidth of $\theta_{FWHM}$, we define the effective
beam solid angle as $\omega_{beam}= (\pi/(4\ln2))\,\theta_{FWHM}^2 =
1.13\theta_{FWHM}^2$, and then define the confusion limited flux
$S_{conf}$ to be such that $N(>S_{conf}) = 1/({\cal
N}_{beam}\omega_{beam})$, where $N(>S)$ is the number per solid angle
of sources brighter than flux $S$. This estimate ignores any
clustering of the sources. We choose ${\cal N}_{beam}=20$ for the
number of beams per source at the confusion limit, which gives similar
results to more detailed analyses \citep{Vaisanen01,Dole04b}. We have
calculated confusion limits in the different \HERSCHEL\ bands using
our predicted number counts, together with the assumed beamsizes given
in Table~\ref{tab:conf_limits}. The predicted confusion limited fluxes
are given in the table and plotted as vertical black dashed lines in
Fig.~\ref{fig:ncts}. Since \HERSCHEL\ images will be essentially
diffraction-limited, confusion sets in at a lower source density at
longer wavelengths, which typically implies a brighter flux.

\begin{table}
\begin{center}
\begin{tabular}{lccccc}
\hline
$\lambda$ & $\theta_{FWHM}$ & $S_{conf}$ & $\gamma_{conf}$ & $z_{50}$
& $z_{90}$ \\
\protect{[$\mum$]}  & [arcsec]     & [$\mJy$] & & & \\
\hline
\hline
70  & 5.2 & 0.24 & 0.88 & 1.15 & 2.53 \\
100 & 7.7 & 1.6 & 0.95 & 0.80 & 1.76 \\
160 & 12  & 9.5 & 1.28 & 0.72 & 1.57 \\
250 & 19  & 22 & 1.78 & 0.70 & 1.64 \\
350 & 24  & 21 & 2.17 & 0.93 & 1.93 \\
500 & 35  & 19 & 2.58 & 1.21 & 1.93 \\
\hline
\end{tabular}
\caption{Predicted confusion limits for \HERSCHEL\ imaging. $\lambda$
  is the wavelength, $\theta_{FWHM}$ is the angular resolution of PACS
  or SPIRE imaging at that wavelength, $S_{conf}$ is the predicted flux at the
  confusion limit, and $\gamma_{conf}$ is the slope of the
  differential source counts at that flux. $z_{50}$ and $z_{90}$ are
  the predicted median and 90-percentile redshifts for galaxies
  brighter than $S_{conf}$.}
\label{tab:conf_limits}
\end{center}
\end{table}

We also indicate in Fig.~\ref{fig:ncts} by vertical and horizontal
dashed coloured lines the regions of flux and surface density that are
planned to be probed by the main cosmological galaxy surveys with
\HERSCHEL. For each survey, the vertical line indicates the nominal
flux limit $S_{min}$ set by integration time and signal-to-noise
(ignoring possible source confusion), and the horizontal line
indicates the minimum surface density of sources that can be probed
given the solid angle $A$ of the survey, which we estimate as
$(dN/d\ln S)_{min} = 1/A$. We consider here the following four planned
surveys, which are all \HERSCHEL\ Key Programmes\footnote{
http://herschel.esac.esa.int/Key\_Programmes.shtml}.  The first three
are deep surveys, while the last is a shallower wide-area survey. The
deep surveys have various tiers, but for simplicity we here consider
only the deepest blank-field tier in each survey, since this sets the
limit for how far down in luminosity the survey can probe at each
redshift, except for HERMES, where we also consider a shallower tier.

\smallskip

\noindent{\em GOODS-Herschel\footnote{http://herschel.esac.esa.int/Docs/KPOT}:}\\
This survey using PACS and SPIRE will have 2 tiers, ``ultradeep'' and
``superdeep'', and will include the deepest imaging in the PACS 100
and 160$\mum$ bands of any of the cosmological key programmes. We
consider here the ultradeep tier, which is in the GOODS-S field.

\noindent{\em PACS Evolutionary Probe
(PEP)\footnote{http://www.mpe.mpg.de/ir/Research/PEP}:}\\ This deep
imaging survey using PACS, which is coordinated with the HERMES
survey, will include blank fields covering a range of 4 in limiting
flux and 50 in area, mainly in the 100 and 160$\mum$ bands. We
consider here the deepest blank field, which is the GOODS-S field.

\smallskip

\noindent{\em Herschel Multi-tiered Extragalactic Survey
(HERMES)\footnote{http://astronomy.sussex.ac.uk/\~{ }sjo/Hermes}:}\\
This survey, using both PACS and SPIRE, and coordinated with PEP, will
be the largest of the cosmological surveys in terms of observing
time. It will include 6 tiers of blank field surveys, covering a range
of 6 in limiting flux and 400 in area. We consider here the deepest
tier, Level-1 (the CDFS/GOODS-S field), and also a shallower tier,
Level-5 (the XMM, ELAIS-N1-SCUBA, Bootes-SCUBA2, EGS-SCUBA2, CDFS and
Lockman fields), which is intermediate in area and in depth in the
SPIRE bands between Level-1 and the shallower ATLAS survey.

\smallskip

\noindent{\em Herschel ATLAS\footnote{http://herschel.esac.esa.int/Docs/KPOT}:}\\
This survey, using PACS and SPIRE, will be the shallowest of the
cosmological surveys, but will cover by far the largest area. There is
only a single tier.

\smallskip

Table~\ref{tab:survey_params} lists the basic parameters (wavelengths,
areas and flux limits ignoring confusion) for these planned surveys
(or tiers within surveys) which we will show in subsequent
figures. These surveys are plotted in Fig.~\ref{fig:ncts}.  These
survey parameters may be modified once the inflight performance of the
telescope is known.

\begin{table*}
\begin{center}
\begin{tabular}{lcccccc}
\hline
Survey & $A$ & $\lambda$ & $S_{min}$ & $N_{gal}$ &
$z_{50}$ & $z_{90}$ \\
       & [$\deg^2$] & [$\mum$]       & [$\mJy$]       &           &         &
         \\
\hline
\hline
GOODS-Herschel            & $0.012$ & $100$ & $0.6$ & 270 & 0.99 & 2.22 \\
Ultra-deep                &         & $160$ & $0.9$ & 540 & 1.17 & 2.57 \\
PEP GOODS-S               & $0.042$ & $70$  & $1.6$ & 130 & 0.64 & 1.51 \\
                          &         & $100$ & $1.7$ & 380 & 0.79 & 1.73 \\
                          &         & $160$ & $1.7$ & $1.1\times10^3$ & 1.09 & 2.28 \\
HERMES Level-1            & $0.11$  & $250$ & $4.2$ & $1.9\times10^3$ & 1.25 & 2.42 \\
                          &         & $350$ & $5.7$ & $1.2\times10^3$ & 1.42 & 2.61 \\
                          &         & $500$ & $4.9$ & 930 & 1.66 & 2.98 \\
HERMES Level-5            & $27$    & $100$ & $27$  & $1.0\times10^4$ & 0.32 & 0.78 \\
                          &         & $160$ & $39$  & $1.4\times10^4$ & 0.37 & 0.96 \\
                          &         & $250$ & $14$  & $9.0\times10^4$ & 0.88 & 1.86 \\
                          &         & $350$ & $19$  & $3.4\times10^4$ & 0.97 & 1.99 \\
                          &         & $500$ & $16$  & $1.9\times10^4$ & 1.26 & 2.08 \\
ATLAS                     & $600$   & $100$ & $67$  & $6.9\times10^4$ & 0.21 & 0.50 \\
                          &         & $160$ & $94$  & $7.3\times10^4$ & 0.21 & 0.50 \\
                          &         & $250$ & $46$  & $2.2\times10^5$ & 0.41 & 1.28 \\
                          &         & $350$ & $62$  & $4.6\times10^4$ & 0.27 & 1.33 \\
                          &         & $500$ & $53$  & $1.4\times10^4$ & 0.20 & 1.22 \\ 
\hline
\end{tabular}
\caption{ Basic parameters for the surveys (or tiers within surveys)
being modelled.The area $A$ is the total value for that tier, and may
include multiple fields. Limiting fluxes $S_{min}$ are $5\sigma$ and
ignore confusion. For HERMES Level-5 some of the flux limits vary
between different fields, so we give the values corresponding to the
majority of the survey area. $N_{gal}$ is the predicted total number
of galaxies in that tier, and $z_{50}$ and $z_{90}$ are their
predicted median and 90-percentile redshifts.  }
\label{tab:survey_params}
\end{center}
\end{table*}

It can be seen from Fig.~\ref{fig:ncts} that, taken together, the
\HERSCHEL\ surveys at each wavelength (excepting 70$\mum$) should
contain galaxies covering a huge range ($\sim 10^4 - 10^5$) in flux.
At 70$\mum$, the planned \HERSCHEL\ surveys probe only slightly deeper
than surveys already carried out using \SPITZER. We show in the panel
the regions of flux and source density probed by two of these
\SPITZER\ surveys, SWIRE\footnote{http://swire.ipac.caltech.edu/swire}
( $A=49\deg^2$, $S_{min}=18\mJy$ \citealt{Lonsdale03}) and
FIDEL\footnote{http://irsa.ipac.caltech.edu/data/SPITZER/FIDEL}
($A=0.5\deg^2$, $S_{min}=3.3\mJy$, \citealt{Dickinson07,Huynh07}). The
\HERSCHEL\ PEP GOODS-S survey will probe about two times fainter than
\SPITZER\, but still well above the predicted \HERSCHEL\ confusion
limit. At 100$\mum$, the PEP survey is predicted to be at the
confusion limit, while the  GOODS-HERSCHEL survey will be below
it. Confusion is predicted to be worst in the 160$\mum$ deep surveys,
where the sensitivity of the GOODS-HERSCHEL survey will be 10 times
below the predicted confusion limit, and that for PEP GOODS-S 6 times
below it. At the longer SPIRE wavelengths (250, 350 and 500$\mum$),
the deepest blank field tier (L1) of the HERMES survey will be 4--5
times fainter than the predicted confusion limit, while L5 tier will
be at or slightly below confusion. The ATLAS survey is predicted to be
safely above the confusion limit (by factors of 2--40) at all
wavelengths.

Various techniques will allow one to probe observationally sources
fainter than the formal confusion limit, the principal ones being {\em
pixel brightness distributions}, {\em gravitational lensing} and {\em
multi-wavelength analysis}. In {\em pixel brightness distribution} (or
$P(D)$) methods, one uses the distribution of pixel brightness to make
statistical inferences about the slope and amplitude of the source
counts at and below the confusion limit, without trying to identify
the individual sources responsible \citep[e.g.][]{Patanchon09}. In
{\em gravitational lensing}, one uses the gravitational magnification
of the images of faint background galaxies by a foreground galaxy
cluster to reduce confusion effects. Since both source fluxes and
areas get multiplied by the magnification ${\cal A} > 1$, the cumulative
source counts transform as $N(>S) = (1/{\cal A}) N_0(>S/{\cal A})$,
where $N_0(>S)$ is the unlensed distribution. If the source counts
have the power-law form $dN/d\ln S \propto S^{-\gamma}$, then the
intrinsic (unlensed) flux at the confusion limit is reduced by a
factor ${\cal A}^{-1/\gamma}$ due to lensing.
Table~\ref{tab:conf_limits} lists the source count slopes predicted at
the confusion limit for the different \HERSCHEL\ bands. It can be seen
that the predicted slope $\gamma_{conf}$ at the confusion limit
increases with increasing wavelength, implying that gravitational
lensing can potentially allow one to probe further below the
\HERSCHEL\ confusion limits at shorter wavelengths. For example, for a
gravitational magnification ${\cal A}=10$, the effective confusion
limit is lowered by factors of 0.09--0.4 as the wavelength increases
from 100 to 500$\mum$. Finally, in  {\em multi-wavelength analysis},
one combines images at different wavelengths having different angular
resolutions. Variants of this include {\em  multi-wavelength priors},
where one starts from a source list obtained from higher angular
resolution data at some other wavelength, and tries to extract fluxes
for individual confused sources at the target wavelength, and {\em
multi-wavelength stacking}, where one tries to measure mean fluxes
only \citep[e.g.][]{Dole06,Marsden09}.

\subsection{Redshift distributions}
\label{ssec:redshifts}

\begin{figure*}

\begin{center}

\begin{minipage}{7cm}
\includegraphics[width=7cm]{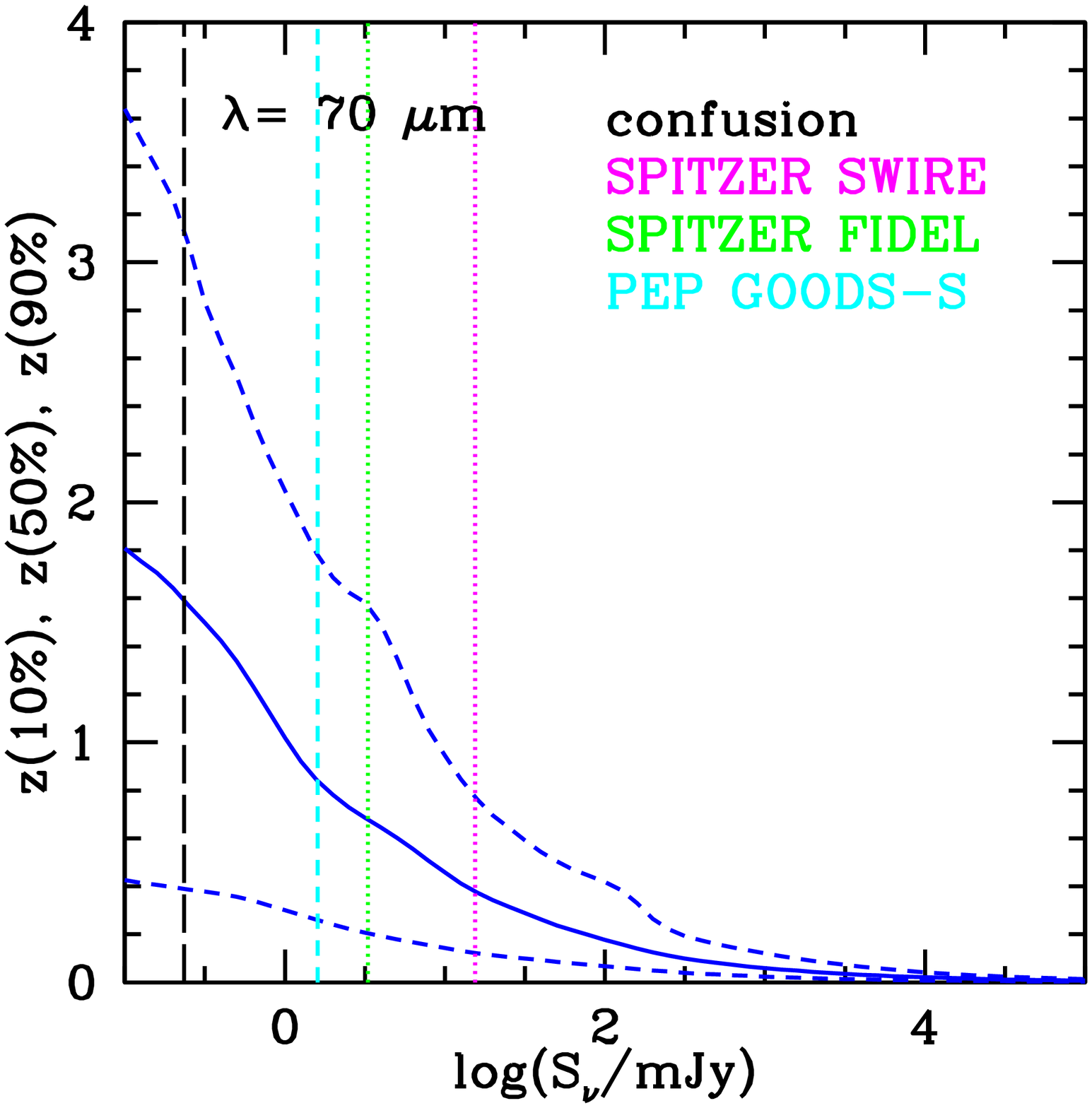}
\end{minipage}
\hspace{1cm}
\begin{minipage}{7cm}
\includegraphics[width=7cm]{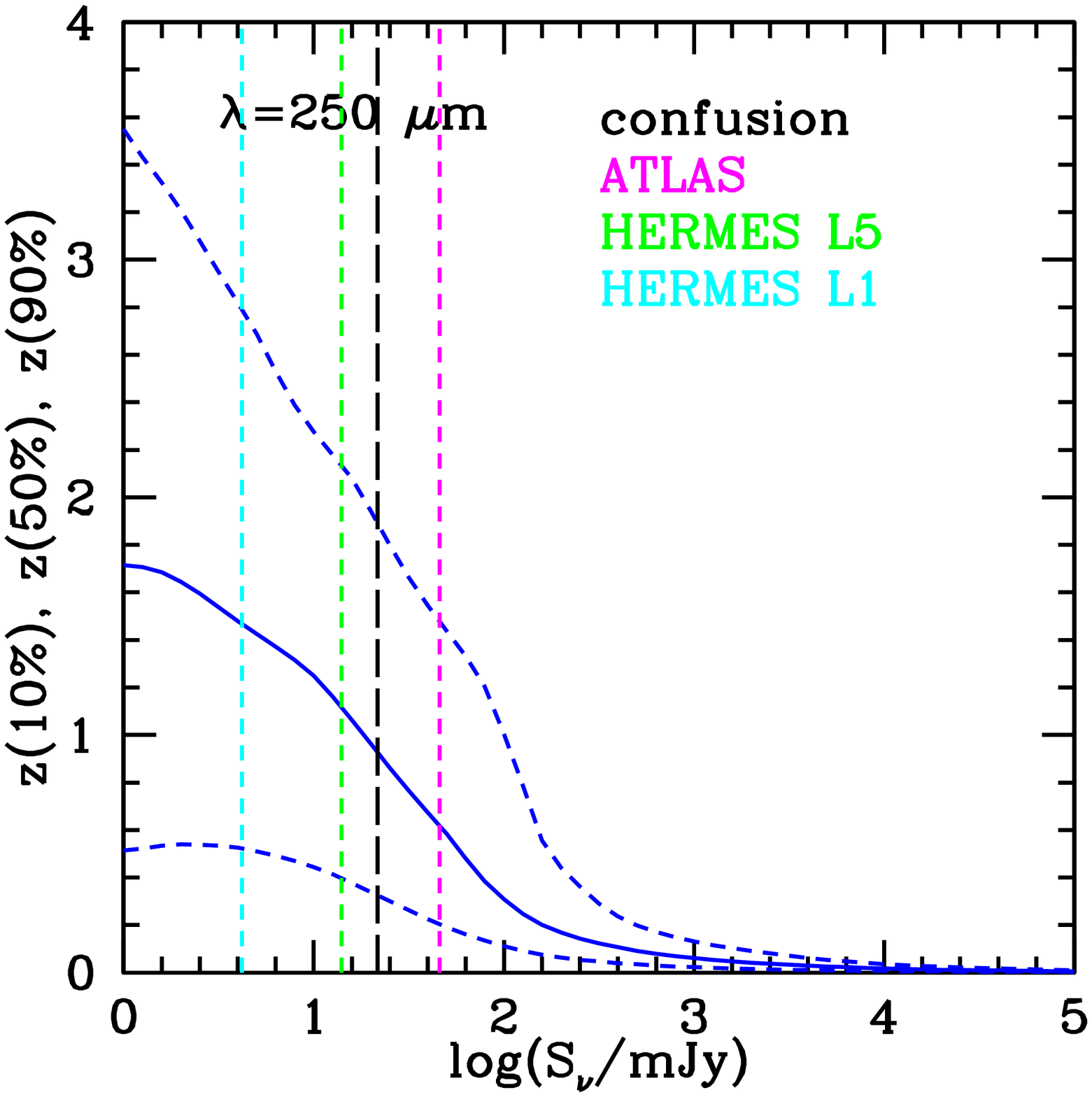}
\end{minipage}

\begin{minipage}{7cm}
\includegraphics[width=7cm]{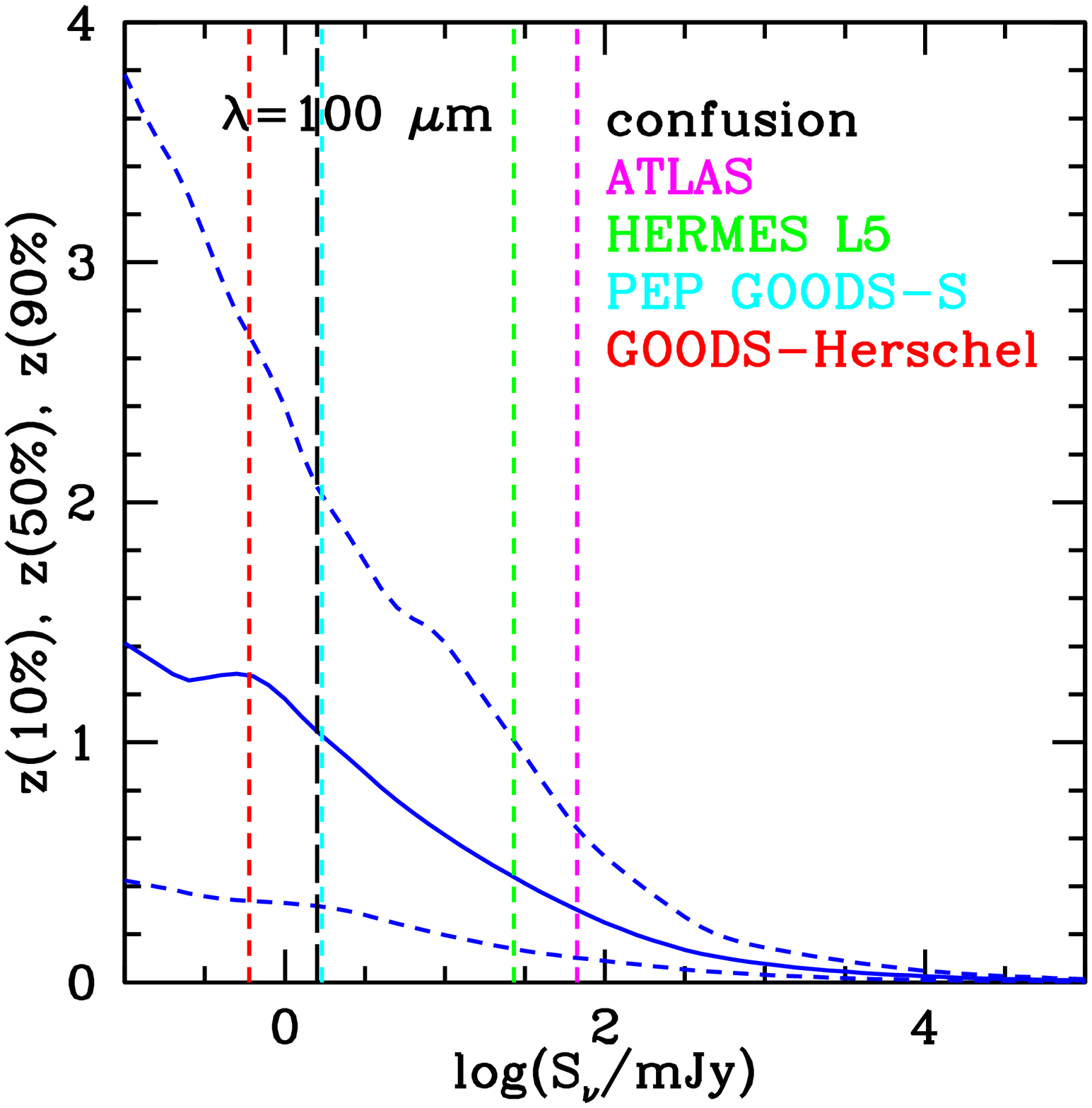}
\end{minipage}
\hspace{1cm}
\begin{minipage}{7cm}
\includegraphics[width=7cm]{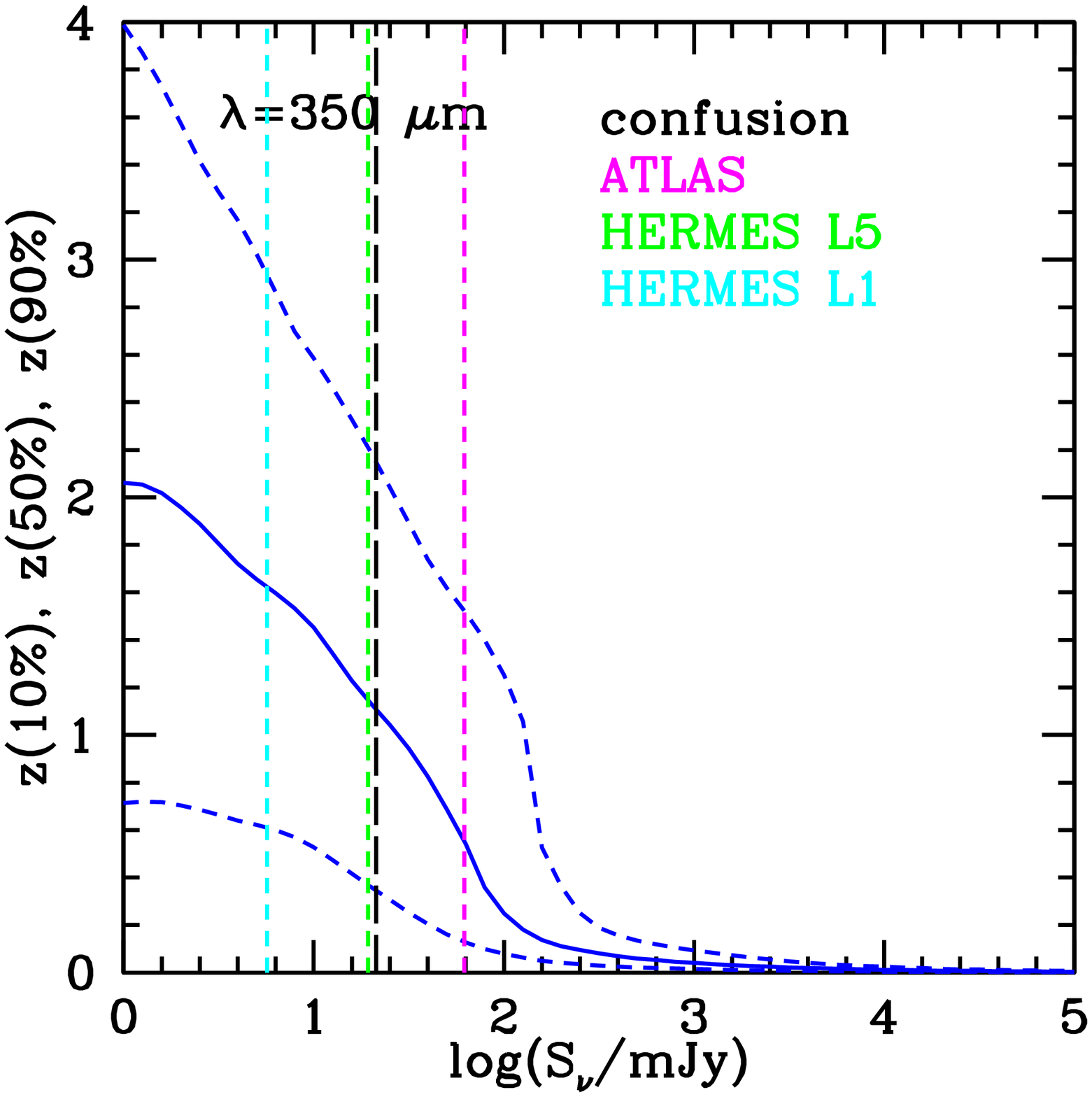}
\end{minipage}

\begin{minipage}{7cm}
\includegraphics[width=7cm]{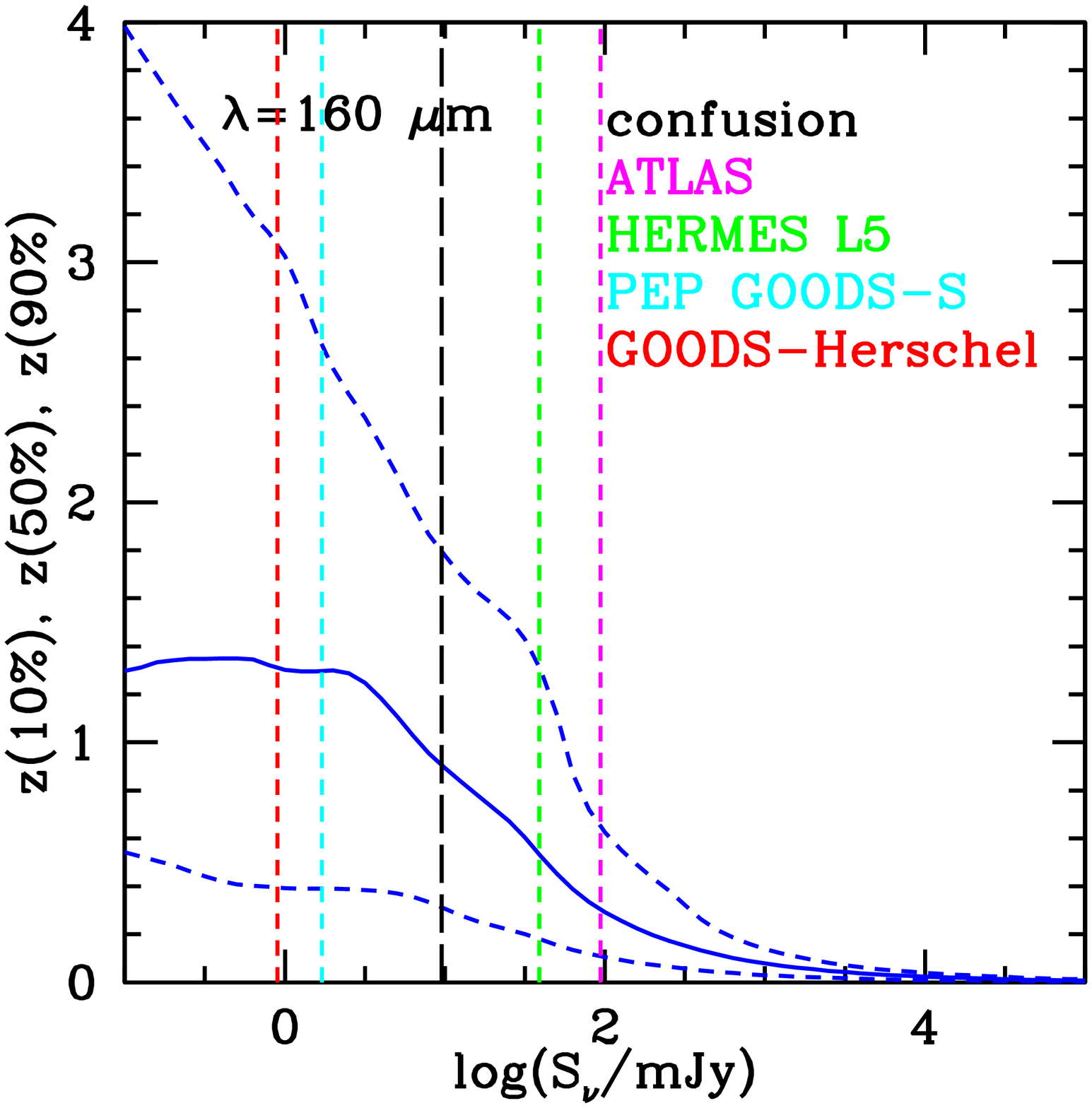}
\end{minipage}
\hspace{1cm}
\begin{minipage}{7cm}
\includegraphics[width=7cm]{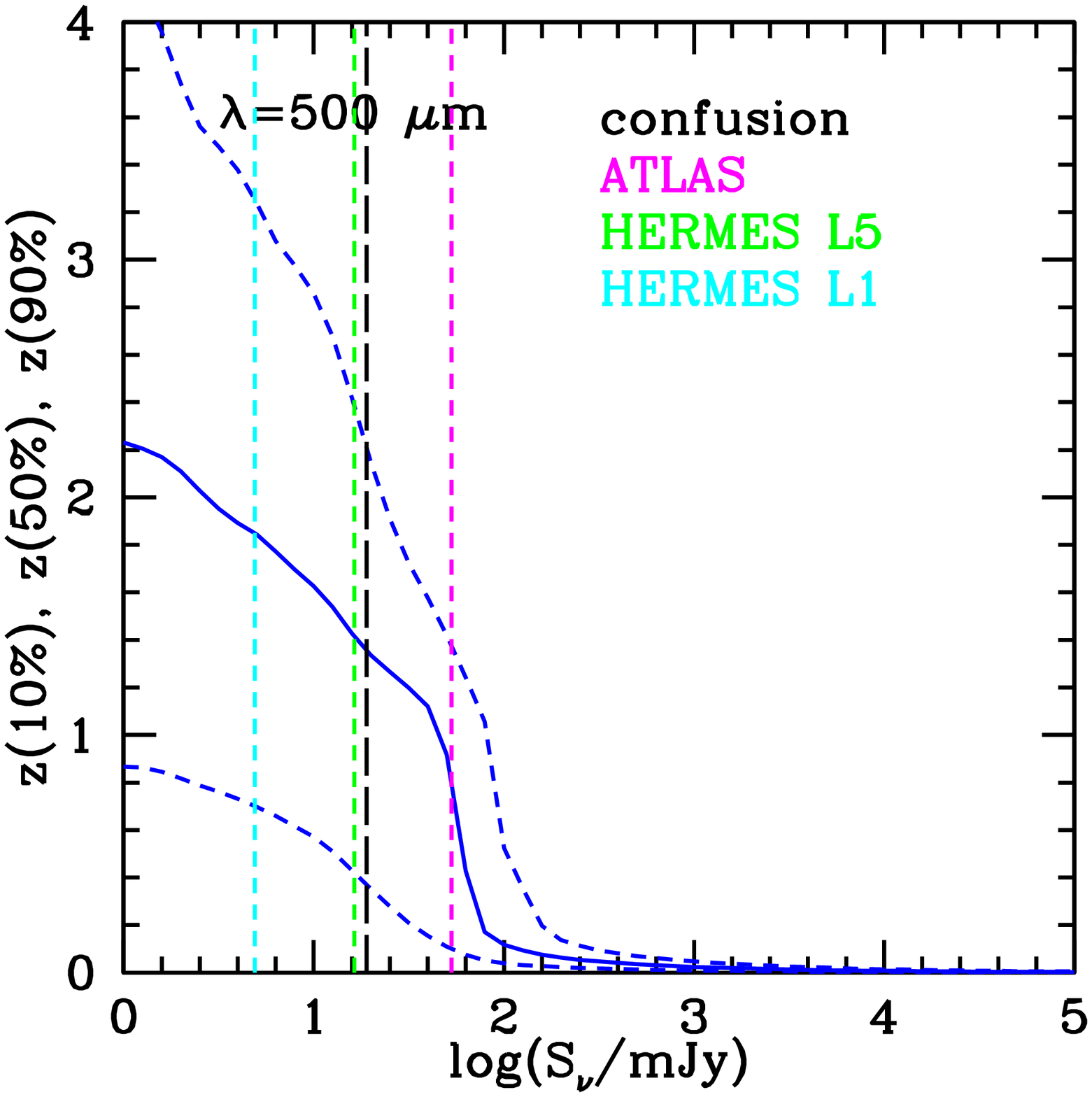}
\end{minipage}

\end{center}

\caption{Median and percentiles of predicted redshift distributions as
  functions of flux in the six \HERSCHEL\ bands. The solid blue lines
  show the median and the dashed lines the 10\% and 90\% percentiles
  for galaxies at each flux. The vertical dashed black line shows the
  estimated confusion limit for the model at each wavelength. The
  coloured lines show the flux limits for some planned Key Project
  surveys, as indicated in the key. }

\label{fig:zmedian}
\end{figure*}

Having identified sources in images and measured their counts as a
function of flux, the next key step observationally is to measure
their redshifts (either spectroscopically or using photometric
redshift methods), and construct redshift distributions in different
flux ranges. We therefore consider the model predictions for redshift
distributions next.

Firstly, in Fig.~\ref{fig:zmedian}, we show the predicted median
redshift (together with the 10\% and 90\% percentiles) as a function
of flux in each of the PACS and SPIRE bands. As in
Fig.~\ref{fig:ncts}, we also indicate the flux at the confusion limit
and at the different Key Project survey limits by vertical dashed
black and coloured lines respectively. The predicted median and 90\%
percentile redshifts for each survey tier are also given in
Table~\ref{tab:survey_params}. We see from the figure that the median
redshifts for sources at the confusion limit are around $z_{50} \sim
1-1.5$, with the highest values at the shortest and longest \HERSCHEL\
wavelengths. If one assumes that the confusion limit can be completely
circumvented, then of the surveys listed, HERMES-L1 at 500$\mum$ will
probe to the highest median redshift ($z_{50} \approx 1.8$). If one
assumes instead that confusion sets a hard limit, then the highest
median redshift is reduced to $z_{50} \approx 1.4$, again achieved in
the HERMES-L1 survey at 500$\mum$.

\begin{figure}
\begin{center}
\includegraphics[width=7cm]{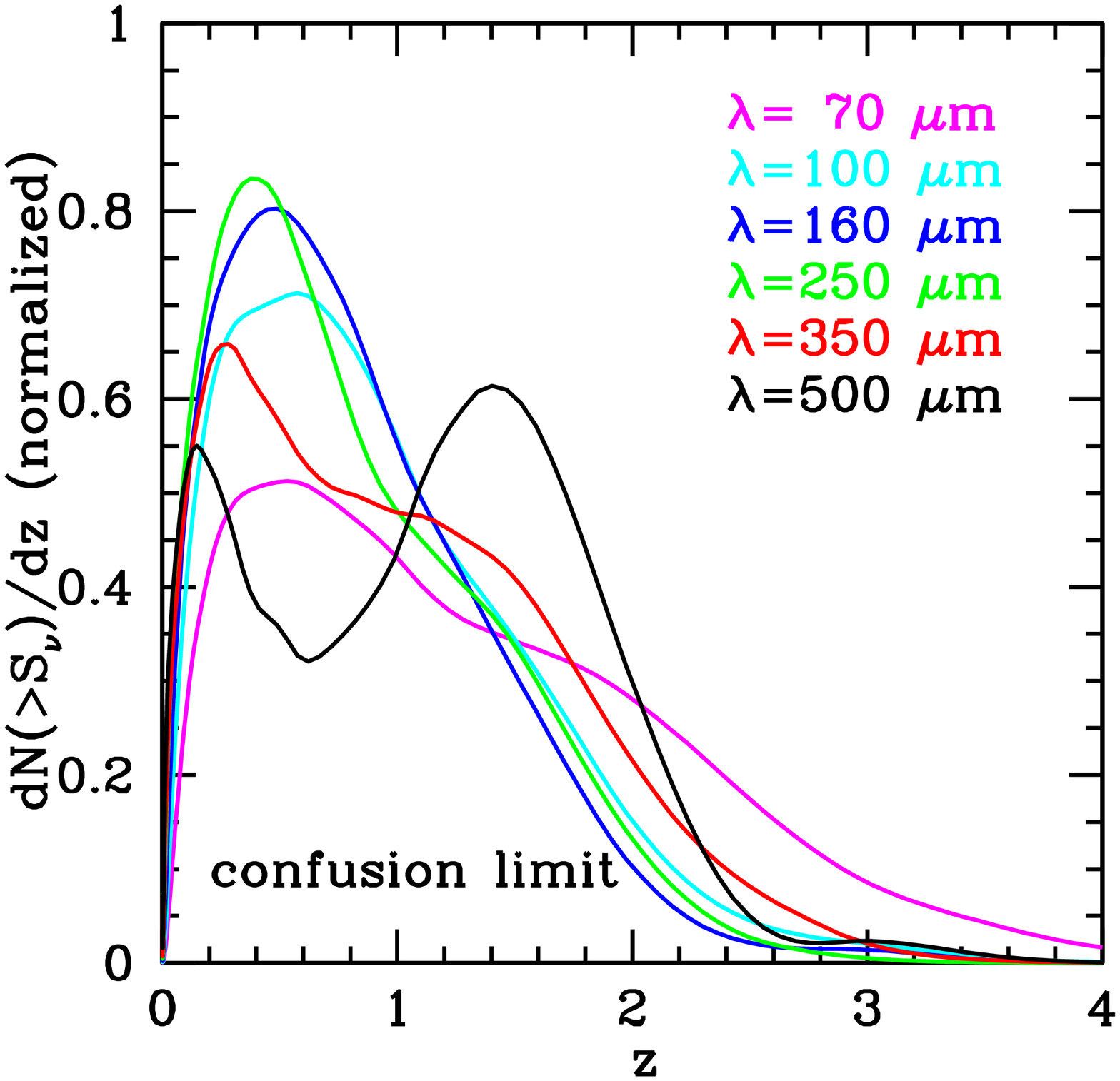}
\end{center}

\caption{Redshift distributions for galaxies brighter than the
  predicted confusion limits in each \HERSCHEL\ band, as indicated in
  the key. The redshift distributions are all normalized to unit area
  under the curve. }
\label{fig:dndz-conf}
\end{figure}

We now look at the redshift distributions in more
detail. Fig.~\ref{fig:dndz-conf} shows the predicted redshift
distributions for galaxies at the confusion limits listed in
Table~\ref{tab:conf_limits}. In this figure, the reshift distributions
have all been normalized to unit area under the curve to allow easier
comparison of their shapes. We have included the redshift distribution
at the 70$\mum$ confusion limit for completeness, even though none of
the planned surveys at 70$\mum$ will go this faint. Apart from
500$\mum$, all of the redshift distributions peak at quite modest
redshifts, $z \sim 0.4-0.8$, although there is a tail of objects to $z
\sim 2$.  The redshift peak gets broader with increasing wavelength,
until at the longest wavelength, 500$\mum$, it splits into two peaks,
with the main peak at $z \approx 1.4$ and a smaller peak at $z \approx
0.2$. This effect at 500$\mum$ is a manifestation of the {\em negative
k-correction}, whereby redshifting of the SED combined with the
negative slope of the dust SED longwards of the peak counteracts the
dimming due to the increasing luminosity distance, and makes higher
redshift galaxies more easily visible than lower redshift
galaxies. This effect is already well known from longer-wavelength
sub-mm observations at 850$\mum$ \citep[e.g.][]{Hughes98}.

\begin{figure*}
\begin{center}

\begin{minipage}{7cm}
\includegraphics[width=7cm]{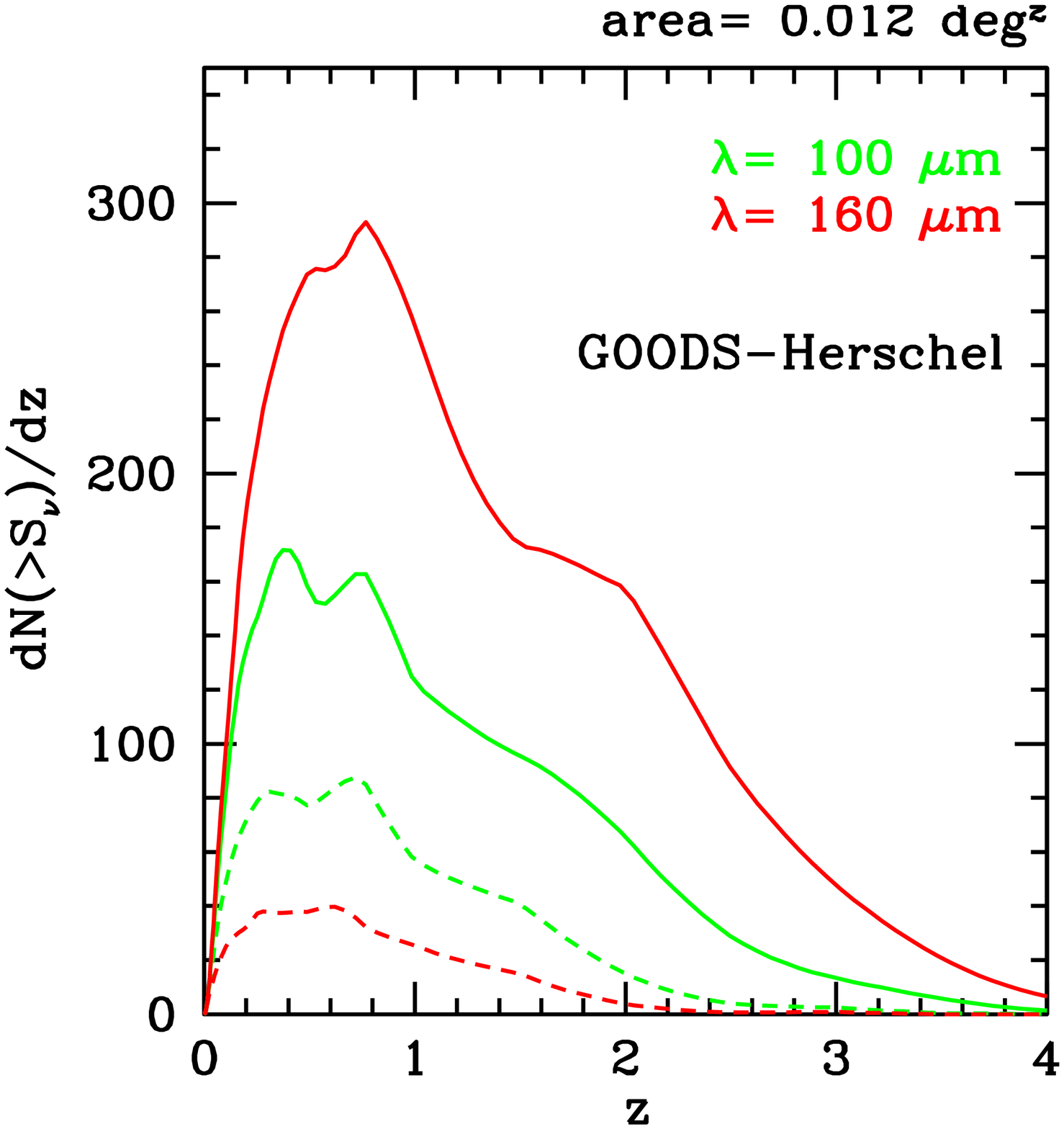}
\end{minipage}
\hspace{1cm}
\begin{minipage}{7cm}
\includegraphics[width=7cm]{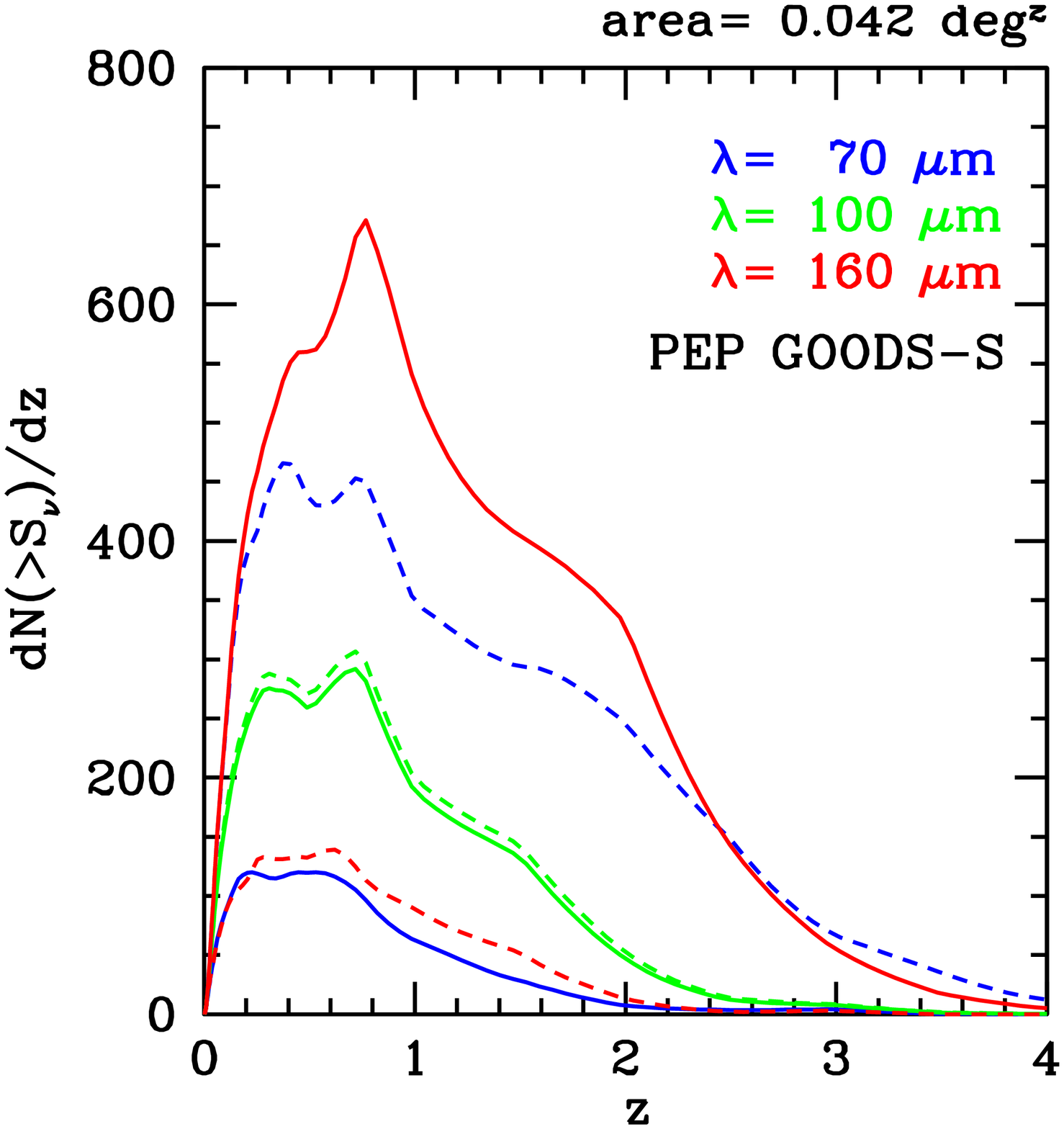}
\end{minipage}

\begin{minipage}{7cm}
\includegraphics[width=7cm]{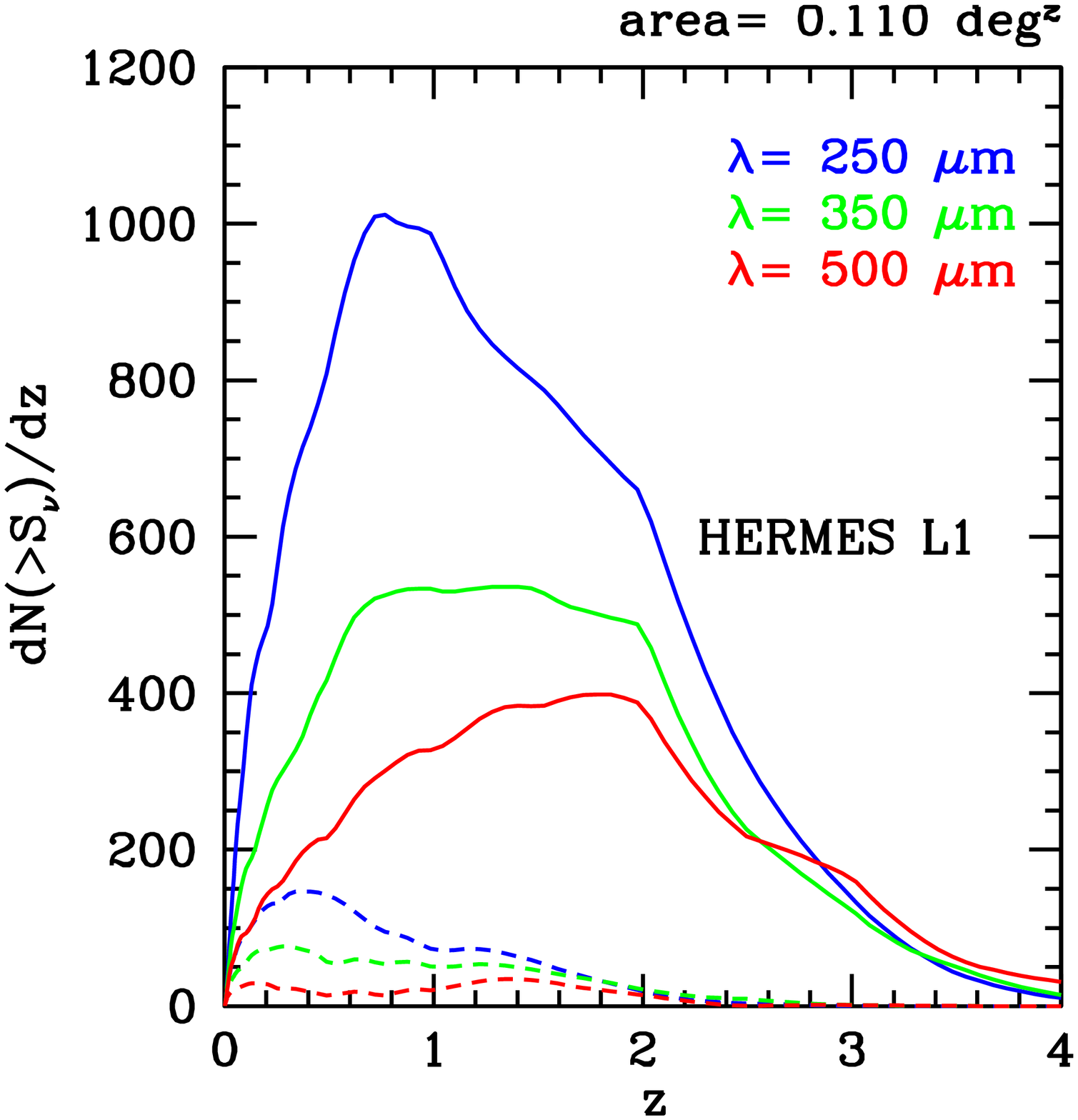}
\end{minipage}
\hspace{1cm}
\begin{minipage}{7cm}
\includegraphics[width=7cm]{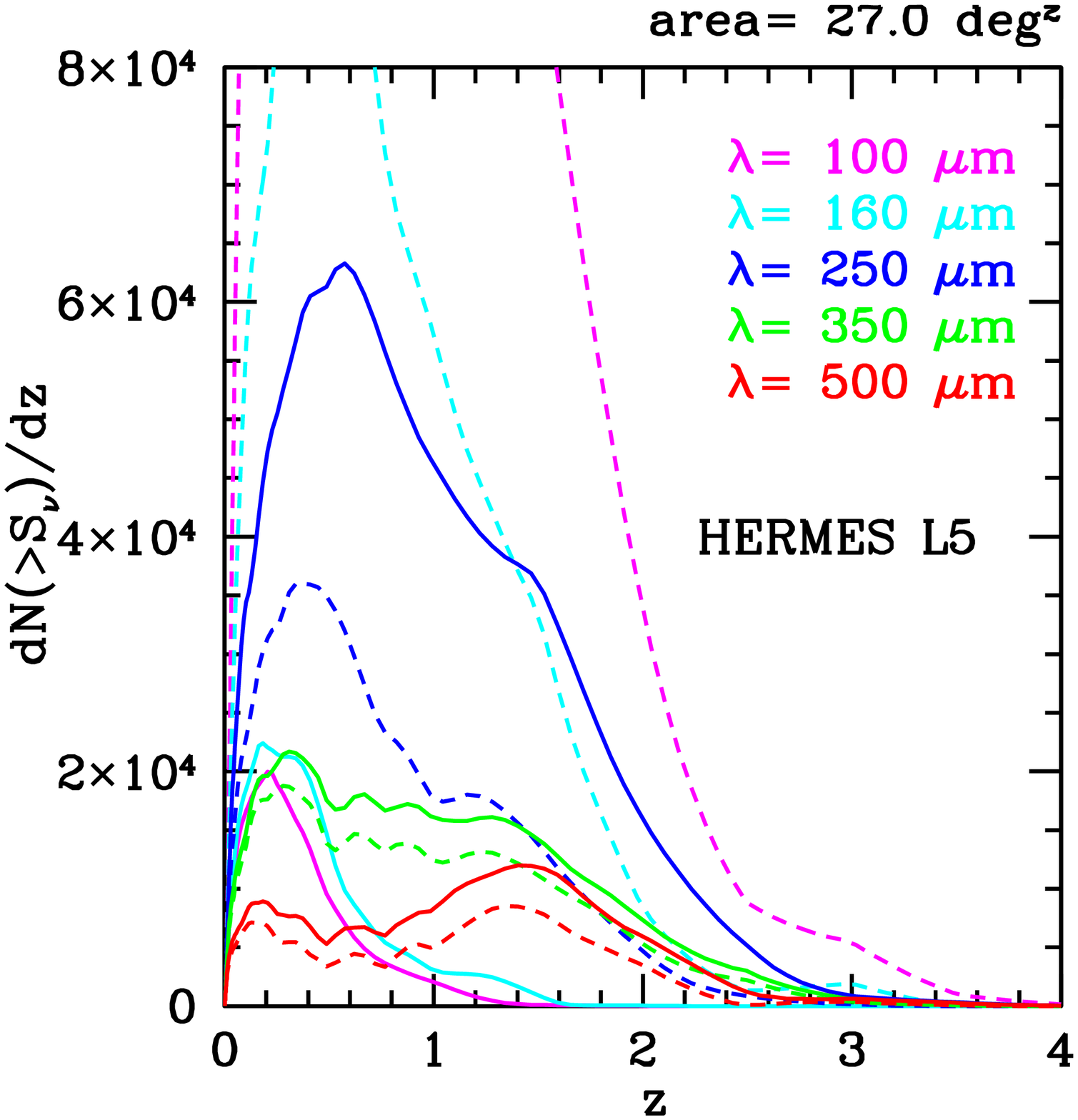}
\end{minipage}

\end{center}

\caption{Redshift distributions in planned deep blank-field
surveys. (a) GOODS-Herschel. (b) PEP GOODS-S. (c) HERMES L1
(CDFS). (d) HERMES L5. In each panel, redshift distributions for the
different wavelengths are shown in different colours, as indicated in
the key. The solid lines show the redshift distribution for galaxies
brighter than the nominal flux limit of the survey, while the dashed lines show the
redshift distribution for galaxies brighter than the predicted
confusion limit. In all panels, the redshift distributions are
normalized so that the area the curve is equal to the predicted number
of galaxies in the survey area.}

\label{fig:dndz-deep}
\end{figure*}

\begin{figure}
\begin{center}
\includegraphics[width=7cm]{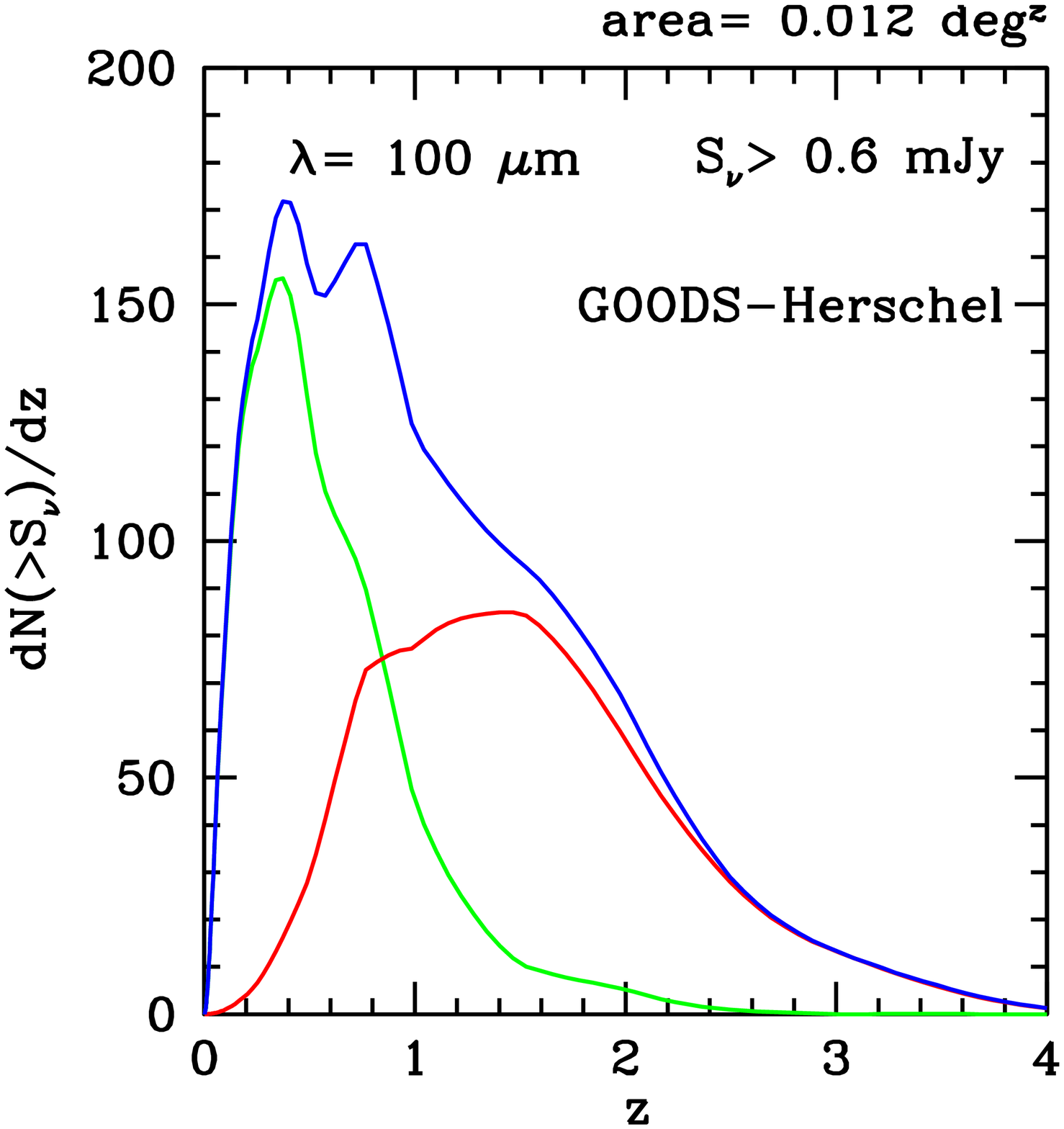}

\includegraphics[width=7cm]{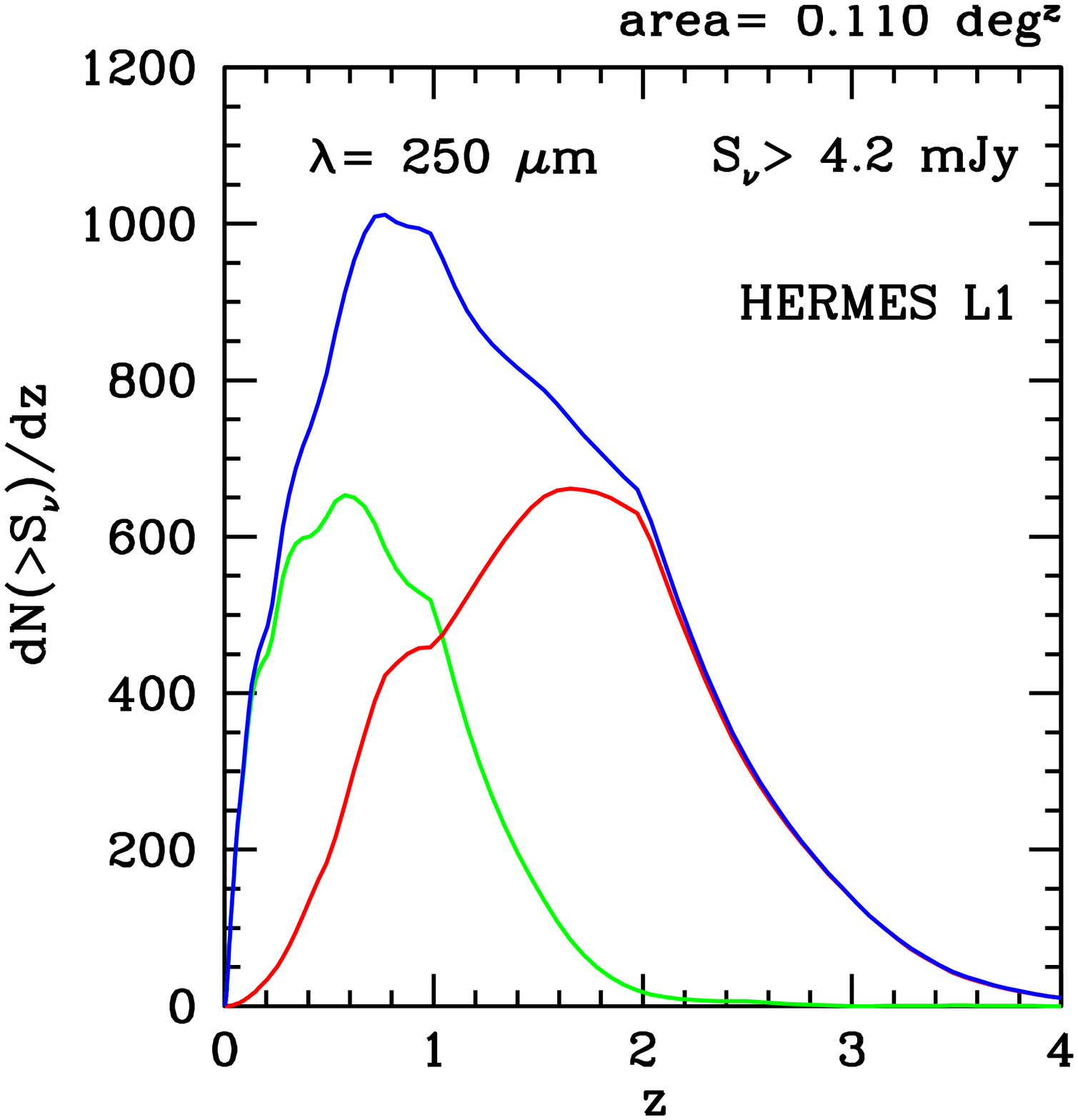}

\includegraphics[width=7cm]{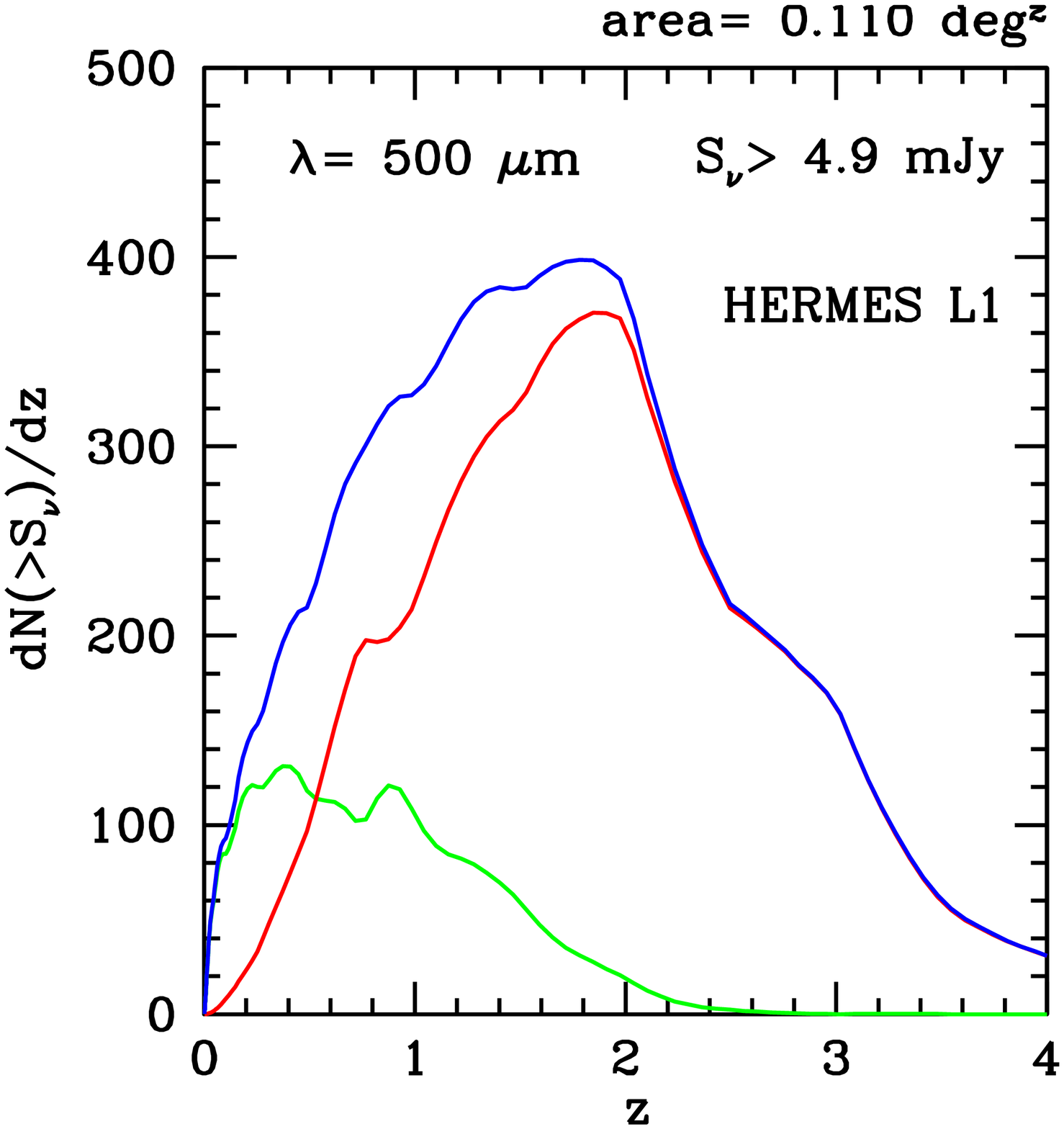}

\end{center}

\caption{Redshift distributions for galaxies brighter than the flux
  limits of the deepest planned blank-field surveys, showing the total
  (blue), together with the contributions of burst (red) and quiescent
  (green) galaxies. (a) GOODS-Herschel at $100\mum$. (b) HERMES-L1 at
  $250\mum$. (c) HERMES-L1 at $500\mum$.}

\label{fig:dndz-deep-detail}
\end{figure}

The next figure, Fig.~\ref{fig:dndz-deep}, shows the predicted
redshift distributions at different wavelengths for all of the deep
survey tiers listed in Table~\ref{tab:survey_params}. In this figure,
the redshift distributions are normalized to the expected number of
galaxies in the survey area, to allow an easier estimate of the number
of galaxies predicted in different redshift ranges for each of the
surveys. Fig.~\ref{fig:dndz-deep-detail} examines the redshift
distributions for these deep surveys in more detail, showing the
separate contributions of quiescent and bursting galaxies to the total
redshift distributions for selected survey tiers at particular
wavelengths. We show two survey tiers (GOODS-Herschel and HERMES-L1)
and three wavelengths (100, 250 and 500 $\mum$) to illustrate the
general behaviour. In all cases. we see that the quiescent galaxies
dominate the distribution at low redshifts and the bursts at high
redshifts, reflecting the higher luminosities of the bursts. We also
see that the bursts become more dominant overall at longer wavelengths
in the deep surveys, due to the effects of the negative k-correction.

\begin{figure}
\begin{center}
\includegraphics[width=7cm]{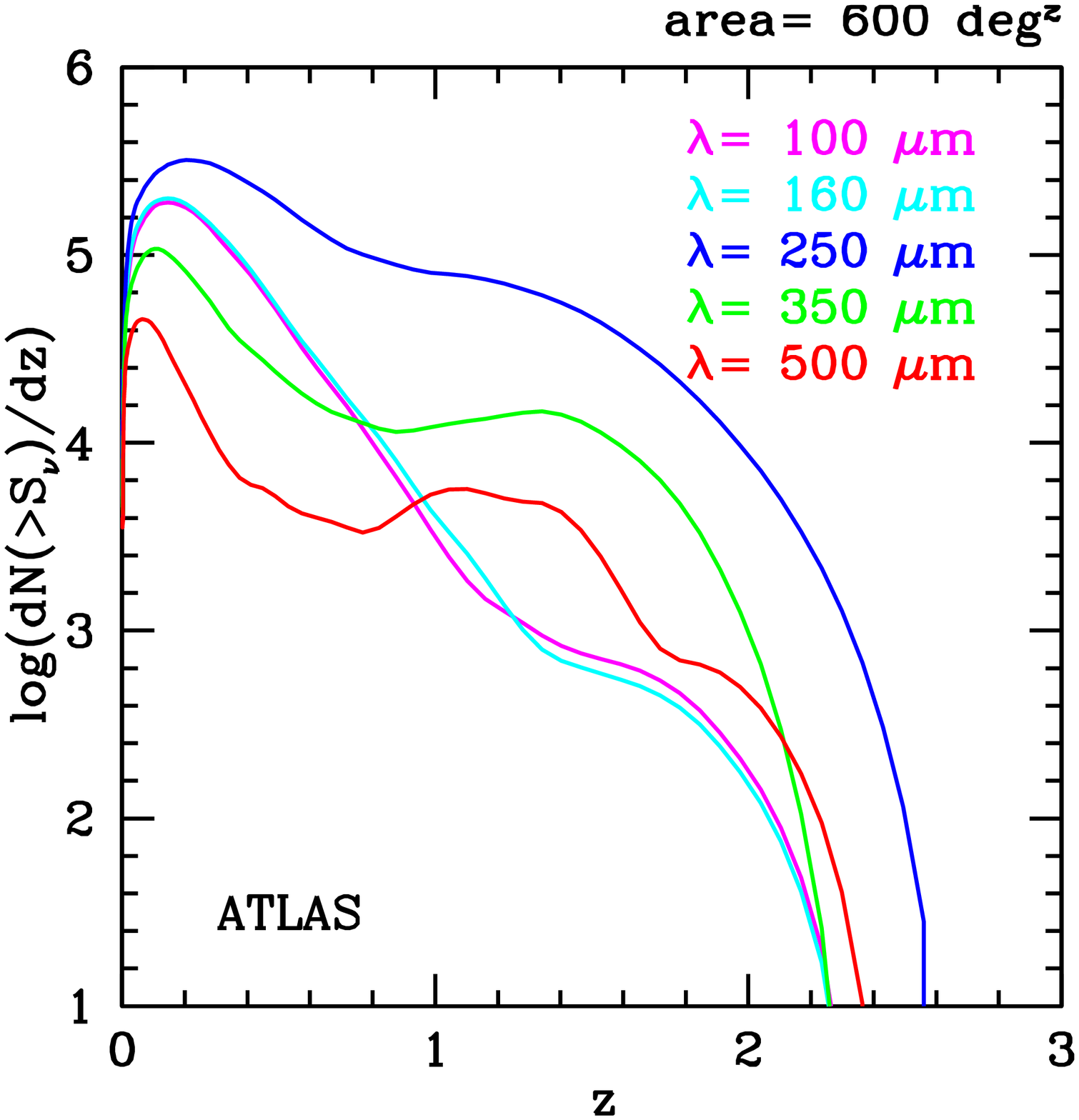}



\end{center}

\caption{Redshift distributions at 100, 160, 250, 350 and 500$\mum$ at
the flux limits of the ATLAS survey. Different wavelengths are shown
by different colours, as indicated in the key. The distributions are
all normalized so that the integral over the distribution is equal to
the expected number of galaxies in the survey area. Note that in this
case the log of the redshift distribution is plotted.}

\label{fig:dndz-ATLAS}
\end{figure}

Finally, in Fig.~\ref{fig:dndz-ATLAS}, we show the predicted redshift
distributions for the shallower but wider-area ATLAS survey. Out of
the 5 wavelengths in this survey, the largest number of galaxies
should be seen at 250$\mum$. The median redshift for galaxies brighter
than the flux limit is also largest at this wavelength ($z_{50}
\approx 0.4$), with $20,000$ galaxies at $z > 1.3$ and $\sim 1000$
galaxies at $z>2$.  The redshift distribution is broader at the longer
wavelengths (250-500 $\mum$) compared to the shorter wavelengths (100
and 160$\mum$), again as a result of the negative k-correction at
longer wavelengths.


\begin{figure}
\begin{center}
\includegraphics[width=7cm]{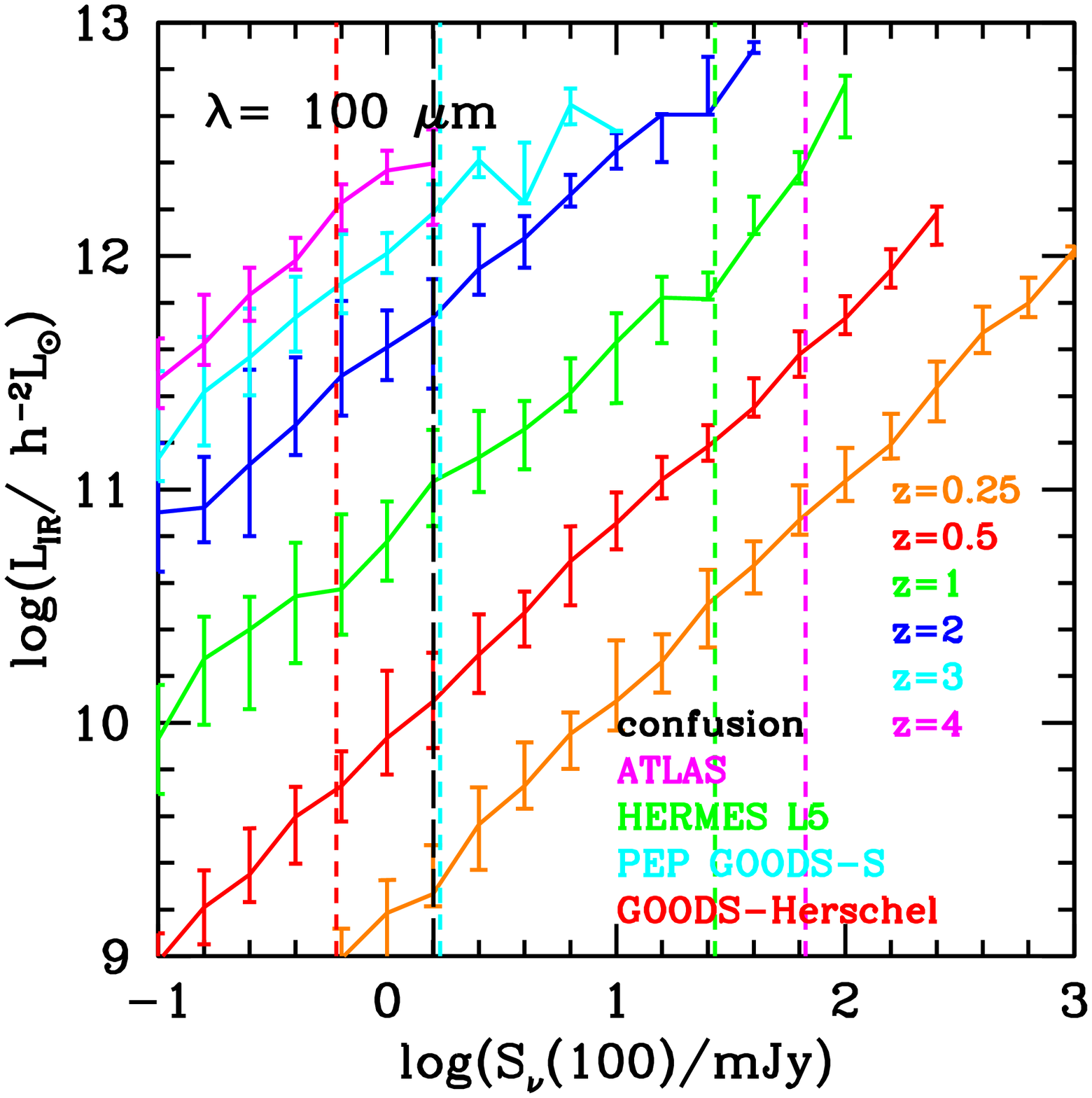}

\includegraphics[width=7cm]{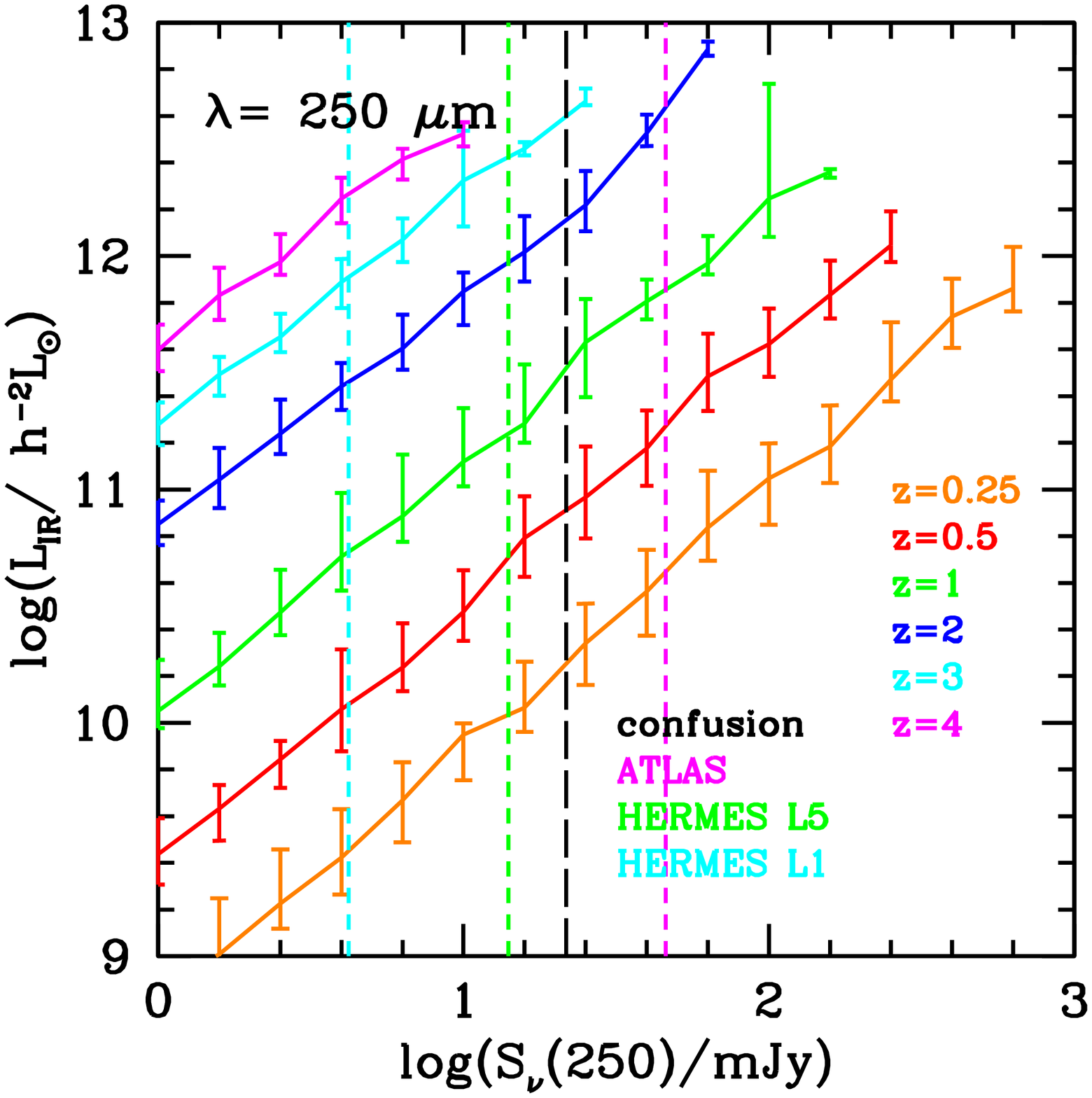}

\includegraphics[width=7cm]{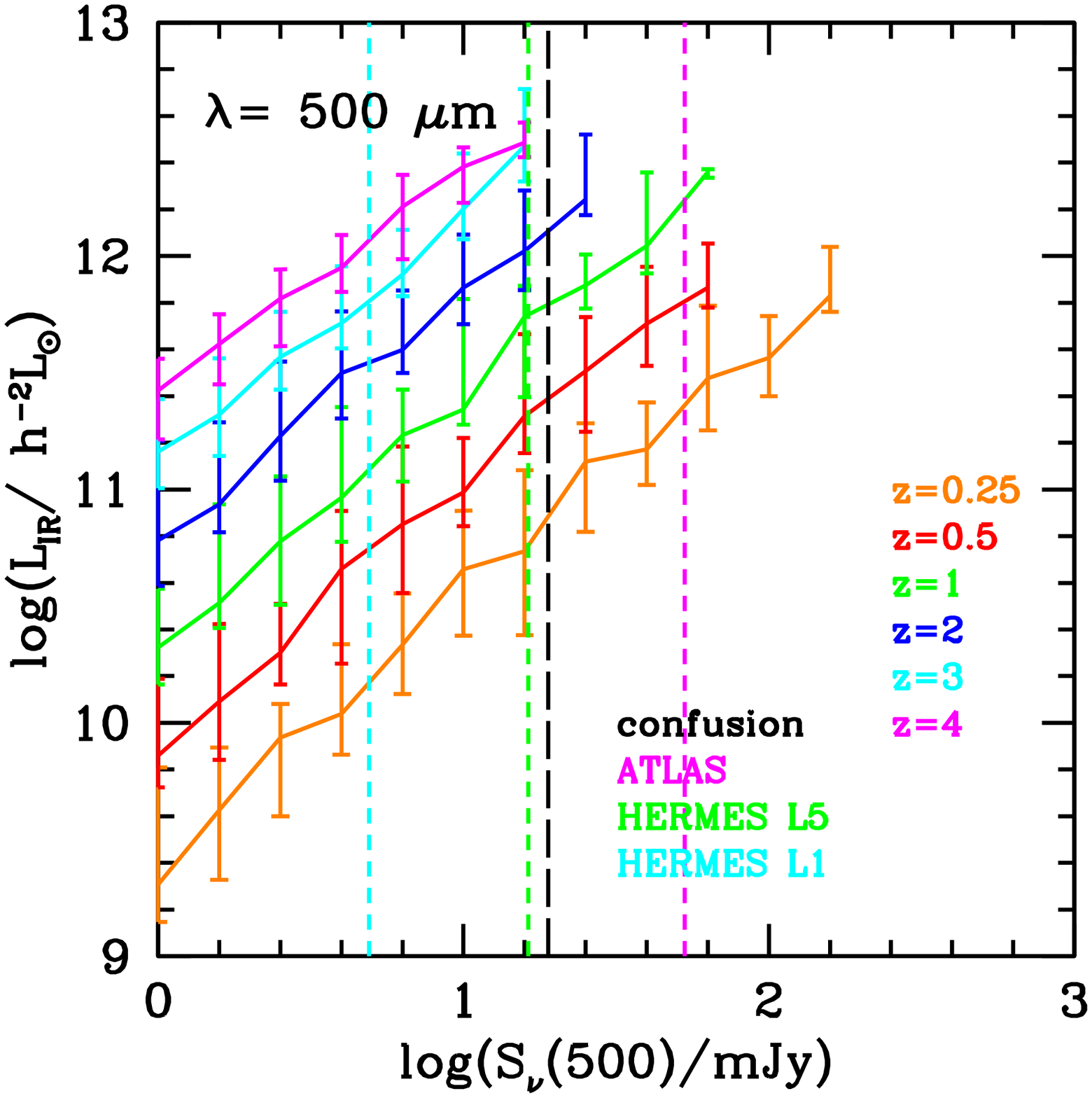}

\end{center}

\caption{Median total IR (8-1000$\mum$) luminosity, $L_{IR}$, {\em vs}
  flux at 100, 250 and 500 $\mum$ for galaxies selected at different
  redshifts, shown by different colours as indicated in the key. The
  ``error bars'' on the lines show the 10-90\% range. The vertical
  dashed black line shows the predicted confusion limit at each
  wavelength, while the vertical dashed coloured lines show the
  nominal flux limits for different planned surveys, as indicated in
  the key.}

\label{fig:LIR_flux}
\end{figure}

\section{Physical properties of \HERSCHEL\ galaxies}
\label{sec:properties}

Having presented predictions for directly observable quantities
(fluxes and redshifts) in the previous section, we now move on to the
predicted physical properties of the galaxies detected in \HERSCHEL\
surveys as a function of flux and redshift. We focus here on four
properties of central physical importance: the total IR luminosity,
$L_{IR}$, the star formation rate, SFR, the stellar mass, $M_{star}$,
and the host dark halo mass, $M_{halo}$.

In Fig.~\ref{fig:LIR_flux}, we plot the median total IR luminosity
$L_{IR}$ (integrated over the wavelength range 8--1000$\mum$) against
flux, for three different \HERSCHEL\ bands (one in each panel), for
galaxies at one of six different redshifts ($z=0.25,0.5,1,2,3,4$, as
indicated by the different colours shown in the key). We have chosen
wavelengths of 100, 250 and 500$\mum$ to be representative of the six
imaging bands. The ``error bars'' on each line show the 10-90\% range
of the distribution at each flux and redshift. We have also plotted
vertical lines showing the nominal flux limits of the different Key
Project surveys discussed in the previous section, as well as the
predicted confusion limit at each wavelength. As mentioned previously,
$L_{IR}$ essentially measures the total luminosity of dust emission
from a galaxy. For galaxies with significant recent star formation,
$L_{IR}$ is powered mostly by far-UV radiation from massive young
stars, and it thus provides a tracer of the dust-obscured star
formation rate for high mass ($m \gsim 5\Msol$) stars. The actual
conversion factor between $L_{IR}$ and the total SFR (integrated over
all stellar masses) depends both on the fraction of the far-UV light
from young stars which is absorbed by dust (which is typically high)
and on the IMF. We see from Fig.~\ref{fig:LIR_flux} that at each
\HERSCHEL\ wavelength and each redshift, there is an approximately
linear relationship between $L_{IR}$ and the flux in that band with
only modest scatter ($\sim 0.2$dex). This simply reflects the fact
that the \HERSCHEL\ bands directly probe the rest-frame far-IR
wavelengths which dominate $L_{IR}$, and that the shape of the far-IR
SED shows only modest galaxy-to-galaxy variation at a given redshift,
with only a weak dependence on $L_{IR}$. The zero-point of this
relation between $L_{IR}$ and flux obviously depends on the SED shape
and on the effects of the luminosity distance and the k-correction;
the zero-point changes with redshift less at longer wavelengths (over
the range $z=0.5-4$, it increases by $\sim 2.5$ dex at 100$\mum$ and
by $\sim 1.5$ dex at 500$\mum$), reflecting the effect of the negative
k-correction at the longer wavelengths.

One of the main goals of \HERSCHEL\ will be to measure the evolution
of the cosmic density of dust-obscured star formation, and for this
purpose it is interesting to know which \HERSCHEL\ wavelength will
probe to the faintest $L_{IR}$ at each redshift, since the latter will
determine what fraction of the total IR luminosity density is actually
resolved into identified objects at that redshift. We see from
Fig.~\ref{fig:LIR_flux} that if source confusion sets a hard limit to
identifying individual sources, then surveys at 100$\mum$ will probe
to the faintest $L_{IR}$, ranging from $\sim 10^{10.1} h^{-2}\Lsol$ at
$z=0.5$ to $\sim 10^{12.4} h^{-2}\Lsol$ at $z=4$. On the other hand,
if all of the planned Key Project surveys manage to resolve objects
down to their nominal flux limits (even where these are below
confusion), then the GOODS-Herschel survey at 160$\mum$ will probe
faintest (down to $L_{IR} \sim 10^{9.5}
h^{-2}\Lsol$ at $z=0.5$ and $\sim 10^{11.9} h^{-2}\Lsol$ at
$z=4$). Of the surveys in the SPIRE bands at 250-500$\mum$, those at
250$\mum$ will probe down to the lowest $L_{IR}$ whether confusion can
be circumvented or not, except at the highest redshifts, $z \gsim
2-3$, for which the 500$\mum$ band becomes more sensitive due to the
negative k-correction effect.

\begin{figure}
\begin{center}
\includegraphics[width=7cm]{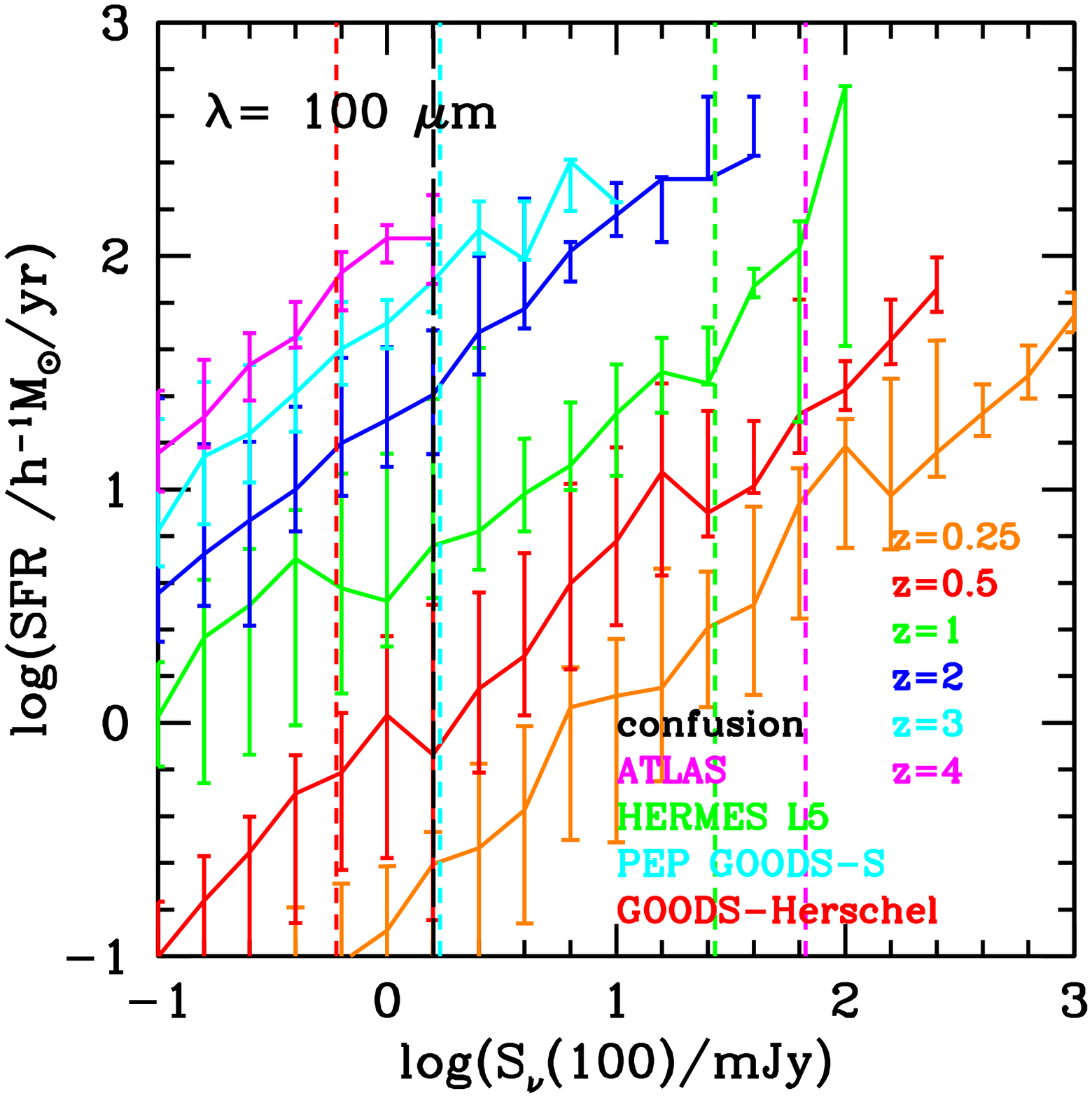}

\includegraphics[width=7cm]{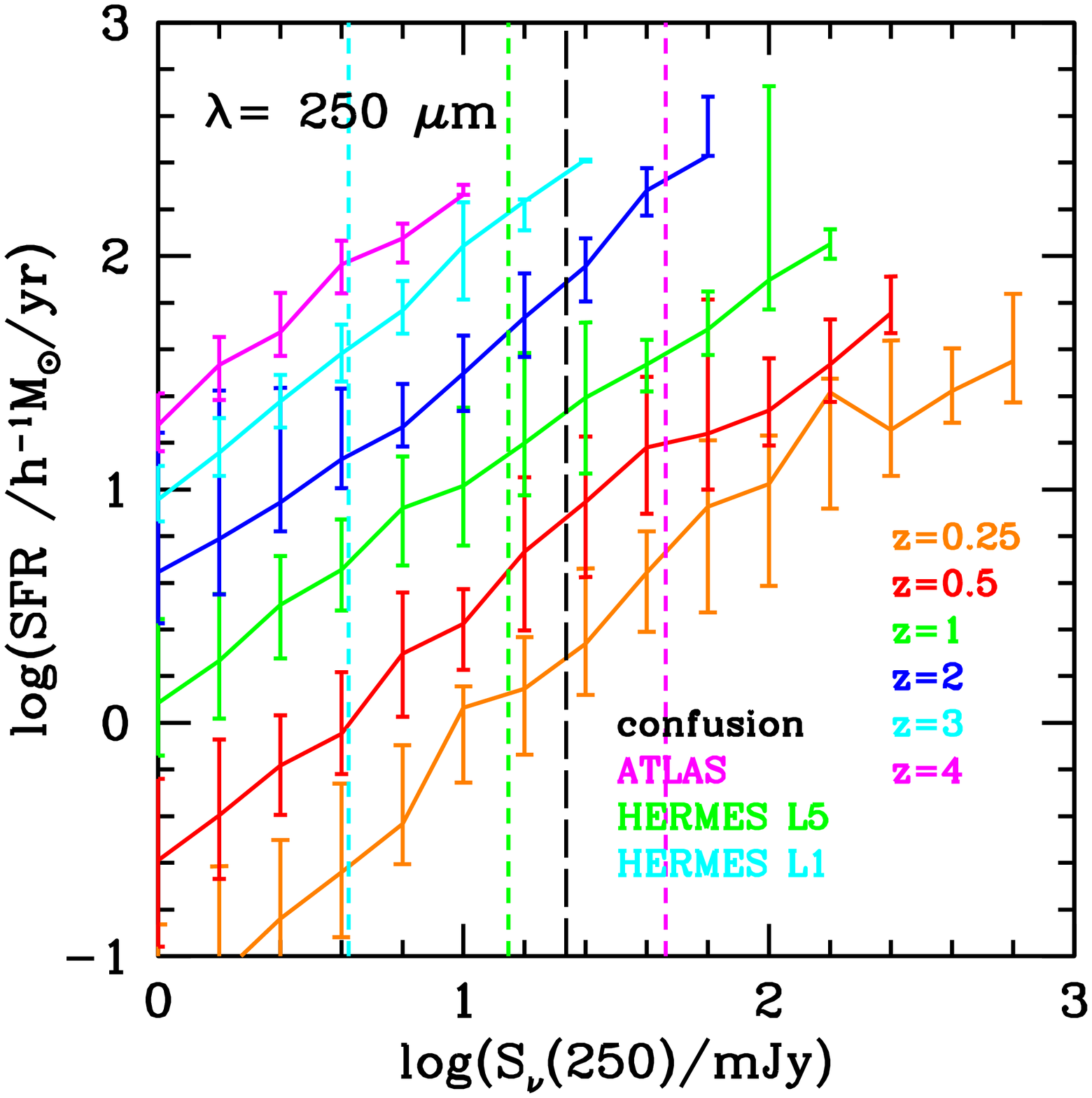}

\includegraphics[width=7cm]{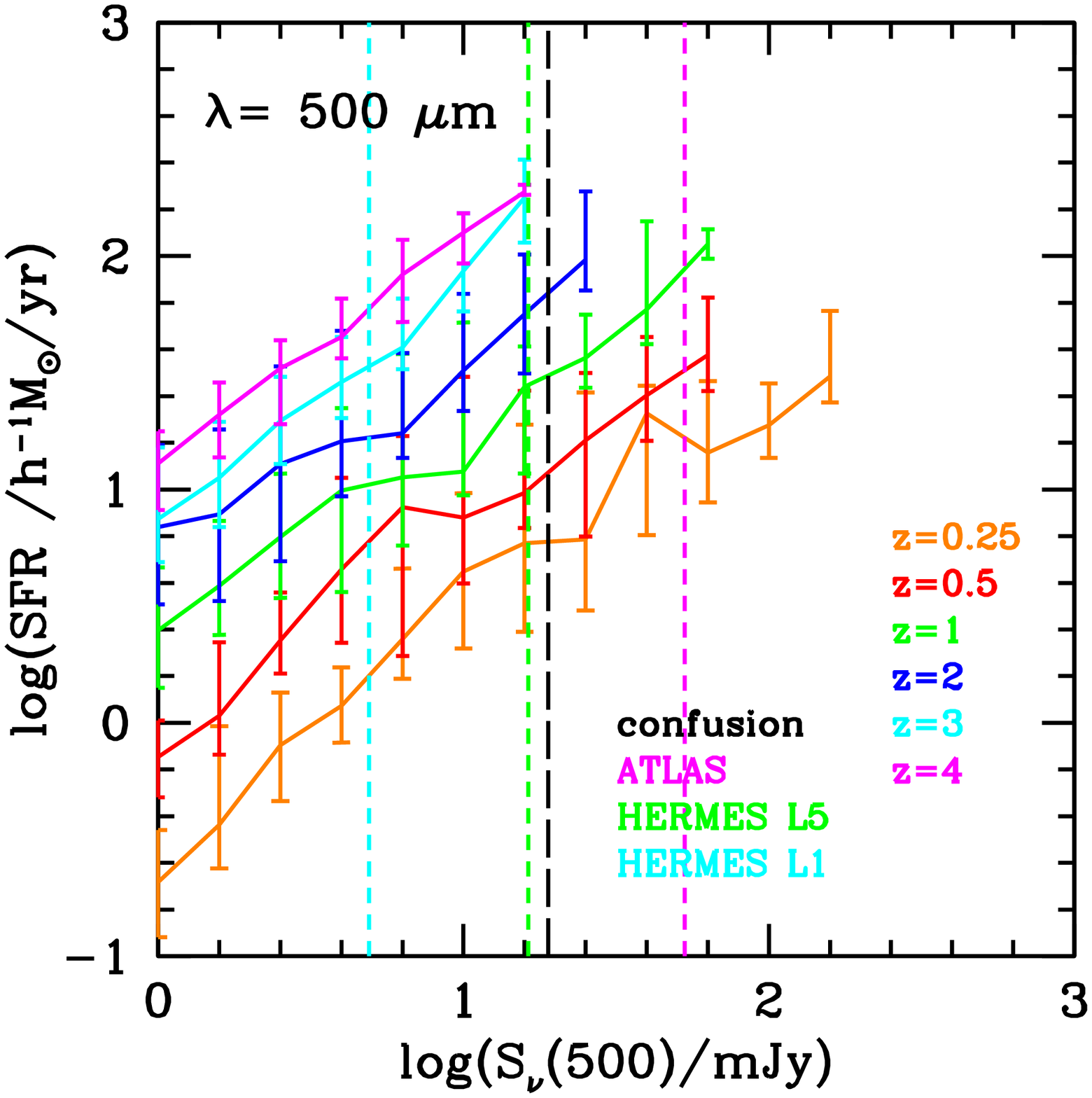}

\end{center}

\caption{Median total SFR {\em vs} flux at 100, 250 and 500 $\mum$ for
  galaxies selected at different redshifts. The different lines are as
  described for Fig.\ref{fig:LIR_flux}.}

\label{fig:SFR_flux}
\end{figure}

Fig.~\ref{fig:SFR_flux} is similar to Fig.~\ref{fig:LIR_flux}, except
that the SFR rather than $L_{IR}$ is plotted against flux. As already
described, the relation between $L_{IR}$ and SFR depends on the
fraction of UV light from young stars absorbed by dust and on the
IMF. In particular, in our model, $L_{IR}/SFR$ is about 4 times larger
for star formation in bursts with the top-heavy ($x=0$) IMF compared
to the stars forming quiescently in disks with the \citet{Kennicutt83}
IMF. The proportion of star formation associated with the burst mode
on average increases with increasing $L_{IR}$, and for this reason the
relation between SFR and flux is shallower than a linear
proportionality. The scatter is also somewhat larger than for the
$L_{IR}$-flux relation. We see from the figure that, at the confusion
limit, \HERSCHEL\ surveys should probe down to SFRs $\sim 1
h^{-1}\Msol\yr^{-1}$ at $z=0.5$ and $\sim 10^2 h^{-1}\Msol\yr^{-1}$ at
$z=4$, the lowest limits being achieved at 100$\mum$. The planned Key
Project surveys may improve on this by factors $\sim 2$ if they can
get below the confusion limit.

\begin{figure}
\begin{center}
\includegraphics[width=7cm]{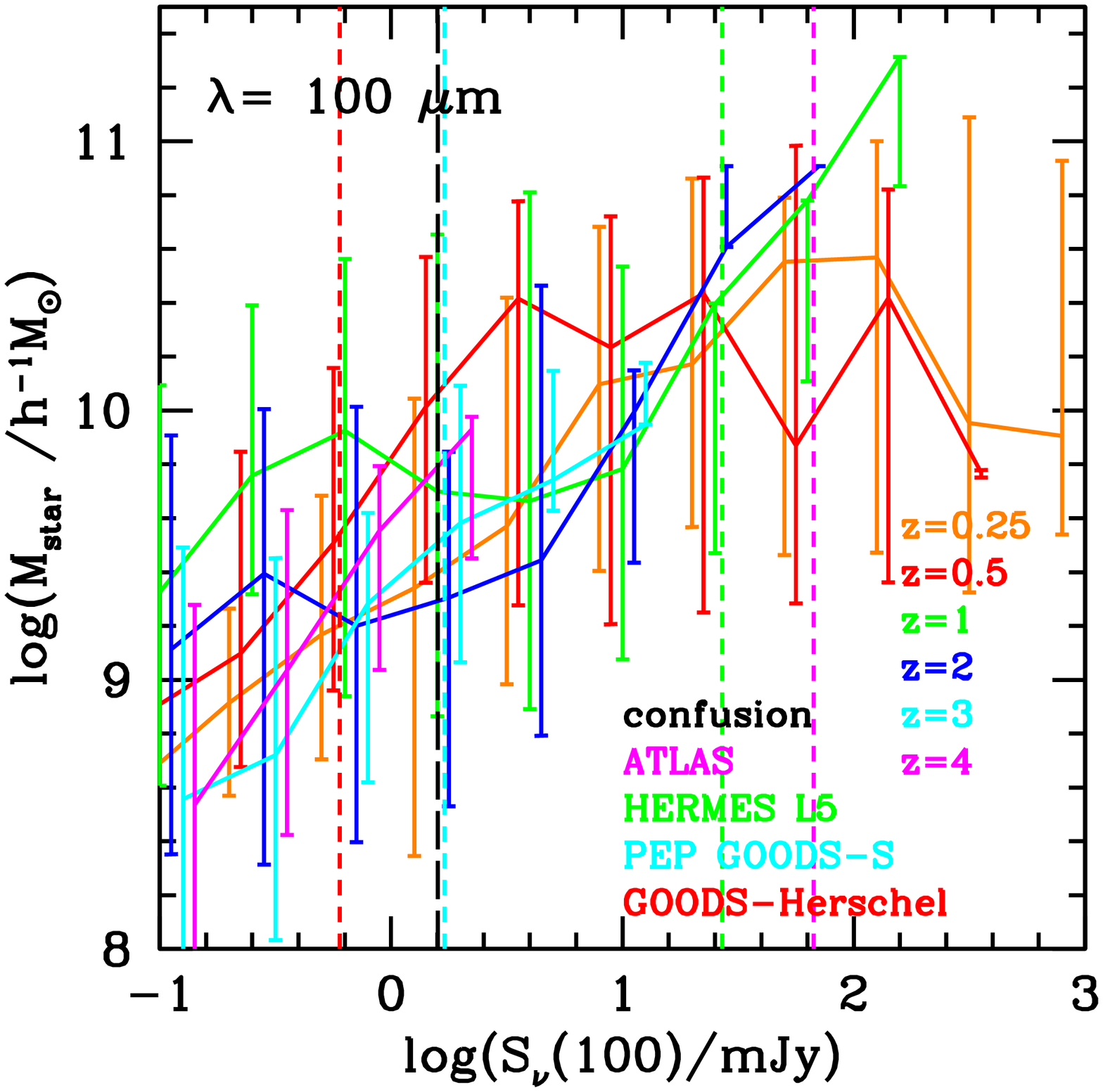}

\includegraphics[width=7cm]{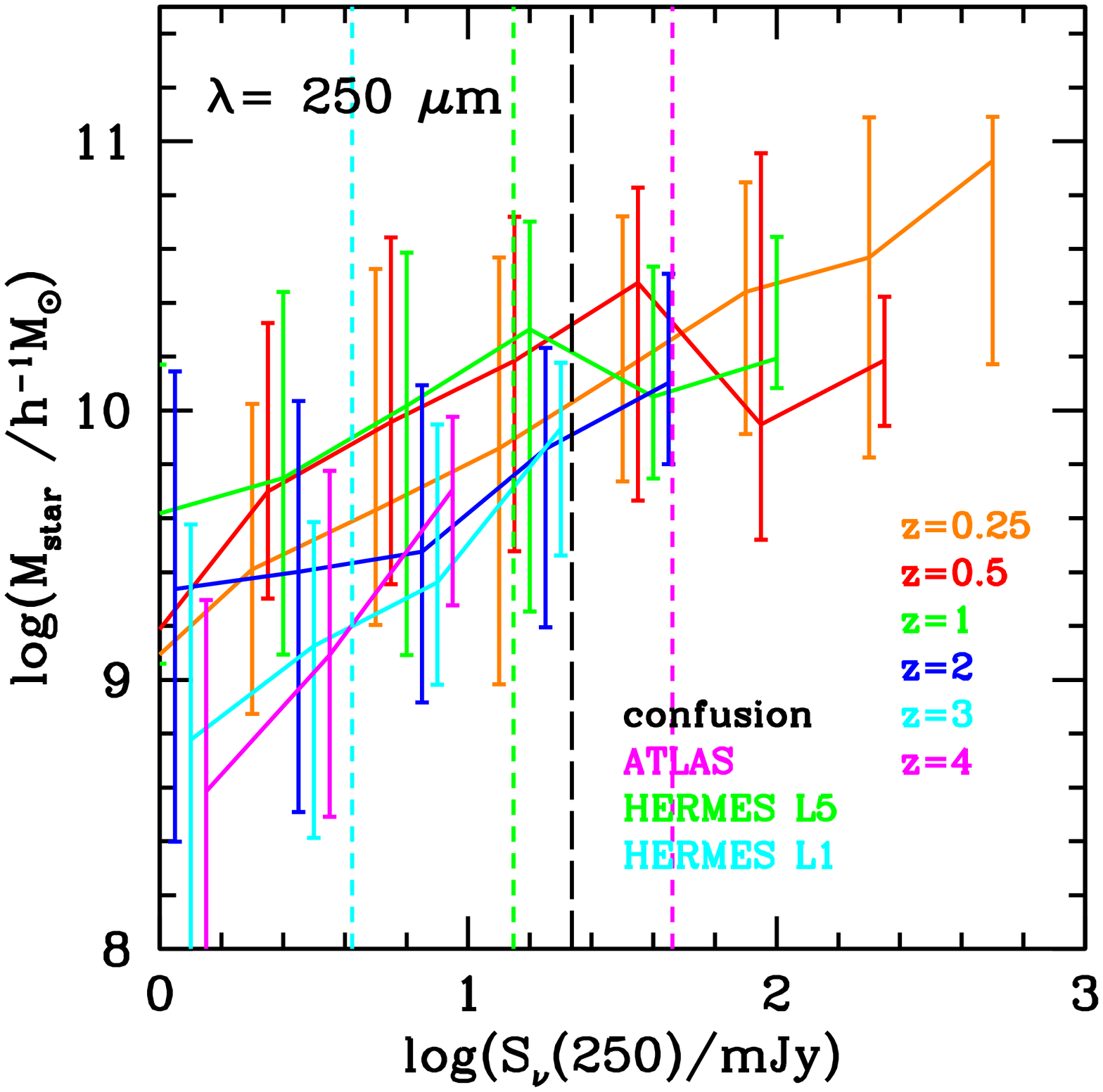}

\includegraphics[width=7cm]{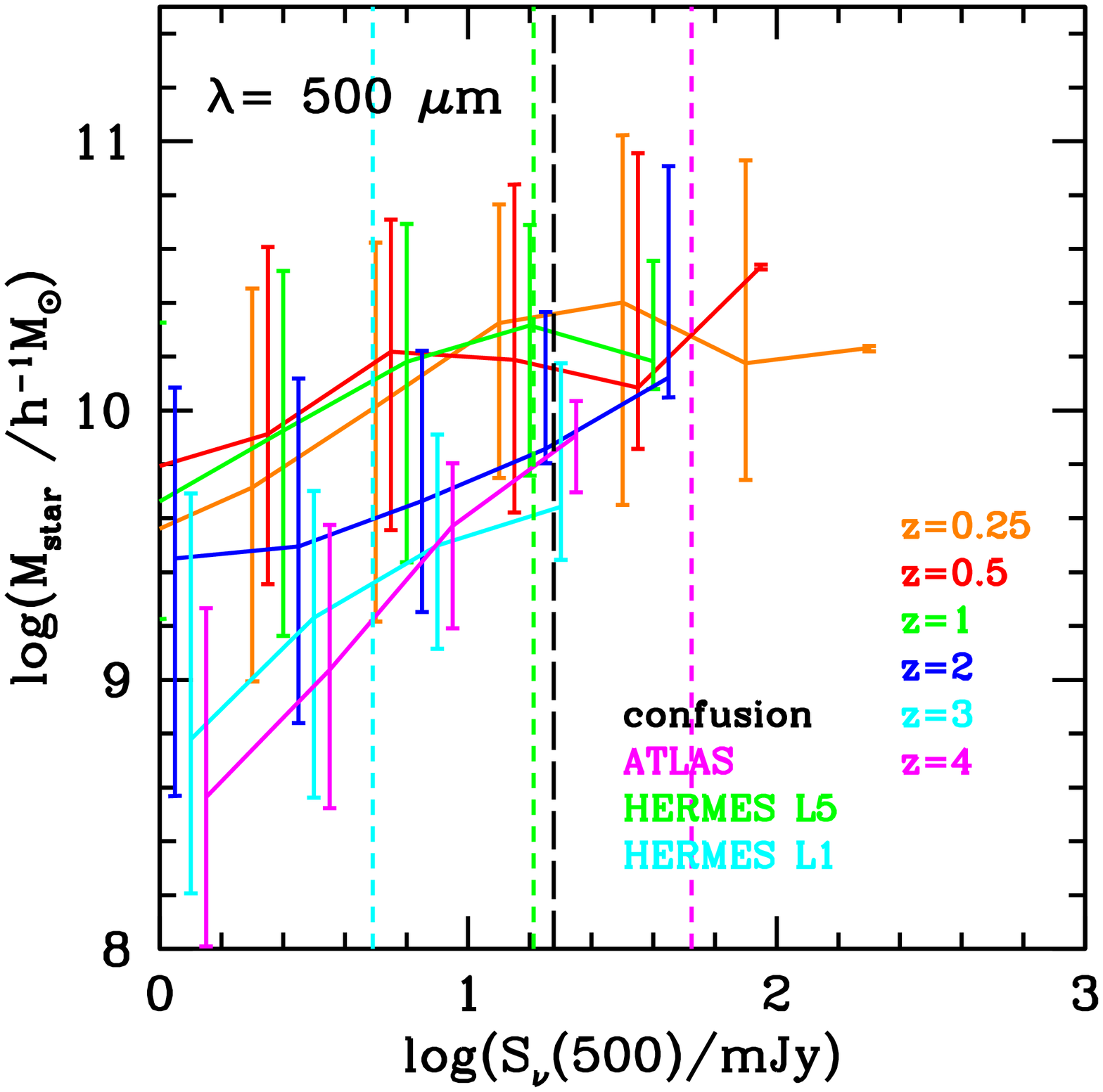}

\end{center}

\caption{Median stellar mass {\em vs} flux at 100, 250 and 500 $\mum$ for
  galaxies selected at different redshifts. The different lines are as
  described for Fig.\ref{fig:LIR_flux}. For clarity, we have
  introduced small horizontal offsets between the lines plotted for
  different redshifts.}

\label{fig:Mstar_flux}
\end{figure}

Fig.~\ref{fig:Mstar_flux} is analogous to the previous two figures,
but now with the stellar mass $M_{star}$ plotted against flux. In this
case, the relation is far from linear, and has much larger scatter
than for $L_{IR}$ or SFR. This is not surprising, since the \HERSCHEL\
flux is proportional to the emission from dust heated mostly by massive young
stars, while the stellar mass includes stars of all ages and
masses. Since the galaxies in our model have complex star formation
histories, there is no simple relation between the current star
formation rate and the total mass of stars formed over the history of
the galaxy. The galaxies found in \HERSCHEL\ surveys should have
stellar masses over the range $M_{star} \sim 10^9-10^{11}
h^{-1}\Msol$, with a weaker dependence on redshift at a given flux
than is typically seen for the SFR.

\begin{figure}
\begin{center}
\includegraphics[width=7cm]{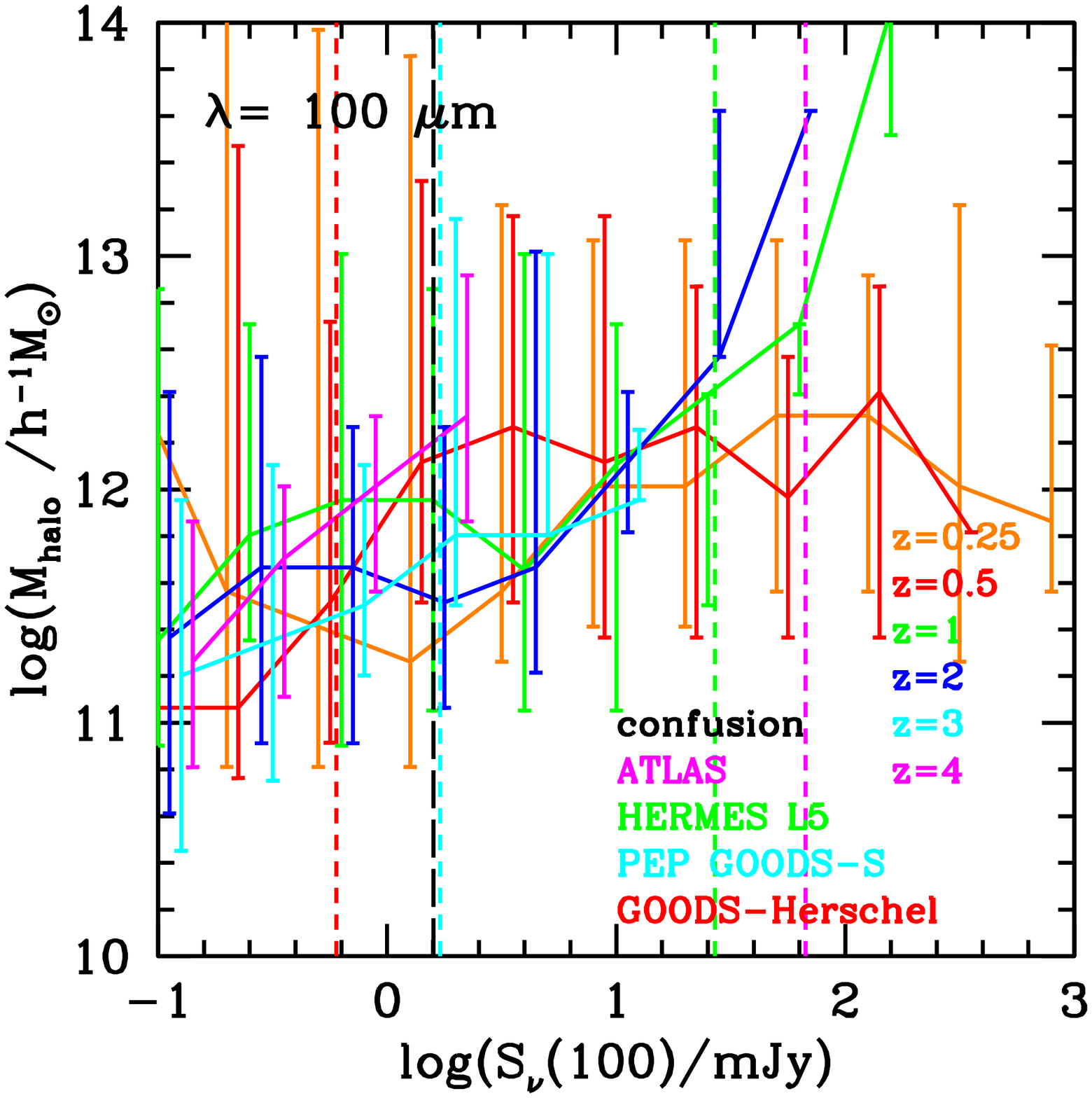}

\includegraphics[width=7cm]{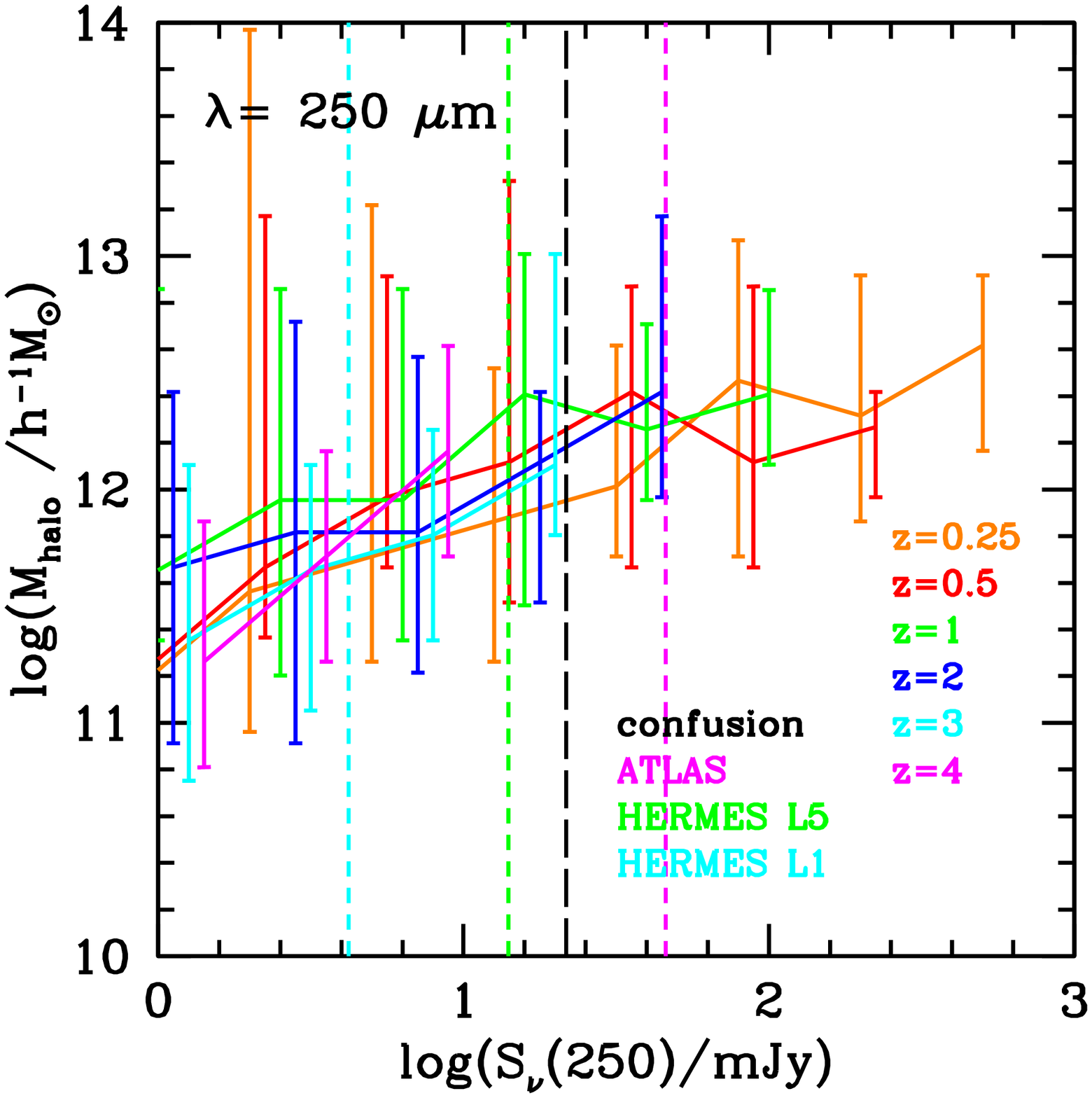}

\includegraphics[width=7cm]{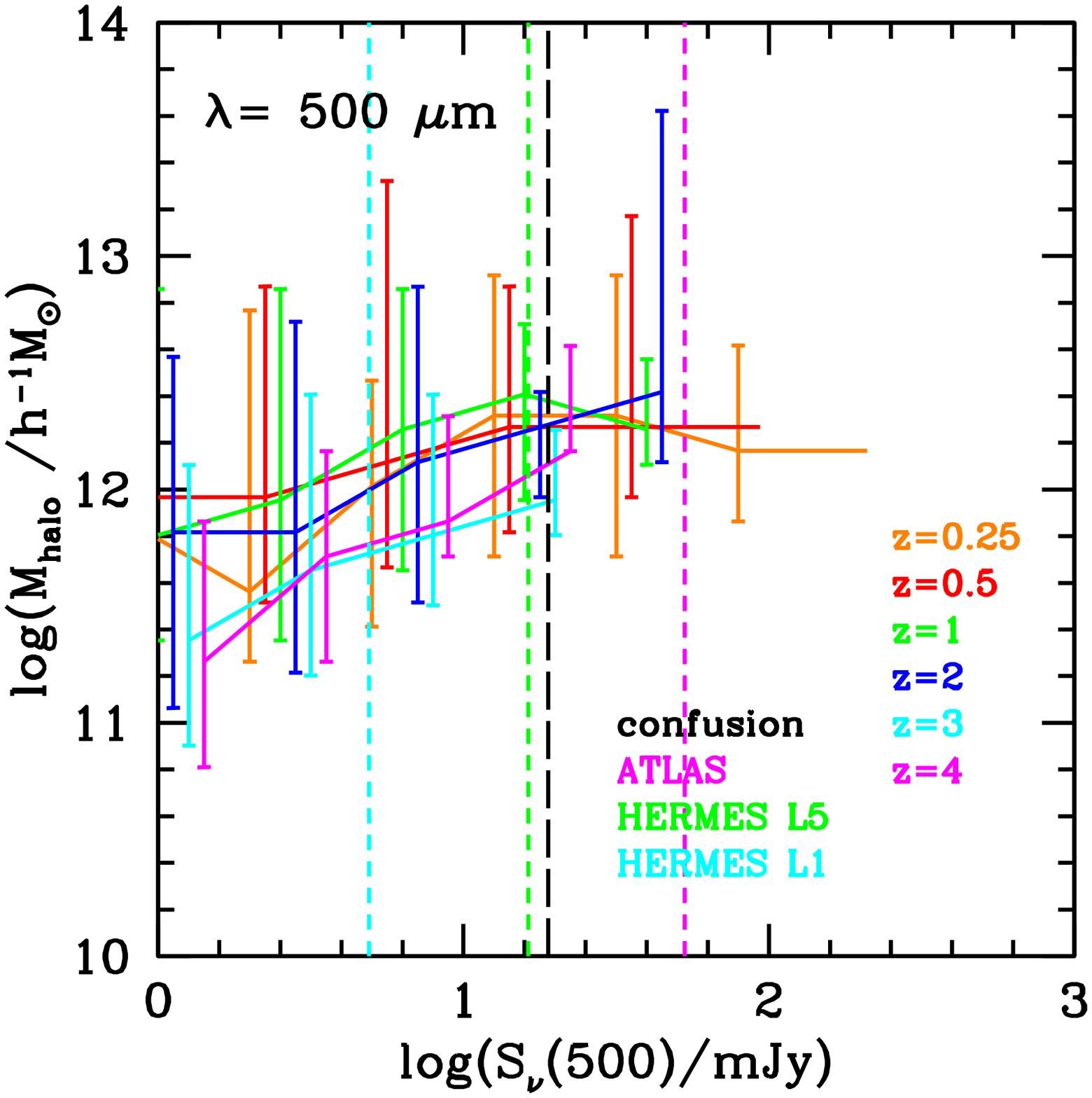}

\end{center}

\caption{Median halo mass {\em vs} flux at 100, 250 and 500 $\mum$ for
  galaxies selected at different redshifts. The different lines are as
  described for Fig.\ref{fig:LIR_flux}. For clarity, we have
  introduced small horizontal offsets between the lines plotted for
  different redshifts. }

\label{fig:Mhalo_flux}
\end{figure}

Finally, in Fig.~\ref{fig:Mhalo_flux}, we show the relation between
the host dark halo mass $M_{halo}$ and flux in the \HERSCHEL\ bands. Here we
see that the median halo mass depends much more weakly on flux or
redshift (and with a larger scatter) than either $M_{star}$ or SFR,
especially at the longer wavelengths. This reflects the fact that in
our model the relation between far-IR luminosity and halo mass is even
more indirect than for the stellar mass, especially due to the
dominance of transient bursts at the higher luminosities. This
produces the relatively flat trend of median halo mass with
\HERSCHEL\ flux. The weak dependence on redshift is because halos of a
given mass on average host higher SFRs at higher $z$, which
compensates for the larger luminosity distance. We see from
Fig.~\ref{fig:Mhalo_flux} that the galaxies found in the Key Project
cosmological surveys should typically have halo masses $M_{halo} \sim
10^{12} h^{-1}\Msol$. This will have important implications when we
consider the clustering of \HERSCHEL\ galaxies in \S\ref{sec:clustering}.

\begin{figure}
\begin{center}

\includegraphics[width=7cm]{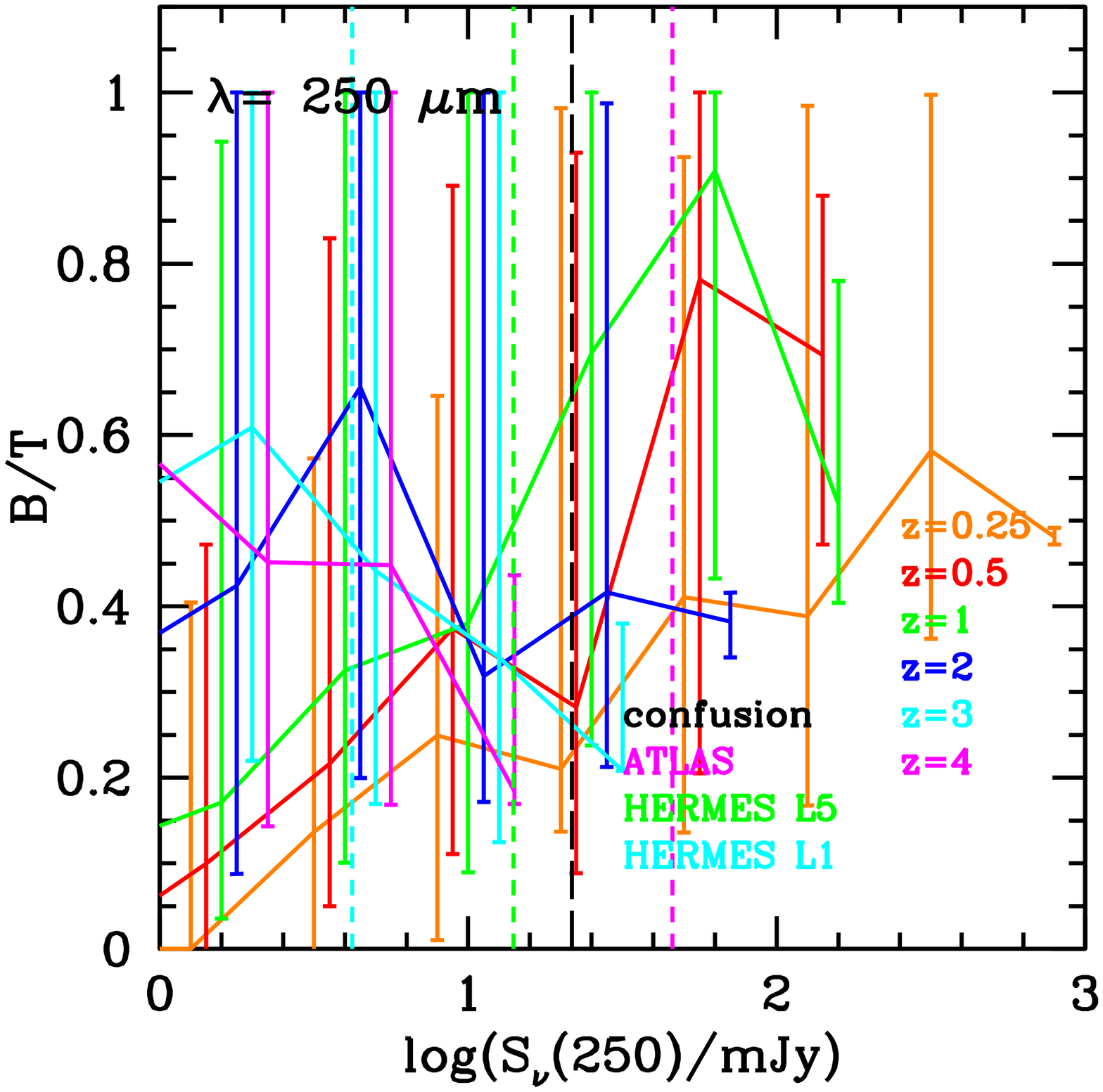}

\includegraphics[width=7cm]{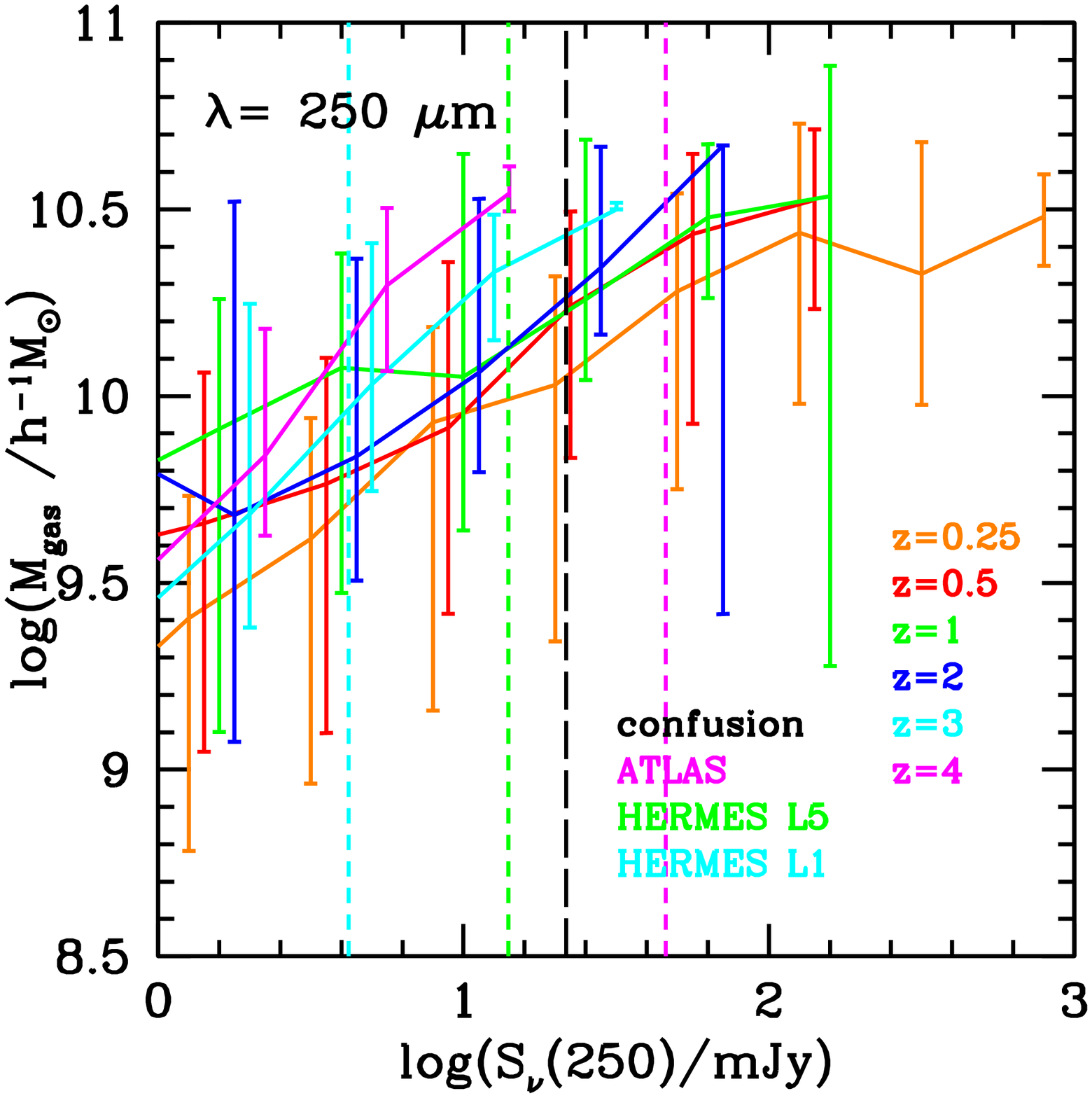}

\end{center}

\caption{Median bulge-to-total stellar mass ratio $B/T$ and cold gas
  mass {\em vs} flux at 250 $\mum$ for galaxies selected at different
  redshifts. The different lines are as described for
  Fig.\ref{fig:LIR_flux}. For clarity, we have
  introduced small horizontal offsets between the lines plotted for
  different redshifts.}

\label{fig:BT_mgas_flux}
\end{figure}

We have also investigated the dependence of the stellar bulge-to-total
mass ratio $B/T$ and the cold gas mass $M_{gas}$ on flux in the
\HERSCHEL\ bands. For brevity, we only show results for the 250$\mum$
band in Fig.~\ref{fig:BT_mgas_flux}. We find that the 10-90\% range
for $B/T$ covers nearly the whole possible range $0\leq B/T \leq 1$ at
most fluxes and redshifts of interest in the planned \HERSCHEL\
cosmological surveys, i.e. there is no strong preference for one
morphological type over another. The exception to this is at low fluxes
and low redshifts, where most sources have $B/T \lsim 0.5$. This
reflects the fact that the galaxies found in these surveys should be a
mixture of quiescently star-forming disk galaxies and starbursts
triggered by galaxy mergers, and even though the most luminous
galaxies in the far-IR are predicted to be bursts, these can be
triggered by either major or minor mergers, producing a bursting
galaxy which can be either bulge or disk dominated. There is a trend
for the median $B/T$ to increase with flux at a given redshift,
presumably reflecting an increase in the fraction of bursts triggered
by major mergers. We illustrate these trends in the top panel of
Fig.~\ref{fig:BT_mgas_flux}. For the cold gas mass $M_{gas}$, shown in
the lower panel of Fig.~\ref{fig:BT_mgas_flux}, we find a correlation
with flux which is weaker than linear, and also fairly weakly
dependent on redshift. We predict that the galaxies found in the
\HERSCHEL\ cosmological surveys should typically have gas masses $\sim
10^{10} h^{-1}\Msol$, which implies that many of them should have CO
emission from their molecular gas which is detectable by current
telescopes \citep[e.g.][]{Solomon05}.


\begin{figure*}
\begin{center}

\begin{minipage}{7cm}
\includegraphics[width=7cm]{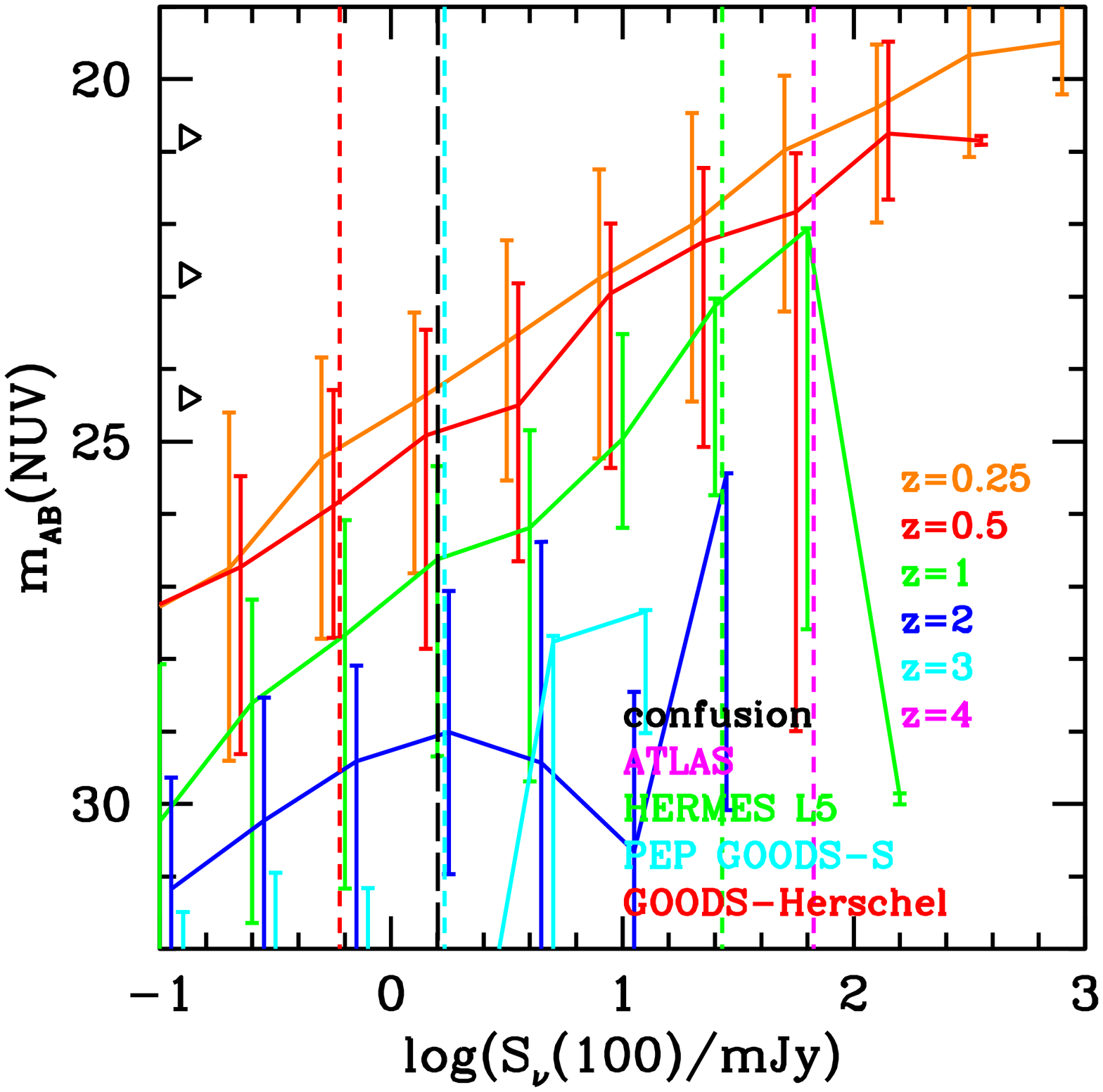}
\end{minipage}
\hspace{1cm}
\begin{minipage}{7cm}
\includegraphics[width=7cm]{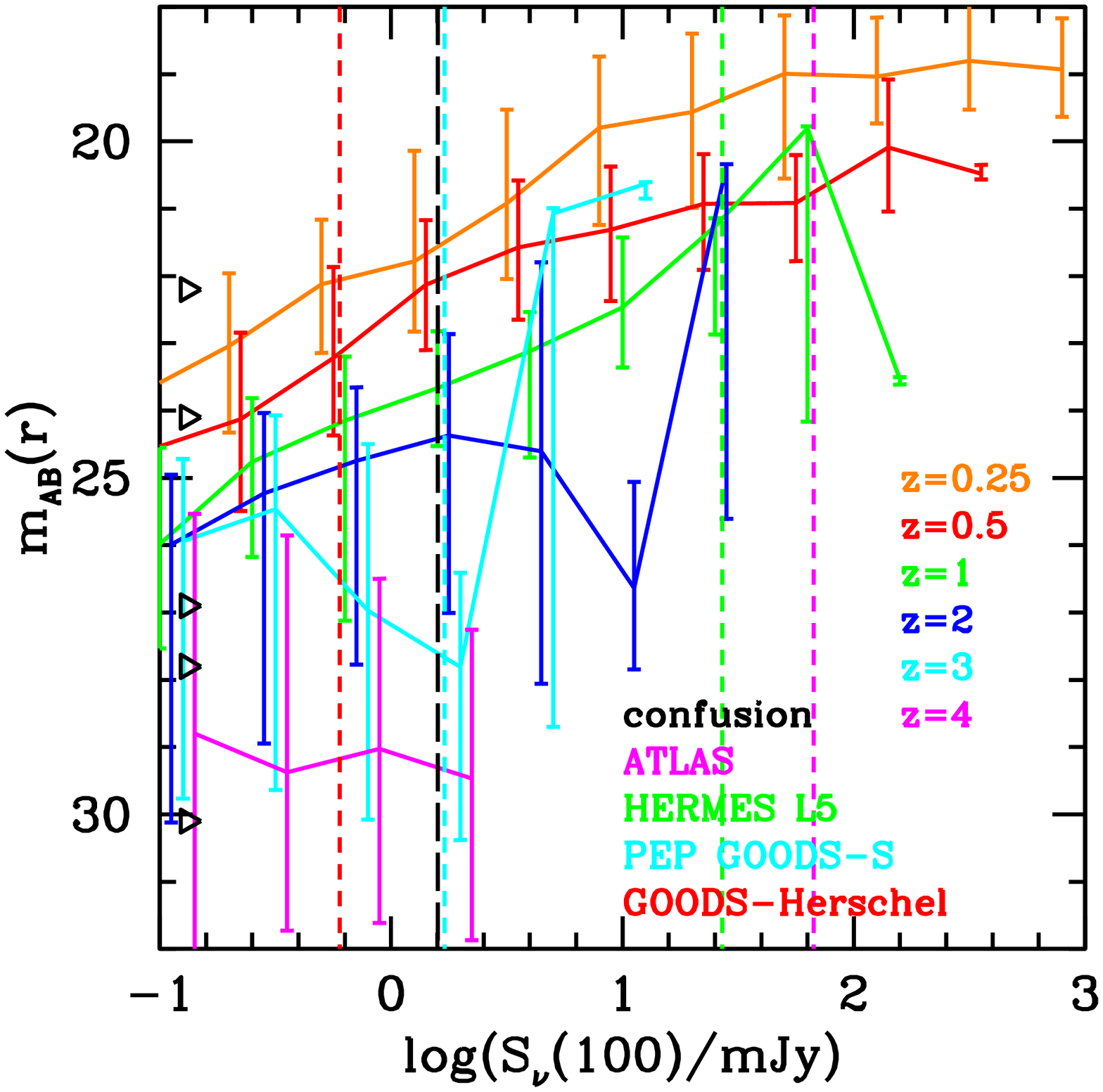}
\end{minipage}

\begin{minipage}{7cm}
\includegraphics[width=7cm]{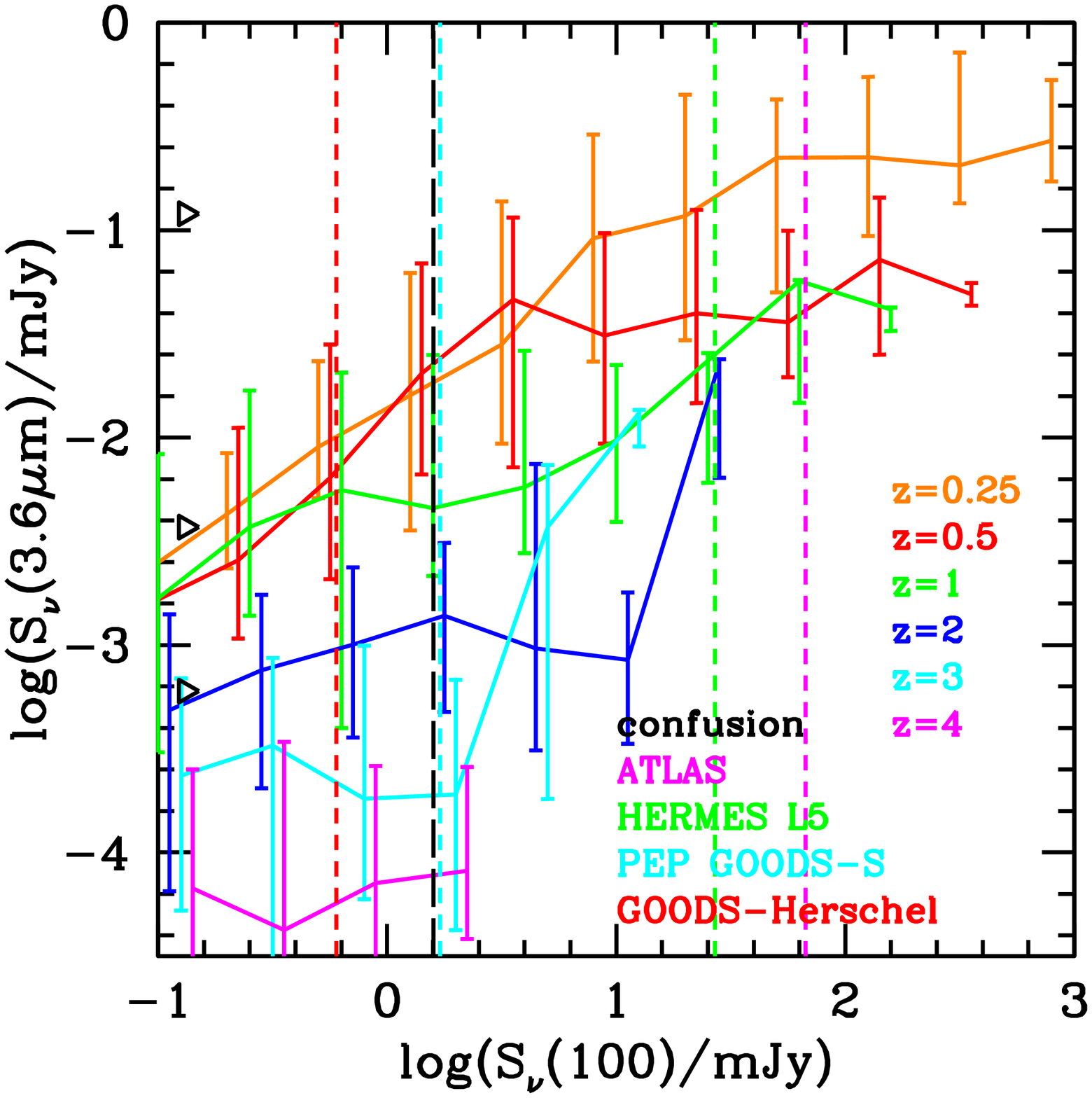}
\end{minipage}
\hspace{1cm}
\begin{minipage}{7cm}
\includegraphics[width=7cm]{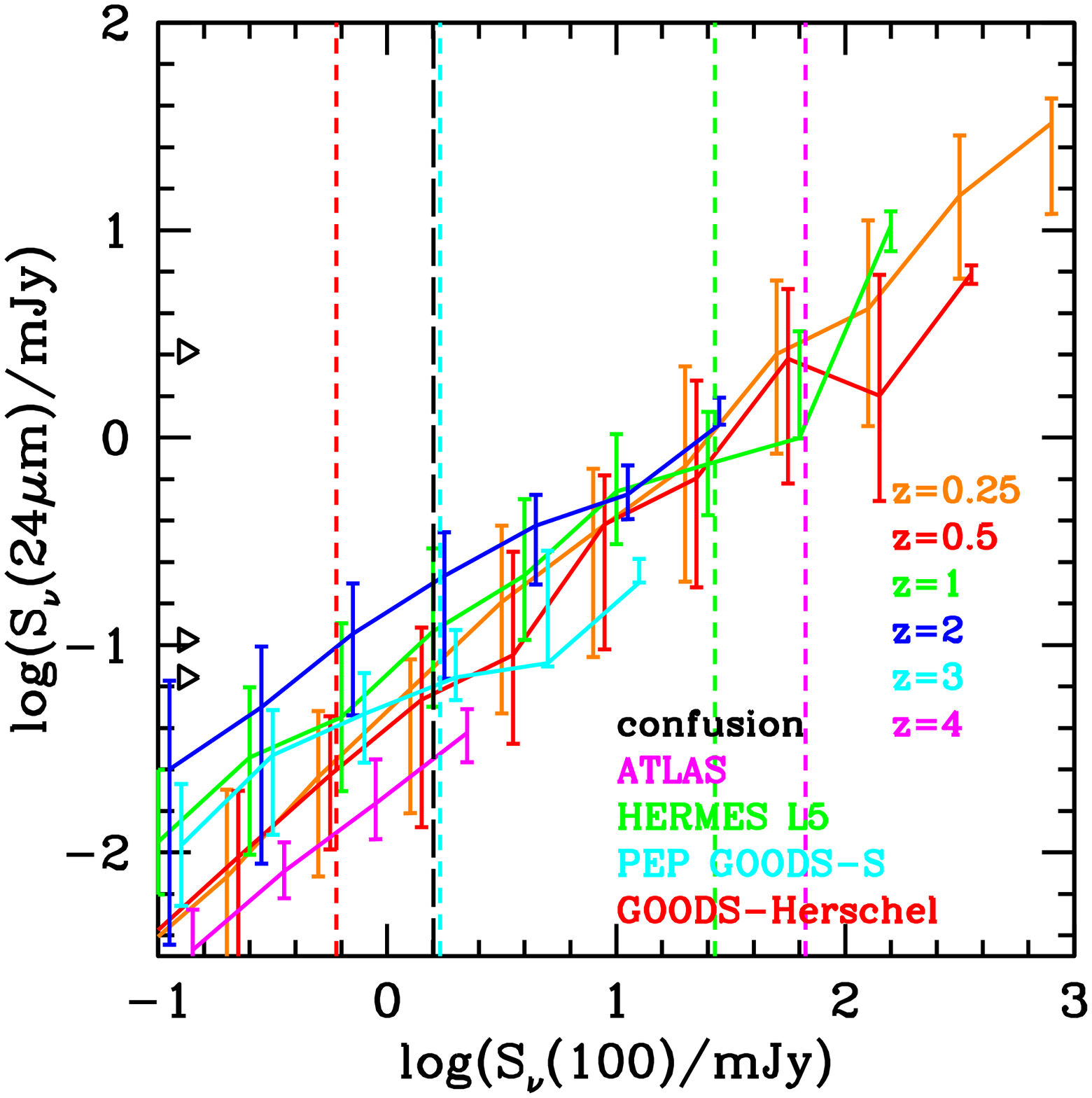}
\end{minipage}

\begin{minipage}{7cm}
\includegraphics[width=7cm]{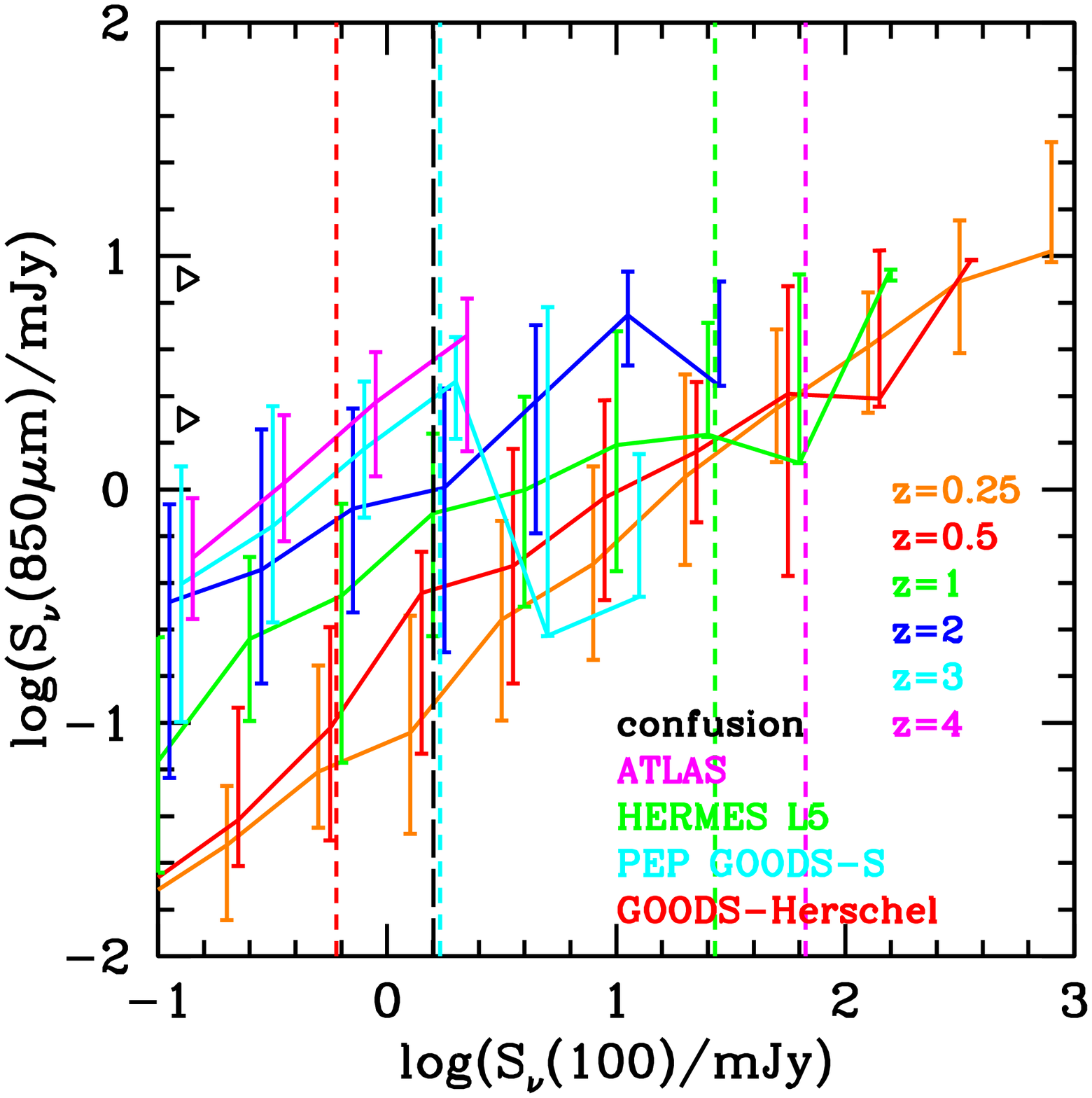}
\end{minipage}
\hspace{1cm}
\begin{minipage}{7cm}
\includegraphics[width=7cm]{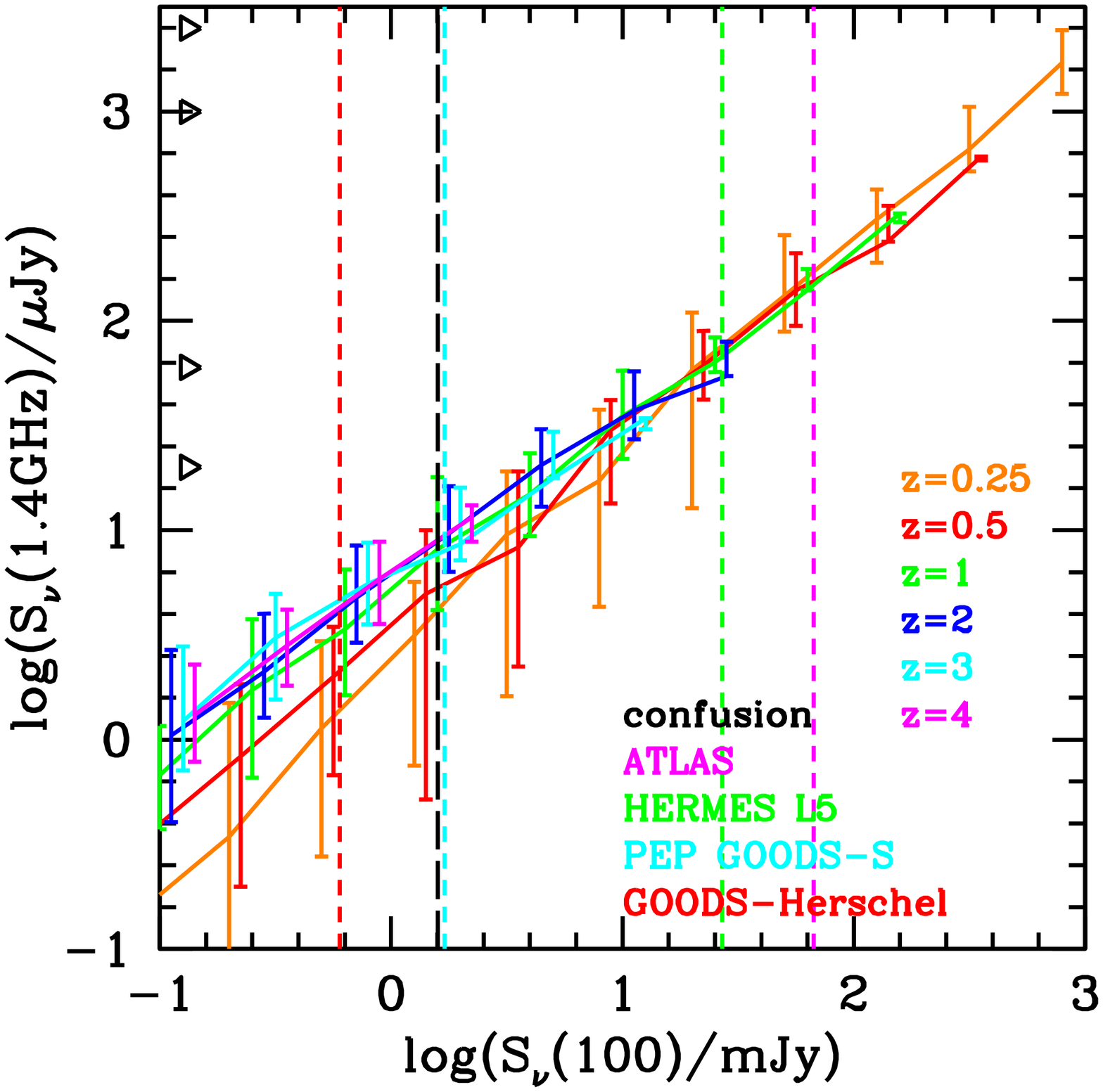}
\end{minipage}

\end{center}

\caption{Median fluxes or magnitudes at other wavelengths for galaxies
  selected at 100$\mum$. (a) GALEX NUV. (b) r-band. (c) IRAC
  3.6$\mum$. (d) MIPS 24$\mum$. (e) SCUBA 850$\mum$. (f) 1.4GHZ. The
  triangles next to the the y-axes in these panels indicate the magnitude
  or flux limits for the different surveys which are discussed in the
  text. For clarity, we have introduced small horizontal offsets
  between the lines plotted for different redshifts.}

\label{fig:multi-wave-100}
\end{figure*}

\begin{figure*}
\begin{center}

\begin{minipage}{7cm}
\includegraphics[width=7cm]{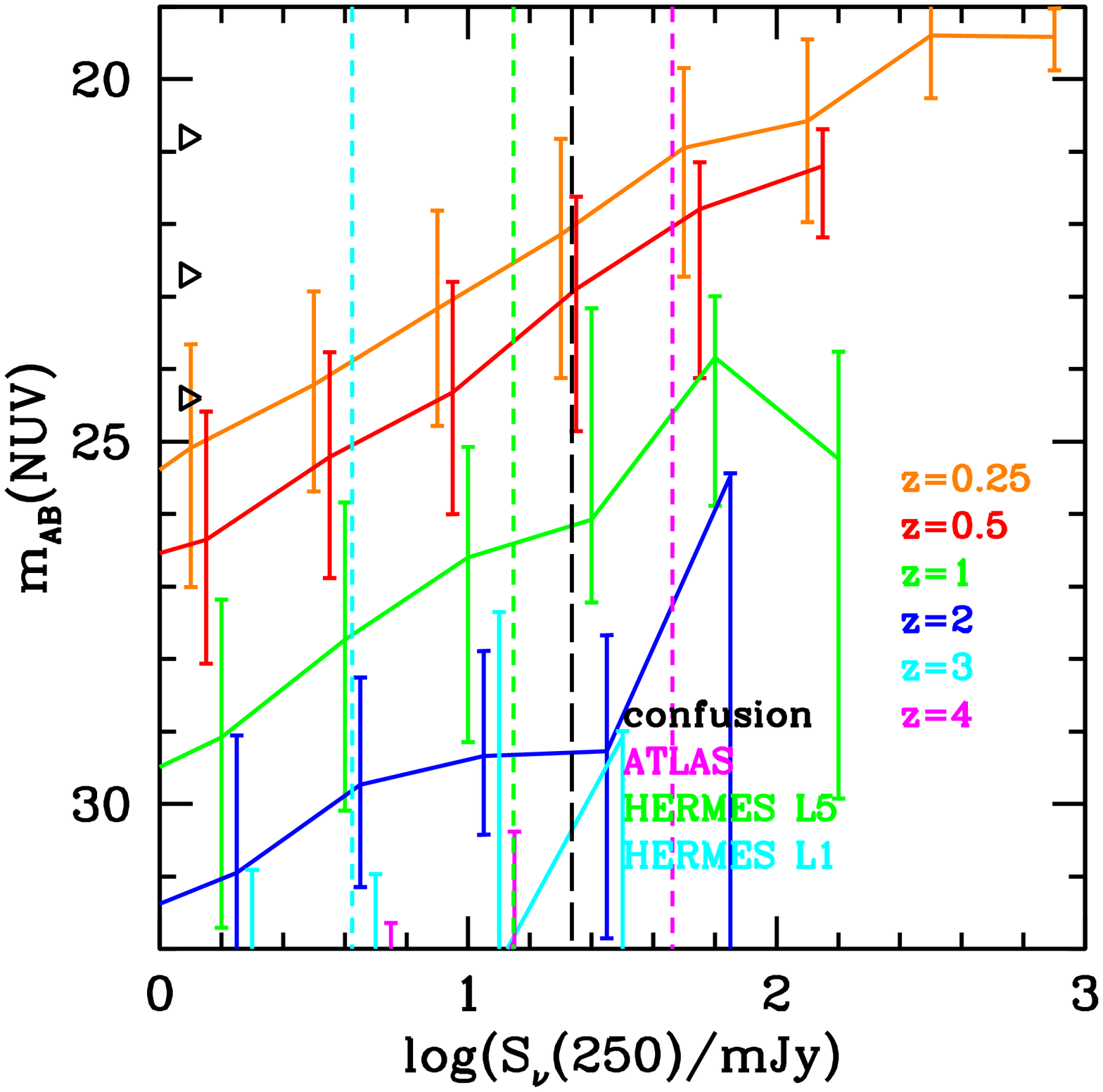}
\end{minipage}
\hspace{1cm}
\begin{minipage}{7cm}
\includegraphics[width=7cm]{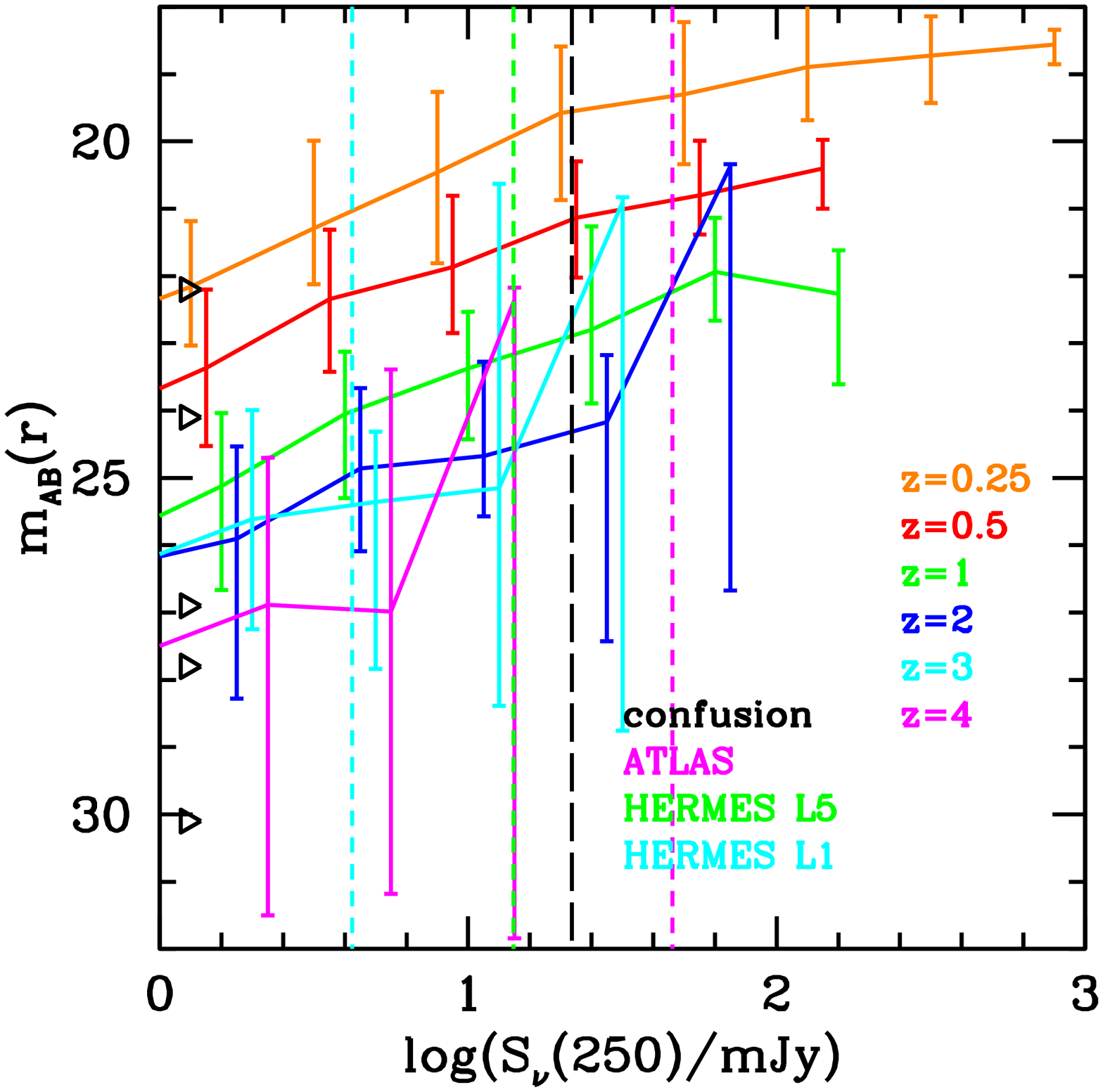}
\end{minipage}

\begin{minipage}{7cm}
\includegraphics[width=7cm]{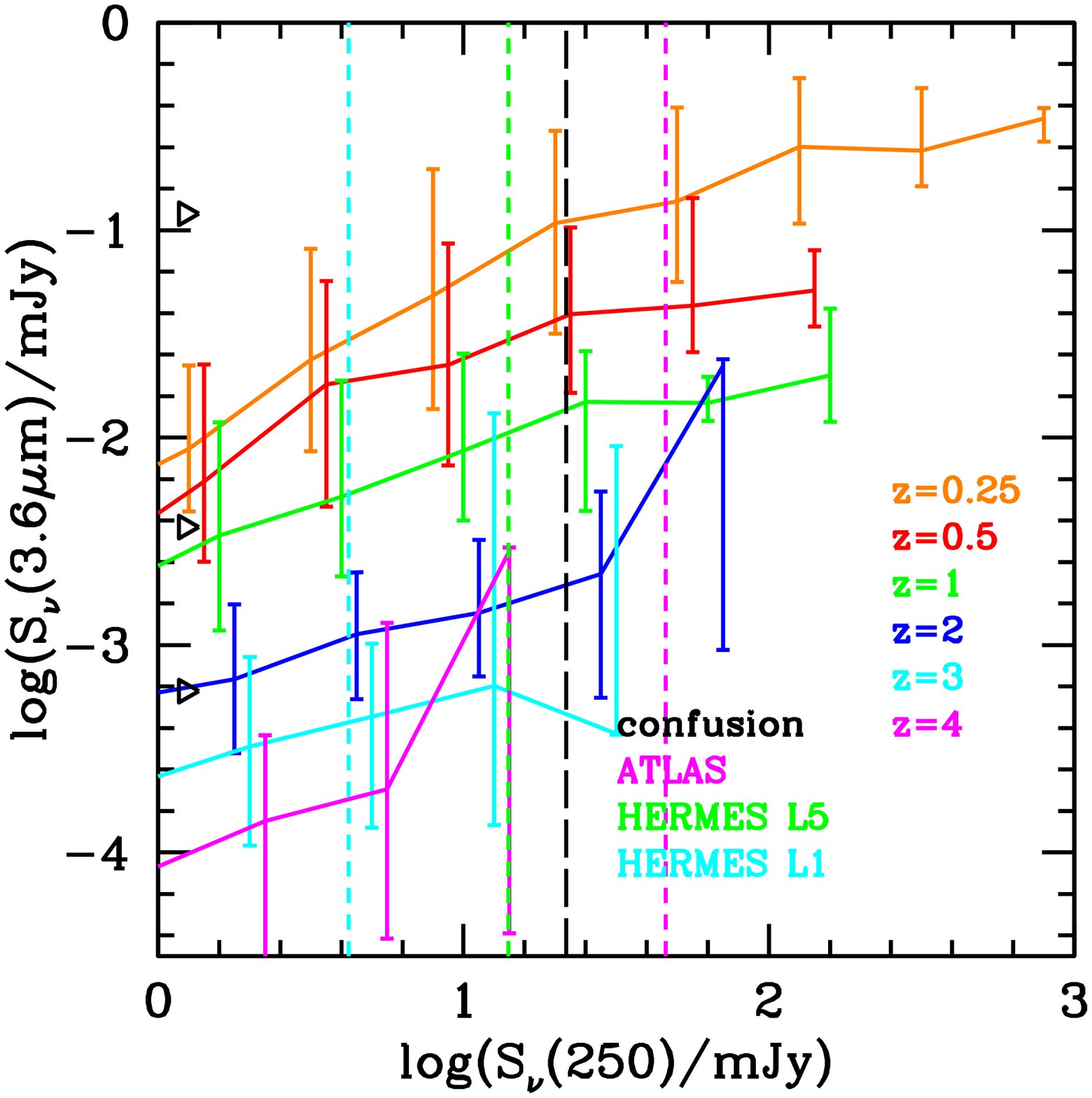}
\end{minipage}
\hspace{1cm}
\begin{minipage}{7cm}
\includegraphics[width=7cm]{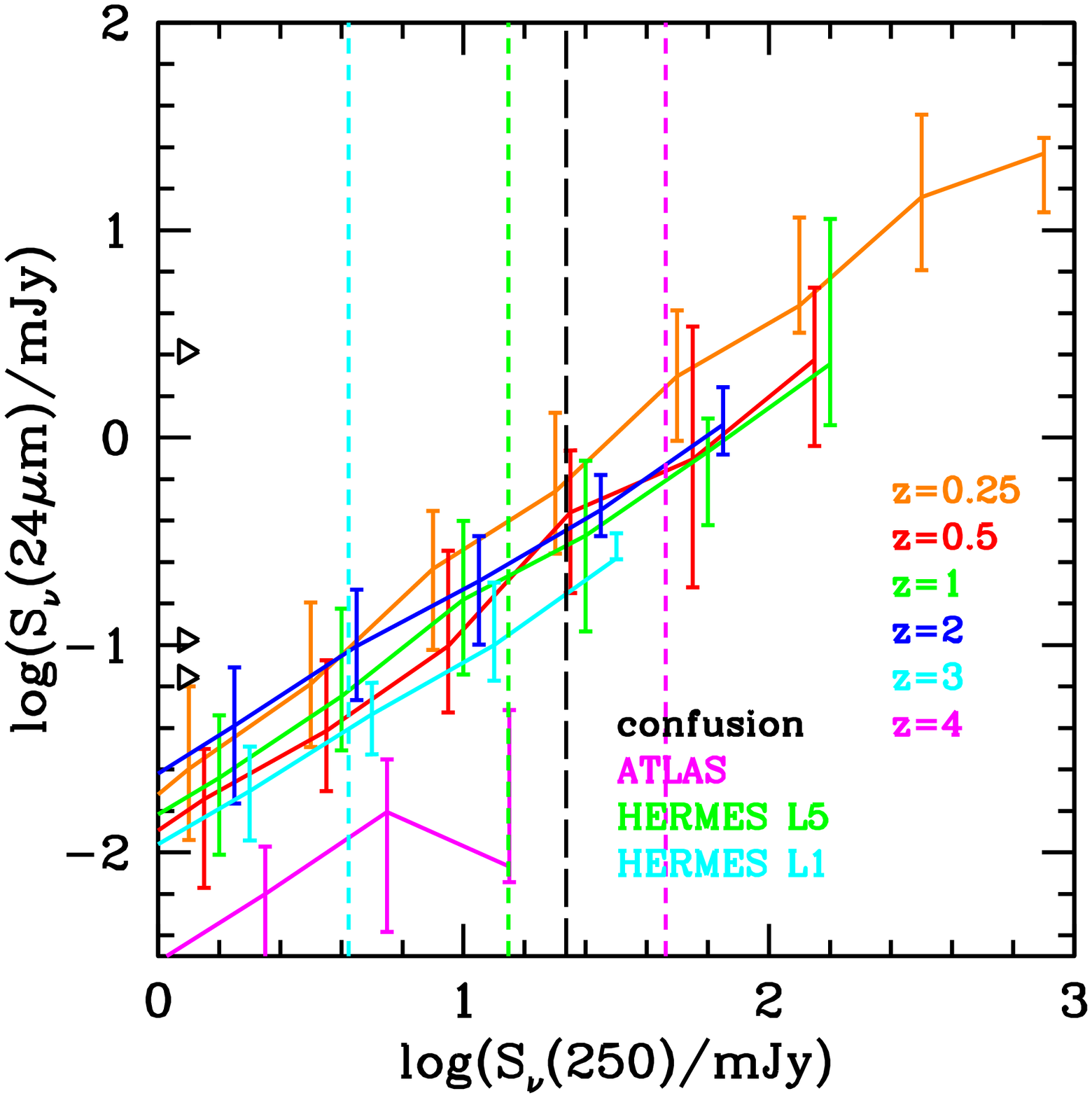}
\end{minipage}

\begin{minipage}{7cm}
\includegraphics[width=7cm]{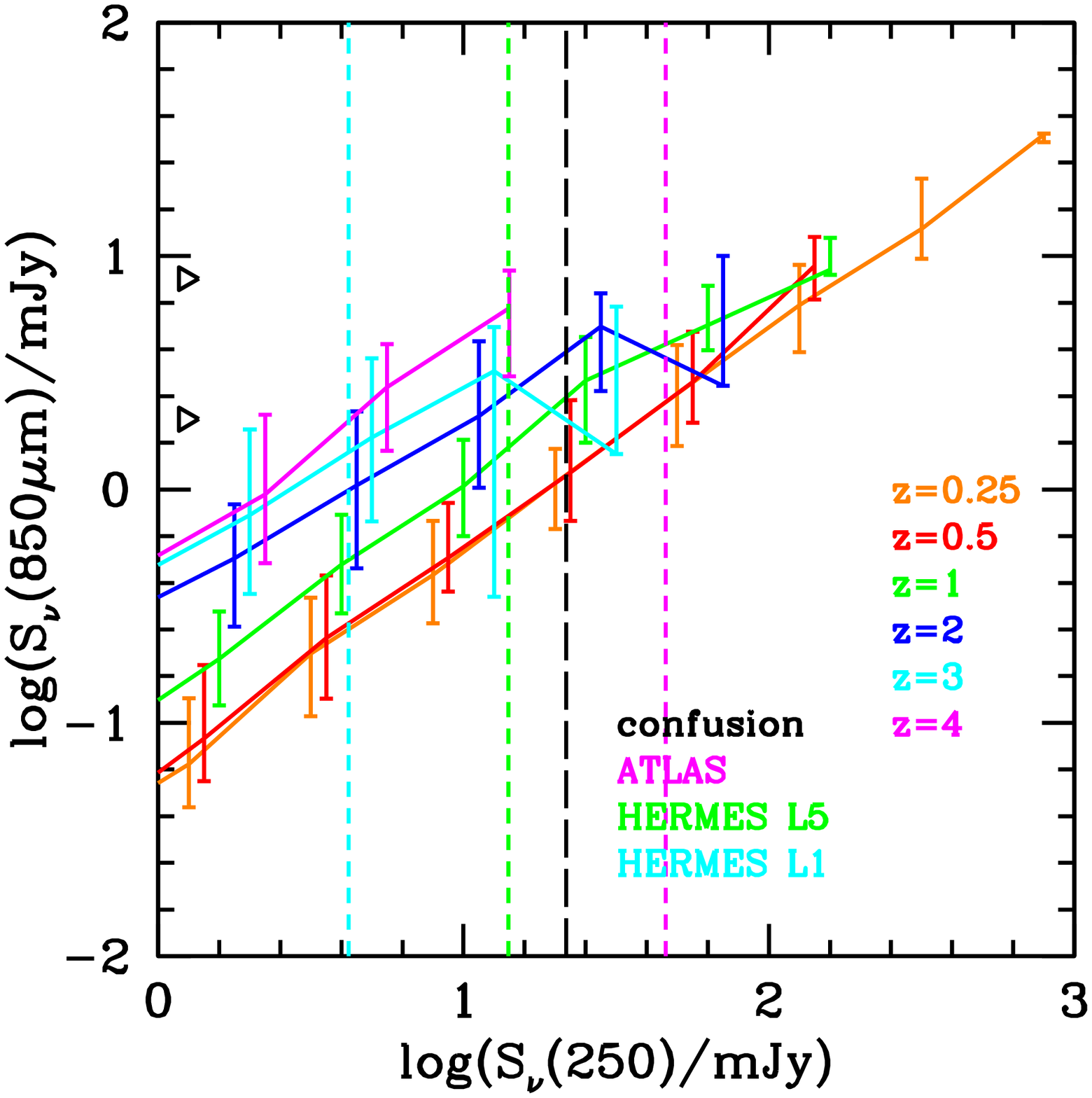}
\end{minipage}
\hspace{1cm}
\begin{minipage}{7cm}
\includegraphics[width=7cm]{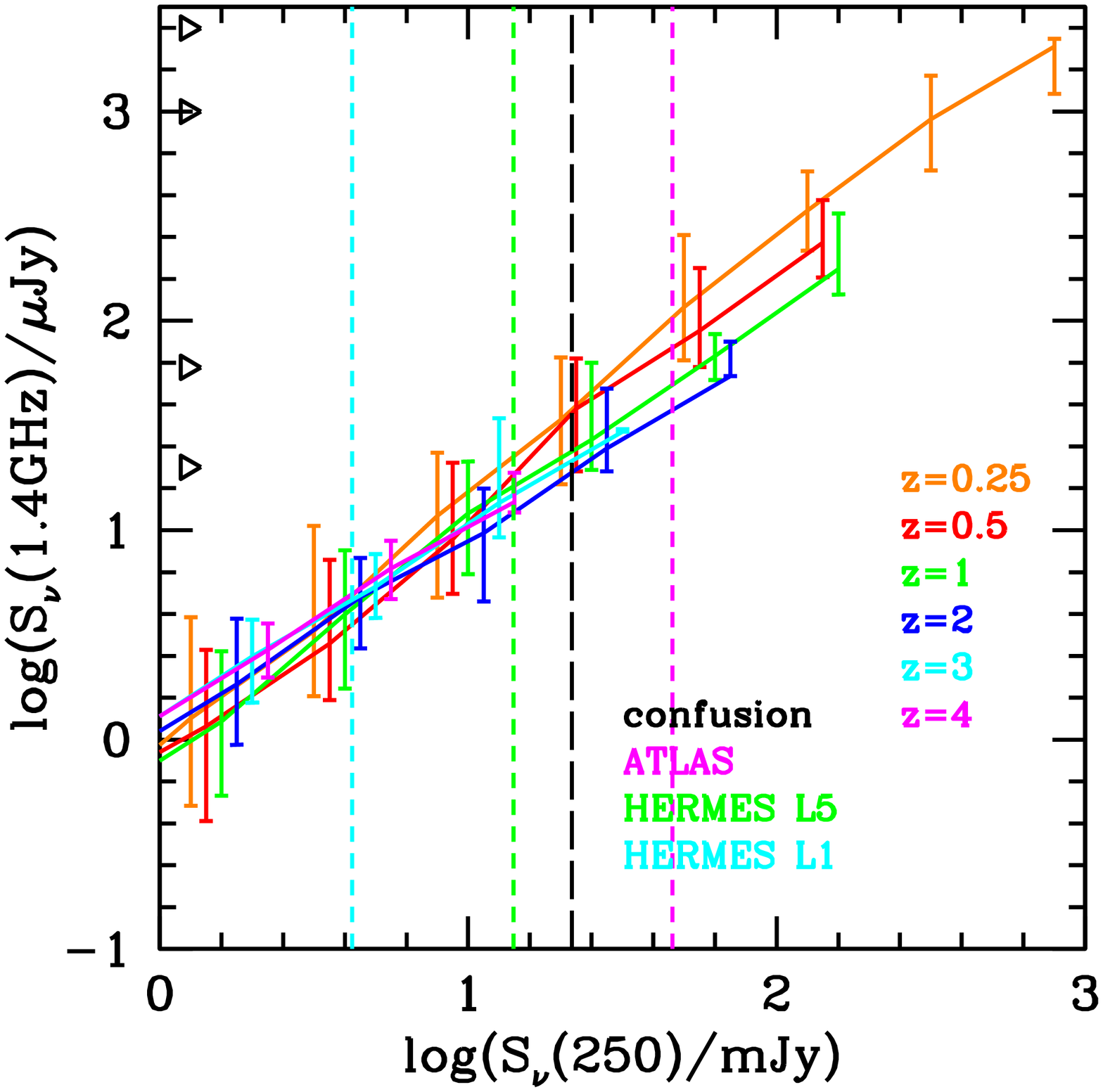}
\end{minipage}

\end{center}

\caption{Median fluxes or magnitudes at other wavelengths for galaxies
  selected at 250$\mum$. (a) GALEX NUV. (b) r-band. (c) IRAC
  3.6$\mum$. (d) MIPS 24$\mum$. (e) SCUBA 850$\mum$. (f) 1.4GHZ. The
  triangles indicate the flux limits for different surveys, as in
  Fig.~\ref{fig:multi-wave-100}. For clarity, we have
  introduced small horizontal offsets between the lines plotted for
  different redshifts.
}

\label{fig:multi-wave-250}
\end{figure*}

\begin{figure*}
\begin{center}

\begin{minipage}{7cm}
\includegraphics[width=7cm]{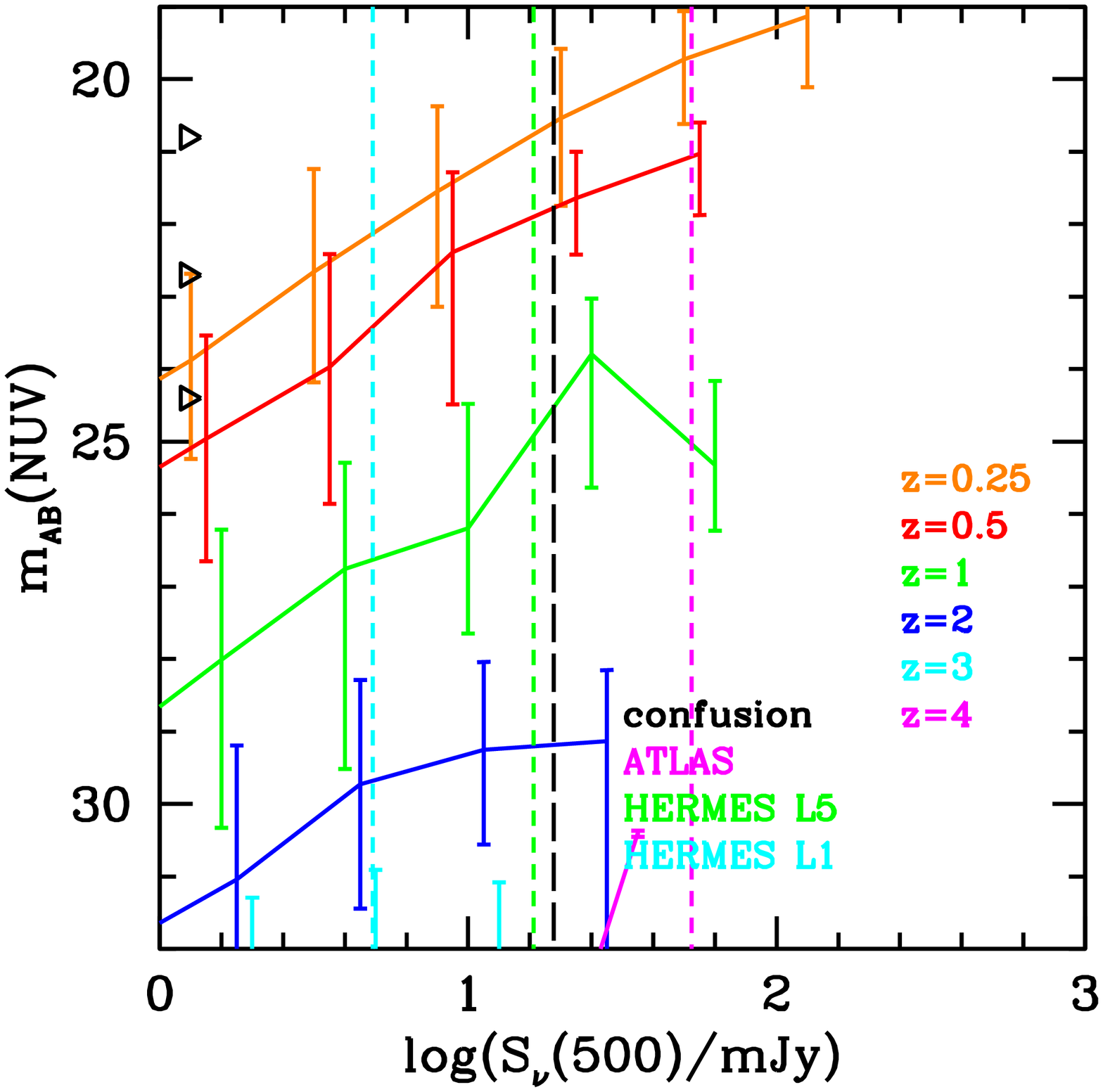}
\end{minipage}
\hspace{1cm}
\begin{minipage}{7cm}
\includegraphics[width=7cm]{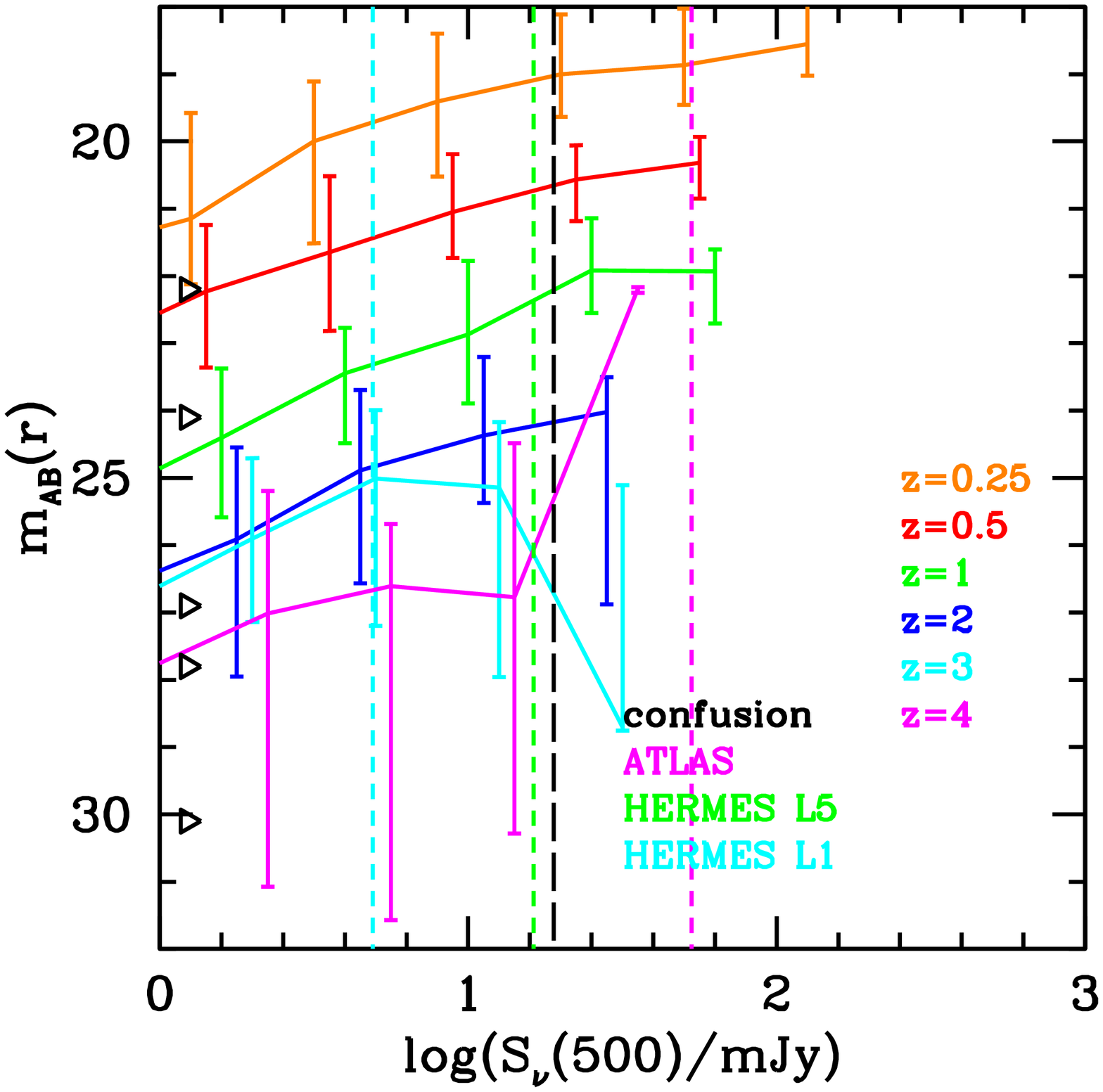}
\end{minipage}

\begin{minipage}{7cm}
\includegraphics[width=7cm]{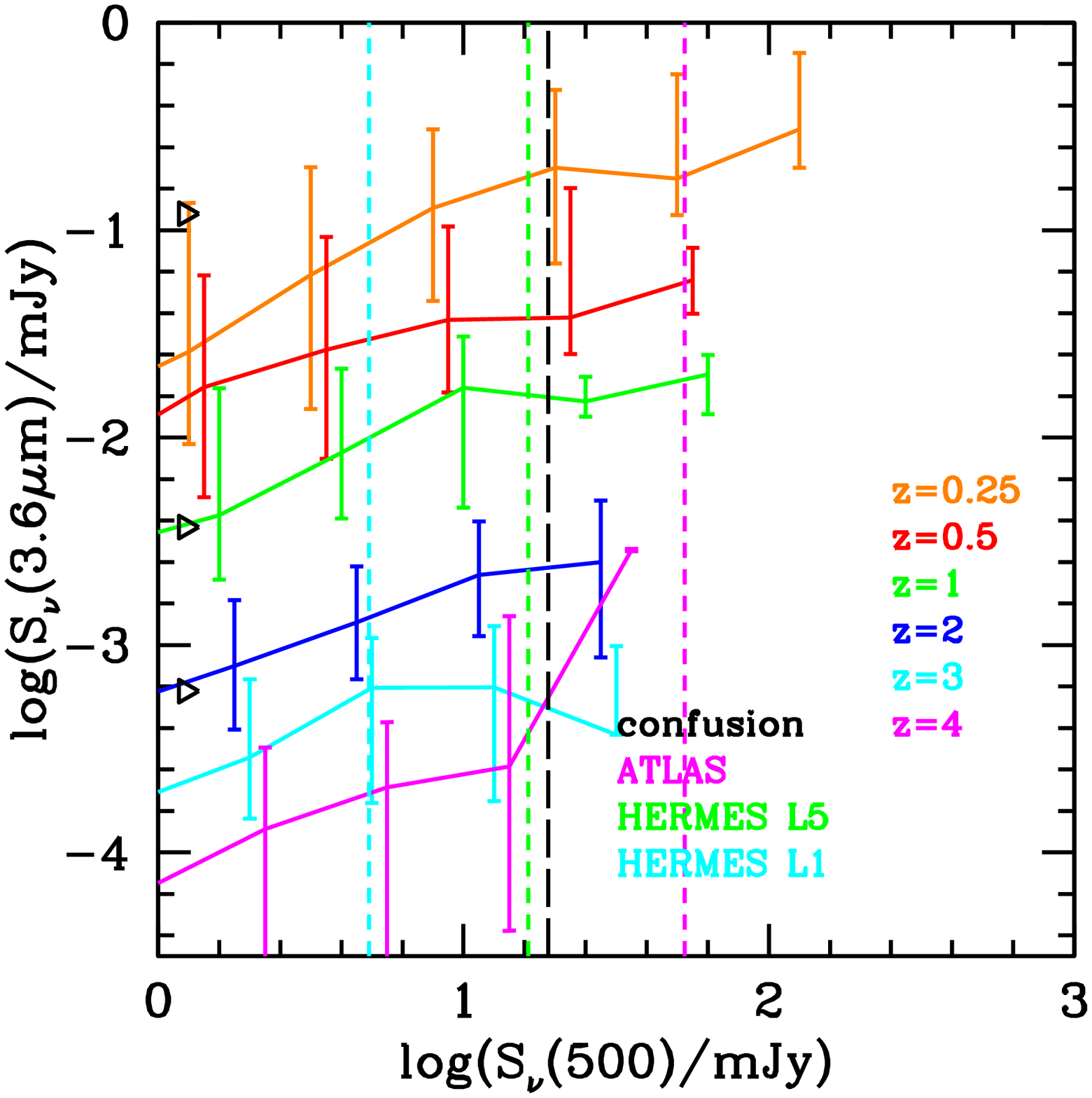}
\end{minipage}
\hspace{1cm}
\begin{minipage}{7cm}
\includegraphics[width=7cm]{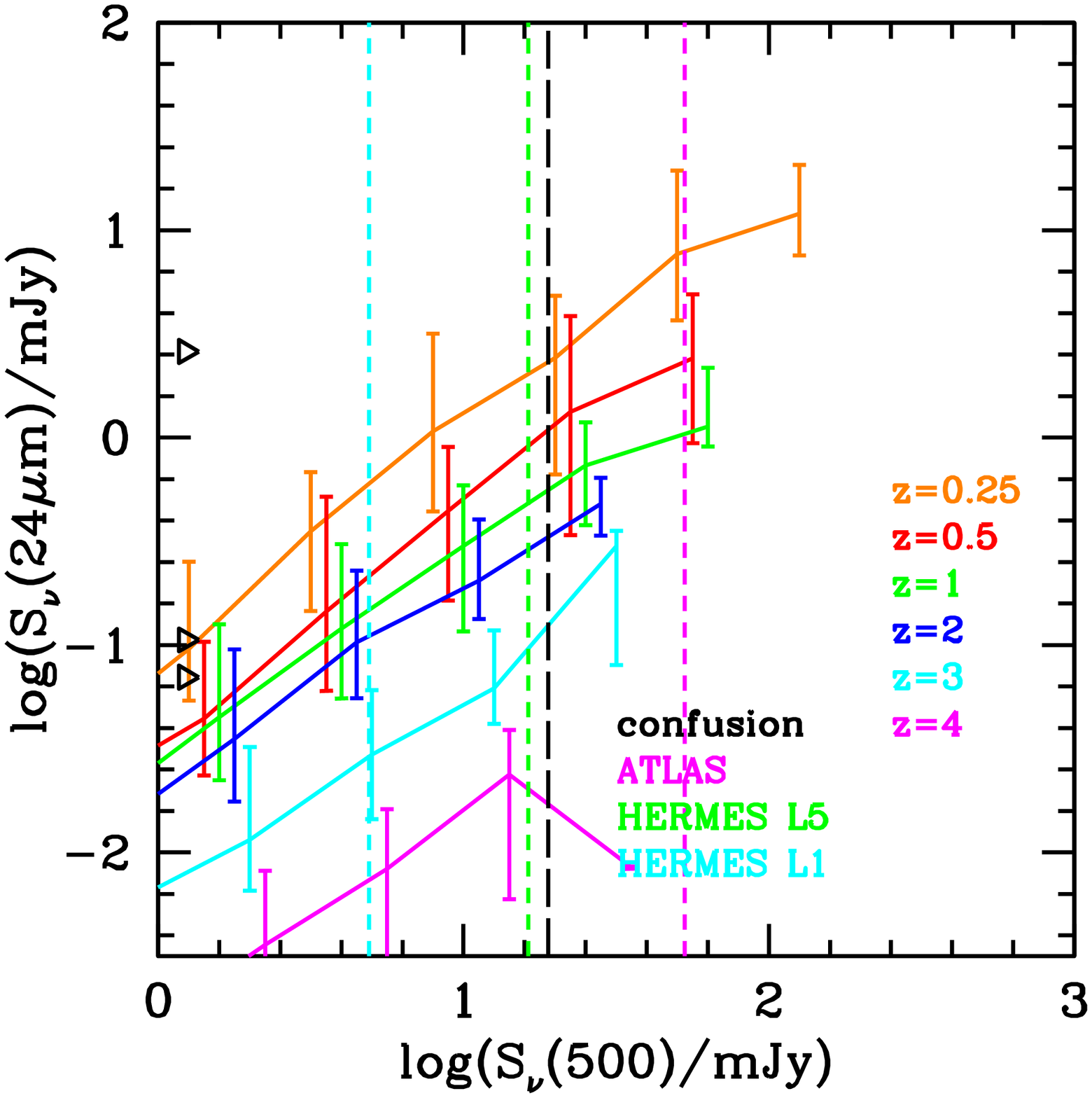}
\end{minipage}

\begin{minipage}{7cm}
\includegraphics[width=7cm]{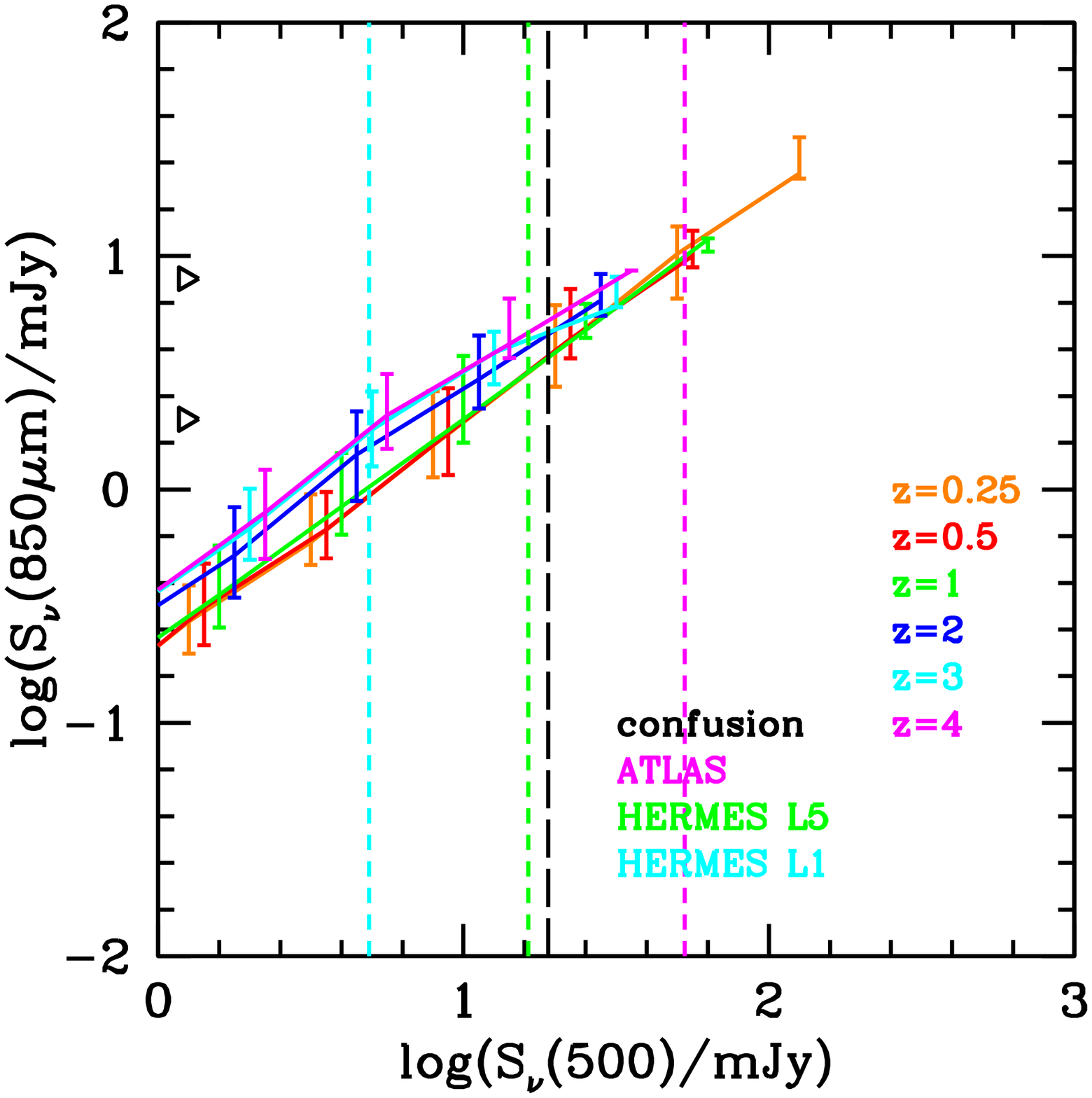}
\end{minipage}
\hspace{1cm}
\begin{minipage}{7cm}
\includegraphics[width=7cm]{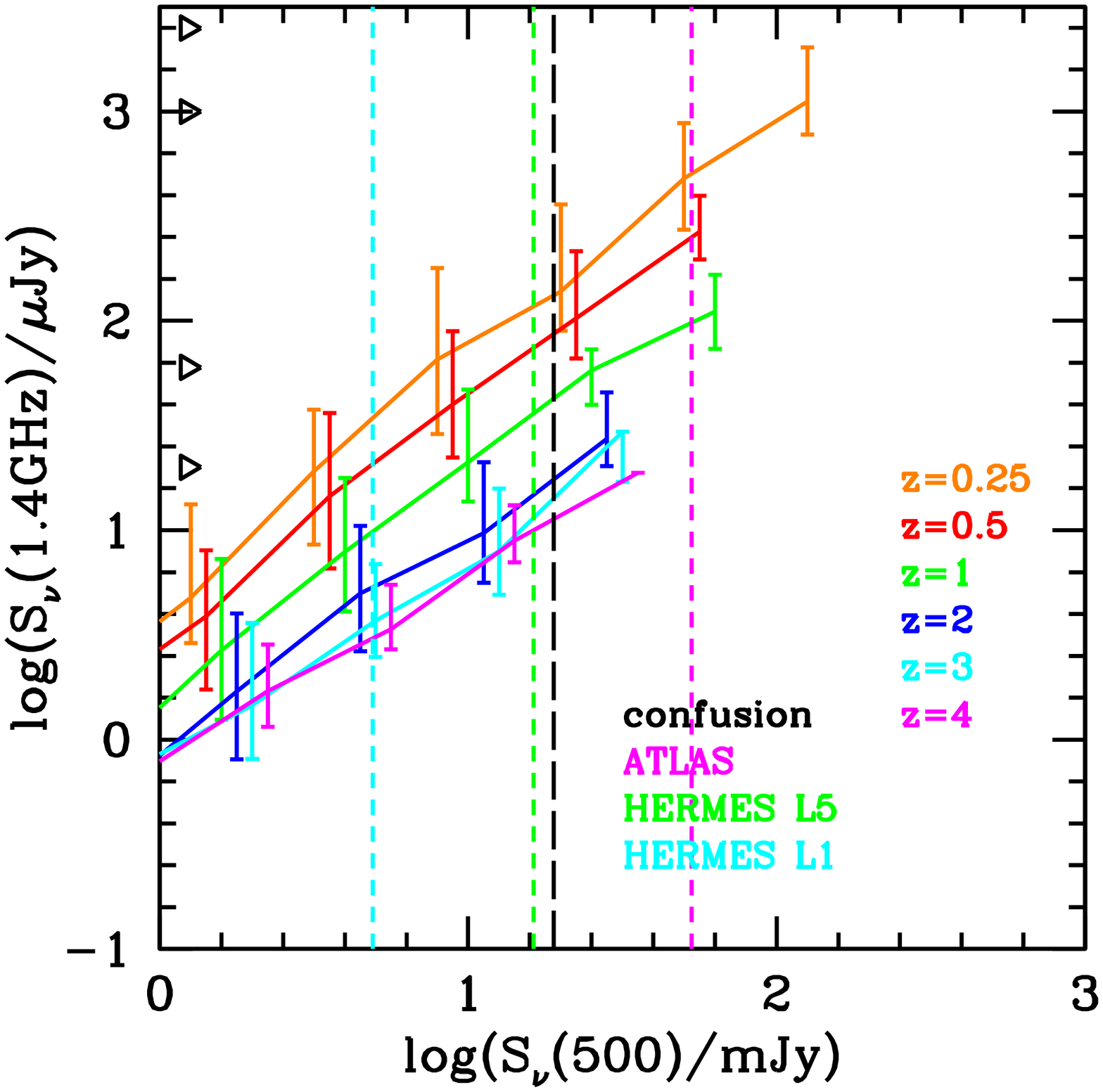}
\end{minipage}

\end{center}

\caption{Median fluxes or magnitudes at other wavelengths for galaxies
  selected at 500$\mum$. (a) GALEX NUV. (b) r-band. (c) IRAC
  3.6$\mum$. (d) MIPS 24$\mum$. (e) SCUBA 850$\mum$. (f) 1.4GHZ. The
  triangles indicate the flux limits for different surveys, as in
  Fig.~\ref{fig:multi-wave-100}. For clarity, we have
  introduced small horizontal offsets between the lines plotted for
  different redshifts.
 }

\label{fig:multi-wave-500}
\end{figure*}

\section{Multi-wavelength predictions}
\label{sec:multi-wavelength}

Obtaining multi-wavelength data complementary to that from \HERSCHEL\
itself will be crucial for achieving the science goals of the
cosmological surveys. Such data will be essential for obtaining
identifications and accurate positions of the sources, for obtaining
accurate redshifts, and for learning more about the physical nature of
these galaxies. It is therefore of interest to see what the model
predicts for the observability of \HERSCHEL\ sources at other
wavelengths. In Figs.~\ref{fig:multi-wave-100},
\ref{fig:multi-wave-250} and \ref{fig:multi-wave-500}, we show
predicted fluxes or AB magnitudes at wavelengths from the UV to the
radio, plotted against the \HERSCHEL\ flux at either 100 or 250 or
500$\mum$. Specifically, we consider the GALEX $NUV$ filter (centred
at $\lambda=0.23\mum$), the SDSS $r$-band (centred at
$\lambda=0.62\mum$), the \SPITZER\ 3.6 and 24$\mum$ bands, the SCUBA
$850\mum$ band, and the VLA $1.4\GHz$ band. The observer-frame NUV,
24$\mum$, $850\mum$ and 1.4GHz bands in the main trace recent star
formation, while the observer-frame 3.6$\mum$ band traces older stars,
and the observer-frame r-band traces older stars at low redshift but
younger stars at high redshift (when it corresponds to the rest-frame
UV). The predicted median fluxes at 0.23, 24 and 850$\mum$ and at 1.4
GHz generally track the fluxes in the \HERSCHEL\ bands fairly well,
but with zero-points which can depend strongly on redshift, depending
on the wavelength. This is particularly the case for the GALEX NUV
band, for which the flux drops rapidly with redshift at $z\gsim 1.5$
when the band falls shortward of the Lyman break in the rest
frame. (Note that we include the effects of absorption by the
intervening IGM when we calculate the UV and optical magnitudes.) The
redshift dependence of the zeropoints is generally least at 1.4 GHz
and 24$\mum$. As expected, the correlation with the \HERSCHEL\ flux is
much weaker for the $r$ and 3.6$\mum$ bands, which are more sensitive
to stellar mass than to the SFR.

For comparison, we note the approximate flux or magnitude limits for
some of the main surveys at these wavelengths (the flux limits for
these surveys are also indicated by triangles along the y-axis in
Figs.~\ref{fig:multi-wave-100}-\ref{fig:multi-wave-500}):

{\em GALEX NUV:} All-sky Imaging Survey (AIS) $m_{AB}=20.8$; Medium
Imaging Survey (MIS)  $m_{AB}=22.7$; Deep Imaging Survey (DIS)
$m_{AB}=24.4$ \citep{Morrissey07}.

{\em SDSS r:} SDSS Legacy Survey $m_{AB}=22.2$ \citep{Abazajian09};
Subaru Deep Field $m_{AB}=27.8$ ($R_C$) \citep{Kashikawa04}; Hubble
Ultra-Deep Field $m_{AB}=30.1$ ($V$ or $i$) \citep{Beckwith06};
Pan-STARRS PS1 $m_{AB}=24.1$ in the $3\pi$ survey and $m_{AB}=26.9$ in the
Medium Deep Survey (MDS) \citep{Chambers06}.

{\em \SPITZER\ 3.6$\mum$:}
SWIRE
$S_{\nu}=3.7\muJy$ \citep{Lonsdale04}; GOODS $S_{\nu}=0.6\muJy$
\citep{Dickinson03}. Also the WISE satellite will survey the whole sky to
$S_{\nu}=120\muJy$ at a wavelength of 3.3$\mum$ \citep{Wright07}.

{\em \SPITZER\ 24$\mum$:} SWIRE
$S_{\nu}=106\muJy$ \citep{Lonsdale04}; GOODS $S_{\nu}=70\muJy$
\citep{Chary04}. Also WISE will survey the whole sky to
$S_{\nu}=2600\muJy$ at a wavelength of 23$\mum$ \citep{Wright07}.

{\em SCUBA 850$\mum$:} HDF $S_{\nu}=2\mJy$ \citep{Hughes98}; SHADES
$S_{\nu}=8\mJy$ \citep{Mortier05}
SASSy\footnote{http://www.jach.hawaii.edu/JCMT/surveys/sassy/}
$S_{\nu}=150\mJy$. 

{\em 1.4 GHz:} NVSS $S_{\nu}=2.5\mJy$ \citep{Condon98}; FIRST
$S_{\nu}=1\mJy$ \citep{Becker95}; Phoenix $S_{\nu}=60\muJy$
\citep{Hopkins03}; 
SSA 13 $S_{\nu}=20\muJy$ \citep{Fomalont06}.


\begin{figure*}
\begin{center}

\begin{minipage}{7cm}
\includegraphics[width=7cm]{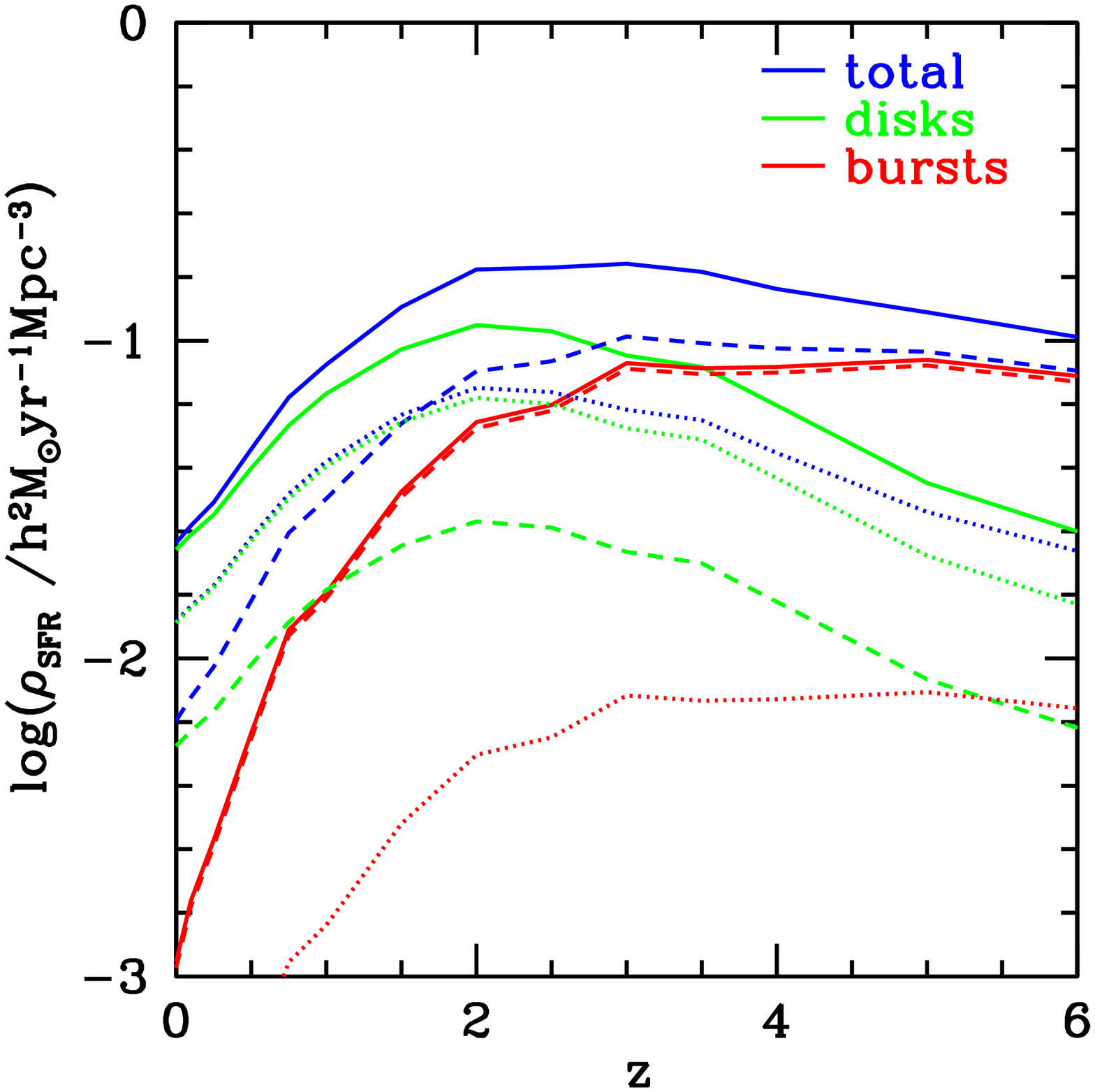}
\end{minipage}
\begin{minipage}{7cm}
\includegraphics[width=7cm]{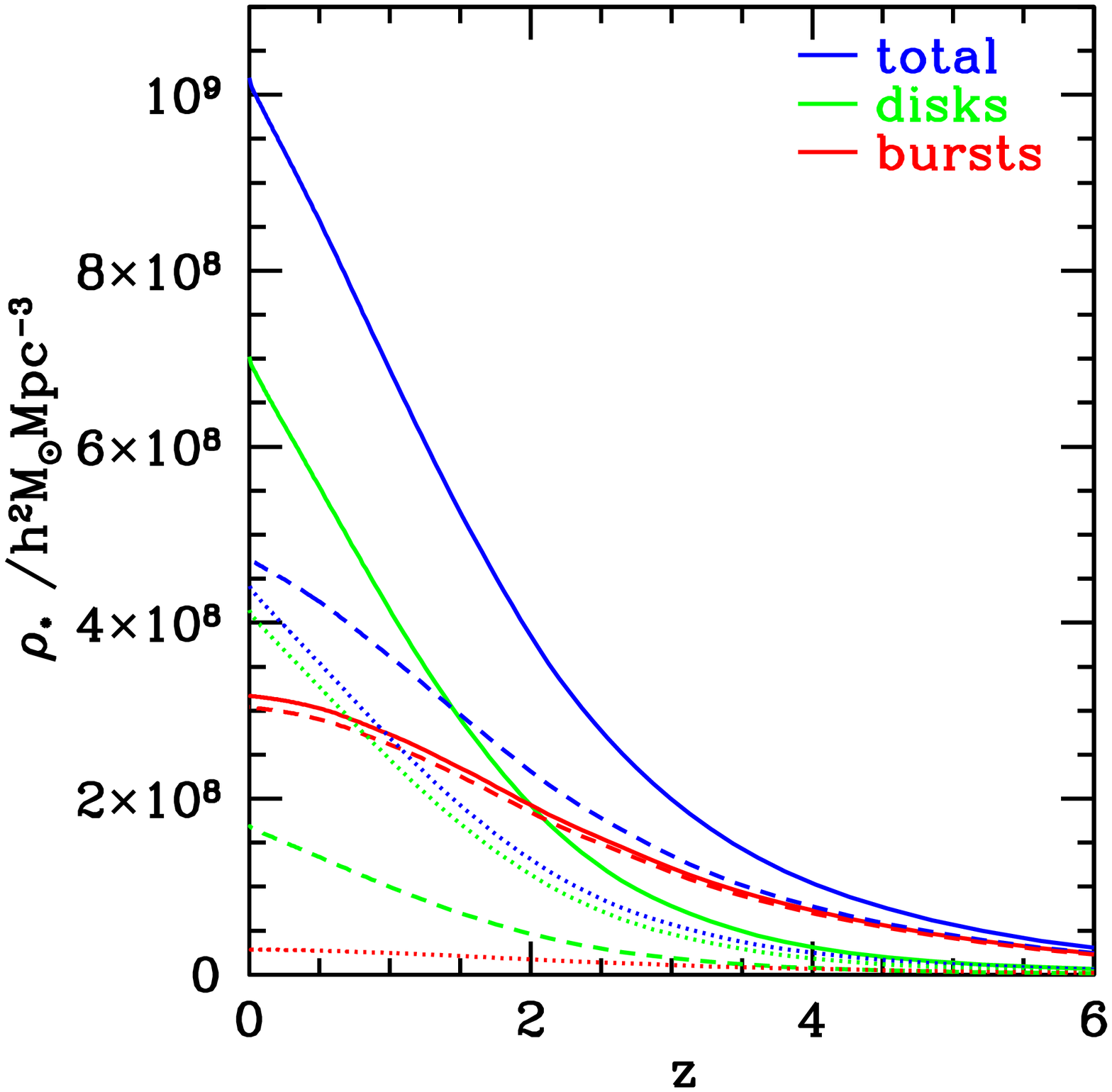}
\end{minipage}

\begin{minipage}{7cm}
\includegraphics[width=7cm]{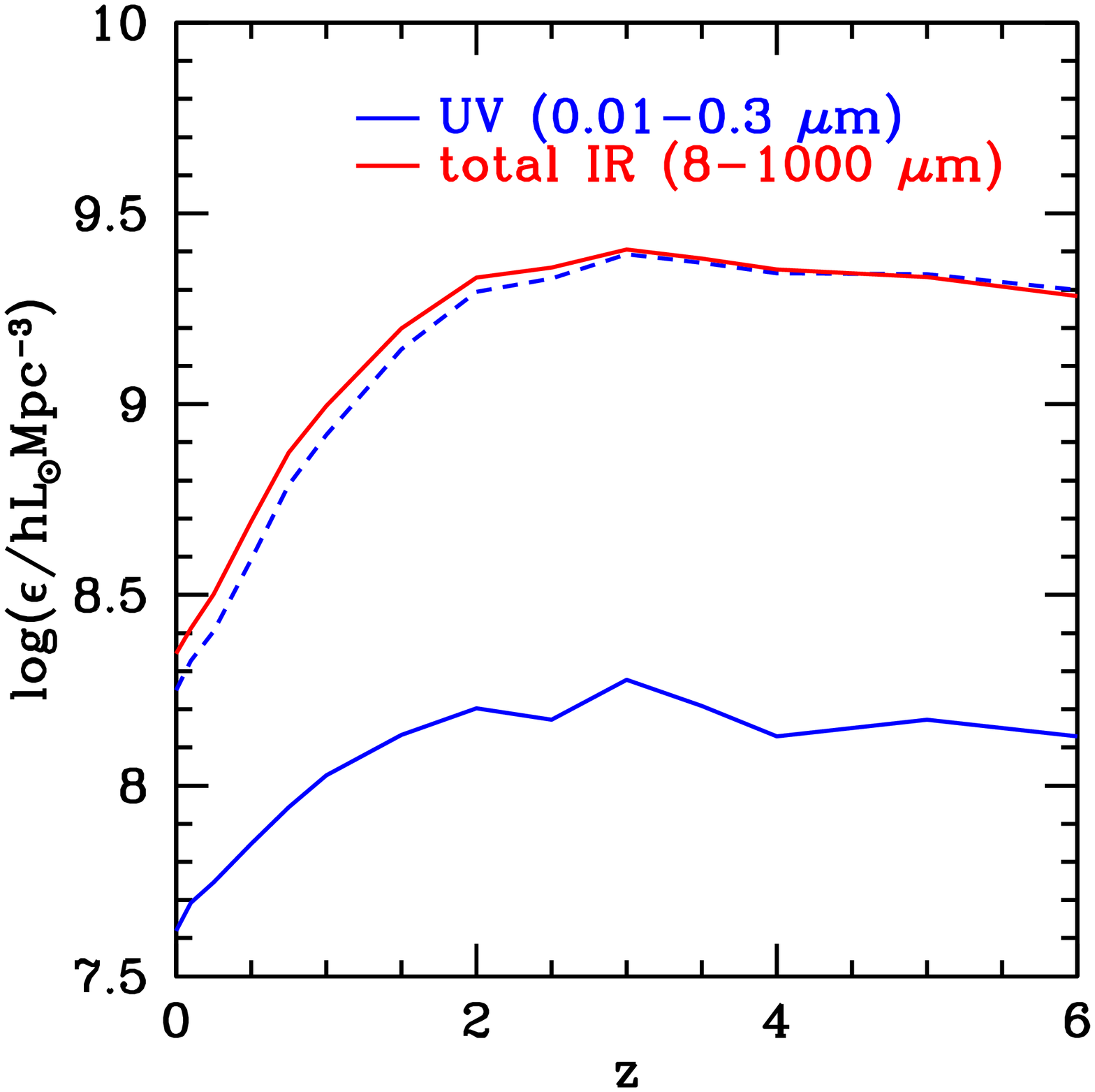}
\end{minipage}
\begin{minipage}{7cm}
\includegraphics[width=7cm]{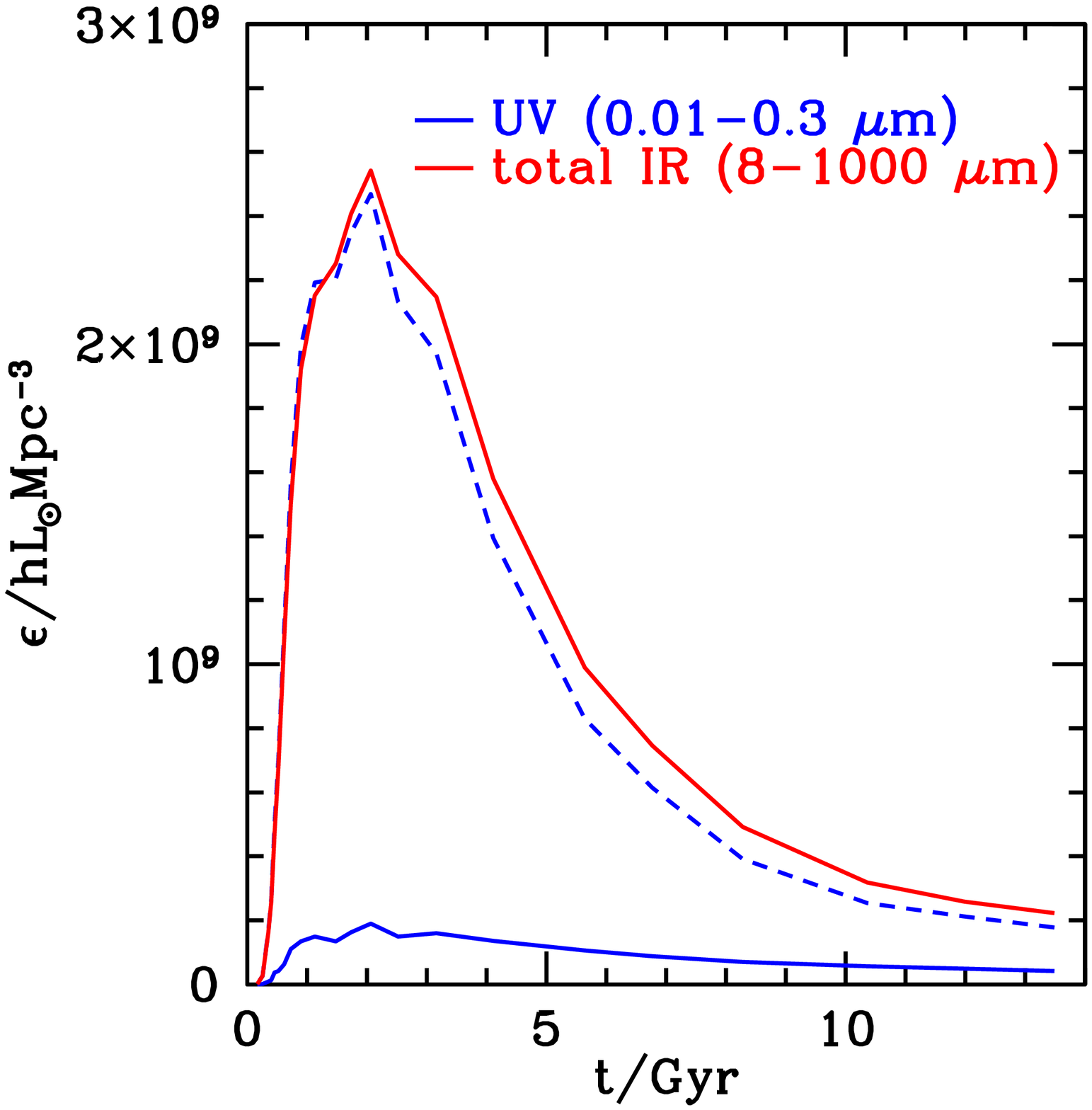}
\end{minipage}

\end{center}

\caption{(a) Cosmic SFR history, showing the total SFR density (blue)
  and the separate contributions from bursts (red) and quiescent disks
  (green). The solid lines show the total SFR density integrated over
  all stellar masses, while the dashed lines show the SFR density in
  massive stars ($m>5\Msol$). The dotted lines show the rate of
  build-up of stellar mass in long-lived stars and remnants, after
  allowing for recycling of gas to the ISM. (b) The time integral of
  SFR density from $t=0$ up to $t(z)$. The line colours and styles are
  identical to (a). (c) Evolution of luminosity density in rest-frame
  UV (with and without dust, shown by the solid and dashed lines) and
  in the mid/far-IR. (d) The same as in (c), except plotted on a
  linear scale against cosmic time. }

\label{fig:SFR-history}
\end{figure*}

\section{Unveiling the cosmic star formation history}
\label{sec:sfrhist}

Since the discovery of the far-IR background by COBE, it has been
known that the bulk of star formation over the history of the Universe
has been obscured by dust. One of the primary goals of \HERSCHEL\ is
to resolve the far-IR background into individual sources and hence
determine the amount of dust-obscured star formation at different
cosmic epochs. How well \HERSCHEL\ will be able to do this depends on
the distribution of the total IR emissivity, $\epsilon_{IR}$, (i.e. the
mean luminosity density per comoving volume) over sources of different
luminosities and over redshift, and how far down in total IR
luminosity, $L_{IR}$, \HERSCHEL\ surveys are able to probe at different
redshifts. This section is devoted to investigating the implications
of our models for this key issue.

We start by showing in the top left panel of
Fig.~\ref{fig:SFR-history} the cosmic star formation history predicted
by our model, both the total star formation density and the separate
contributions to this from ongoing starbursts and from quiescently
star-forming galactic disks (this was earlier shown in
\citealt{Baugh05}). The total SFR density increases by a factor
$\approx 6$ from $z=0$ to $z=2.5$, and then declines very gradually to
higher $z$. We see that the quiescent mode of star formation dominates
the SFR density at redshifts $z \lsim 3$ probed by \HERSCHEL, while
the burst mode dominates at higher redshifts.

We also show by dashed lines in the
same panel the star formation density in massive stars only, which we
define as stars with masses $m > 5\Msol$, which have lifetimes $< 1
\times 10^8 \yr$. We choose this mass range because, when integrated
over the whole galaxy population, such stars dominate  the UV
emissivity from galaxies, and also dominate the heating of the dust
which powers most of the mid- and far-IR emission from galaxies. The
two IMFs in our model (assumed to cover the stellar mass range $0.15 <
m < 120 \Msol$) have very different fractions of their initial stellar
mass in high mass stars: we find $f(m>5\Msol)=0.24$ for the Kennicutt
IMF assumed for quiescent star formation, and $f(m>5\Msol)=0.96$ for
the top-heavy $x=0$ IMF assumed for bursts. (For comparison,
$f(m>5\Msol)=0.22$ for a Salpeter IMF covering the same mass range.)
The SFR density for massive stars therefore evolves more strongly than
that for all stars, increasing by a factor $\approx 15$ from $z=0$ to
a peak at $z\approx 3$. For massive star formation, the burst mode
already dominates the quiescent mode at $z \gsim 1$. Finally, the
dotted lines in the top panel show the rate of build-up of stellar
mass in long-lived stars and stellar remnants, after accounting for
the mass returned to the ISM by dying stars. The fraction of the
initial stellar mass returned in this way is called the recycled
fraction, and depends on the IMF, having values of 0.41 for the
Kennicutt IMF and 0.91 for the $x=0$ IMF. We see that, after allowing
for recycling, the quiescent mode of star formation dominates the
build-up of stellar mass even at the highest redshifts plotted.

We show in the top right panel of Fig.~\ref{fig:SFR-history} the time
integral of the SFR density from $t=0$ up to redshift $z$ for the
different components. Focussing on the integral up to $z=0$, we see
that, integrated over the history of the Universe, 31\% of all star
formation is predicted to occur in the burst mode, but at the present
day this fraction is only 4.8\%. For high mass ($m > 5\Msol$) stars,
64\% of star formation happened in the burst mode over the history of
the universe, while the present-day fraction is 17\%. The burst mode
is thus more important for high mass star formation. However, after
accounting for recycling of mass from dying stars, we find that the
fraction of the present-day stellar mass produced by the burst mode is
only 6.4\%.

In the lower panels of Fig.~\ref{fig:SFR-history} we show how the
model SFR density evolution translates into the evolution of the
cosmic emissivity from galaxies in the UV (defined here as the
integral over the wavelength range $0.01-0.3 \mum$) and the total
mid/far-IR (integrated over the range $8-1000 \mum$, as
previously). The UV emissivity is plotted both with and without dust
extinction. The unextincted UV and total IR emissivities both increase
by factors $\sim 10$ from $z=0$ to $z=3$, and then remain
approximately constant up to $z=6$. This difference in redshift
dependence compared to that for the total SFR density is because both
the UV and total IR emissivities are powered mostly by massive stars
($m \gsim 5\Msol$), and the burst mode of star formation produces a
larger fraction of such stars than the quiescent mode. We see that the
effect of dust extinction on the UV emissivity is predicted to be very
large - integrated over the history of the Universe, 90\% of the UV
energy is predicted to be absorbed by dust. This fraction increases
from 77\% at $z=0$ to 87\% at $z=1$ and 92\% at $z=6$. This emphasizes
how essential it is to measure the cosmic evolution of the total IR
emissivity in order to measure directly the cosmic SFR history, free
of uncertain observational estimates of UV dust extinction. The total
IR emissivity is seen to be very similar to the unextincted far-UV
emissivity at all redshifts, which just reflects the fact that most of
the far-UV radiation is absorbed by dust, and that heating of the dust
by longer wavelength (optical and near-IR) radiation from stars is a
minor contribution to the total when integrated over the whole galaxy
population. (The fraction of the IR emissivity due to heating by
longer wavelength radiation is predicted to be around 20\% integrated
over the history of the Universe, dropping from 40\% at $z=0$ to only
20\% at $z=1.5$ and 10\% at $z=3$.)

\begin{figure}
\begin{center}

\includegraphics[width=7cm]{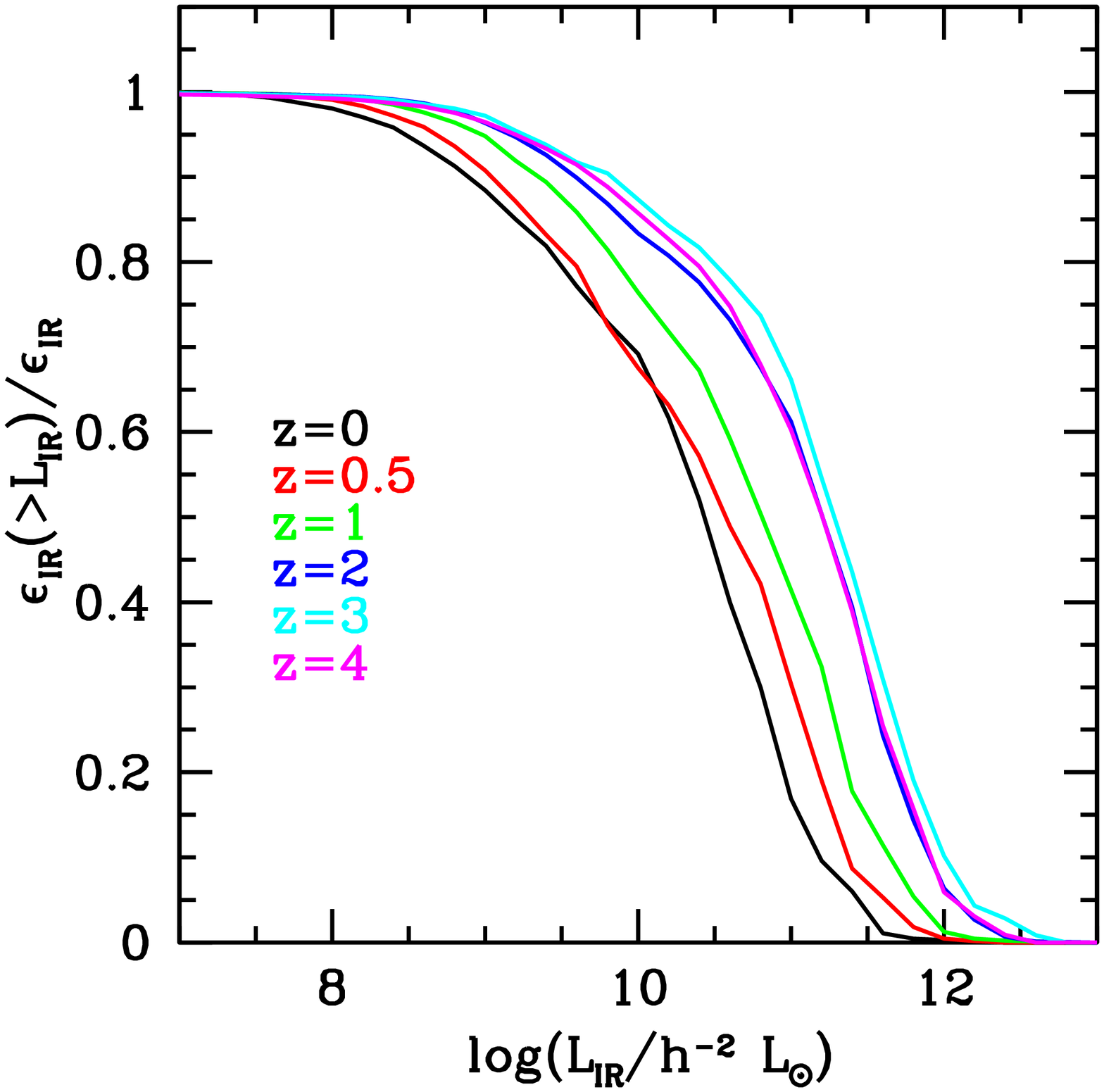}

\includegraphics[width=7cm]{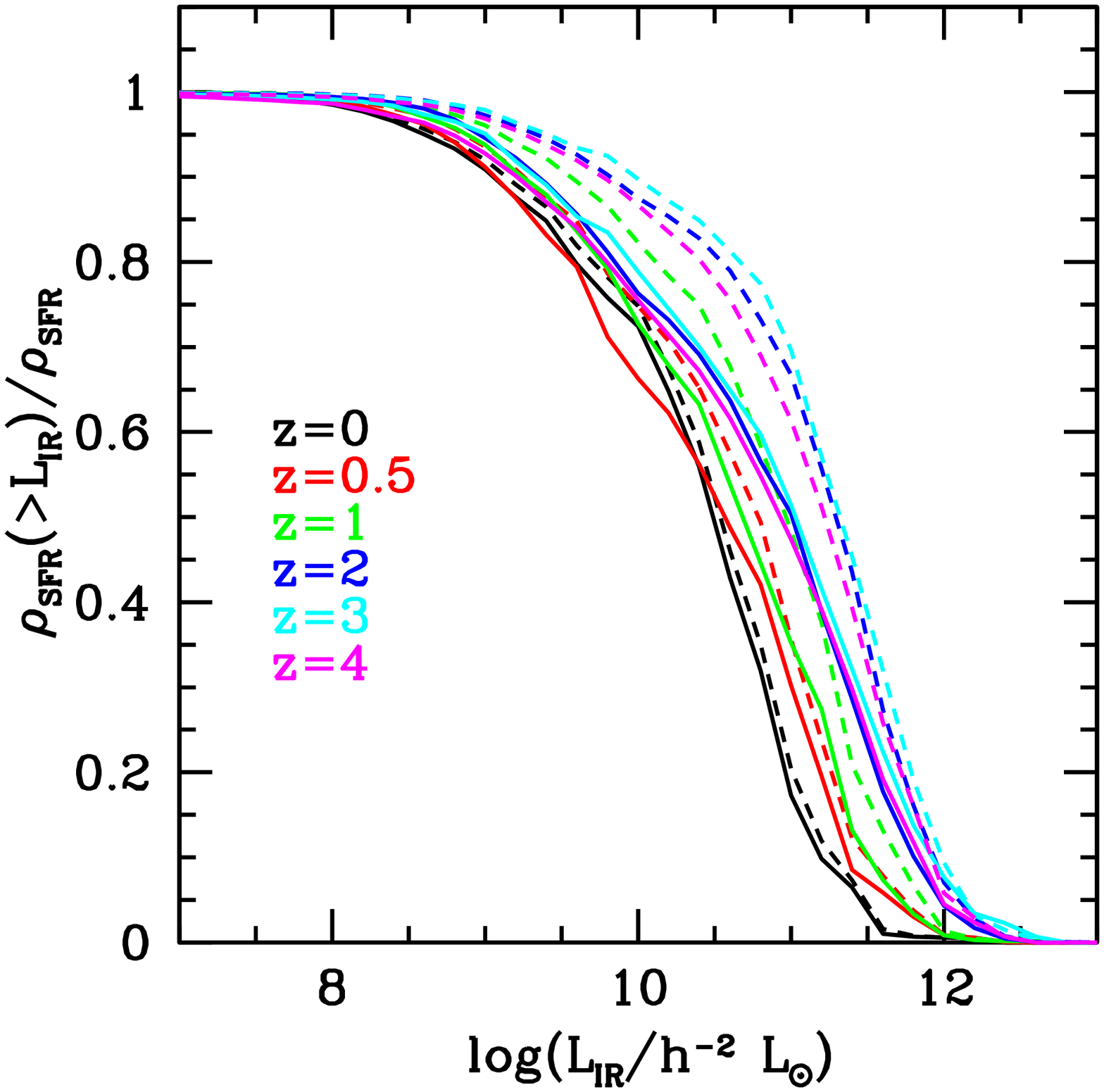}

\includegraphics[width=7cm]{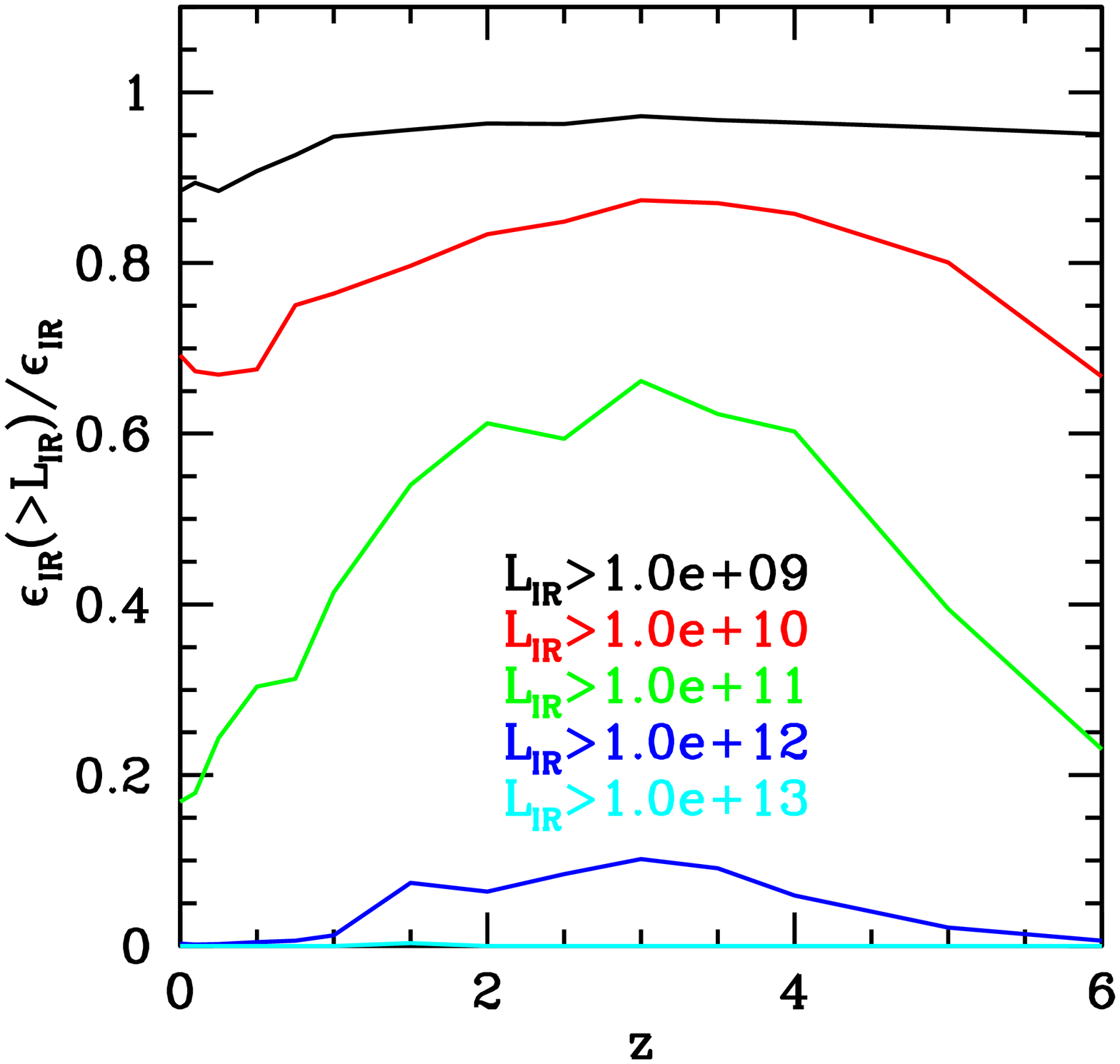}

\end{center}

\caption{(a) Fraction of total IR luminosity density from galaxies
  brighter than $L_{IR}$ at different redshifts. (b) Fractions of
  total SFR (solid lines) and high-mass SFR (dashed lines) from
  galaxies brighter than $L_{IR}$ at different redshifts. (c) Fraction
  of total IR luminosity density from galaxies brighter than $L_{IR}$
  as function of redshift. }

\label{fig:frac_rhoIR}
\end{figure}

\begin{figure}
\begin{center}

\includegraphics[width=7cm]{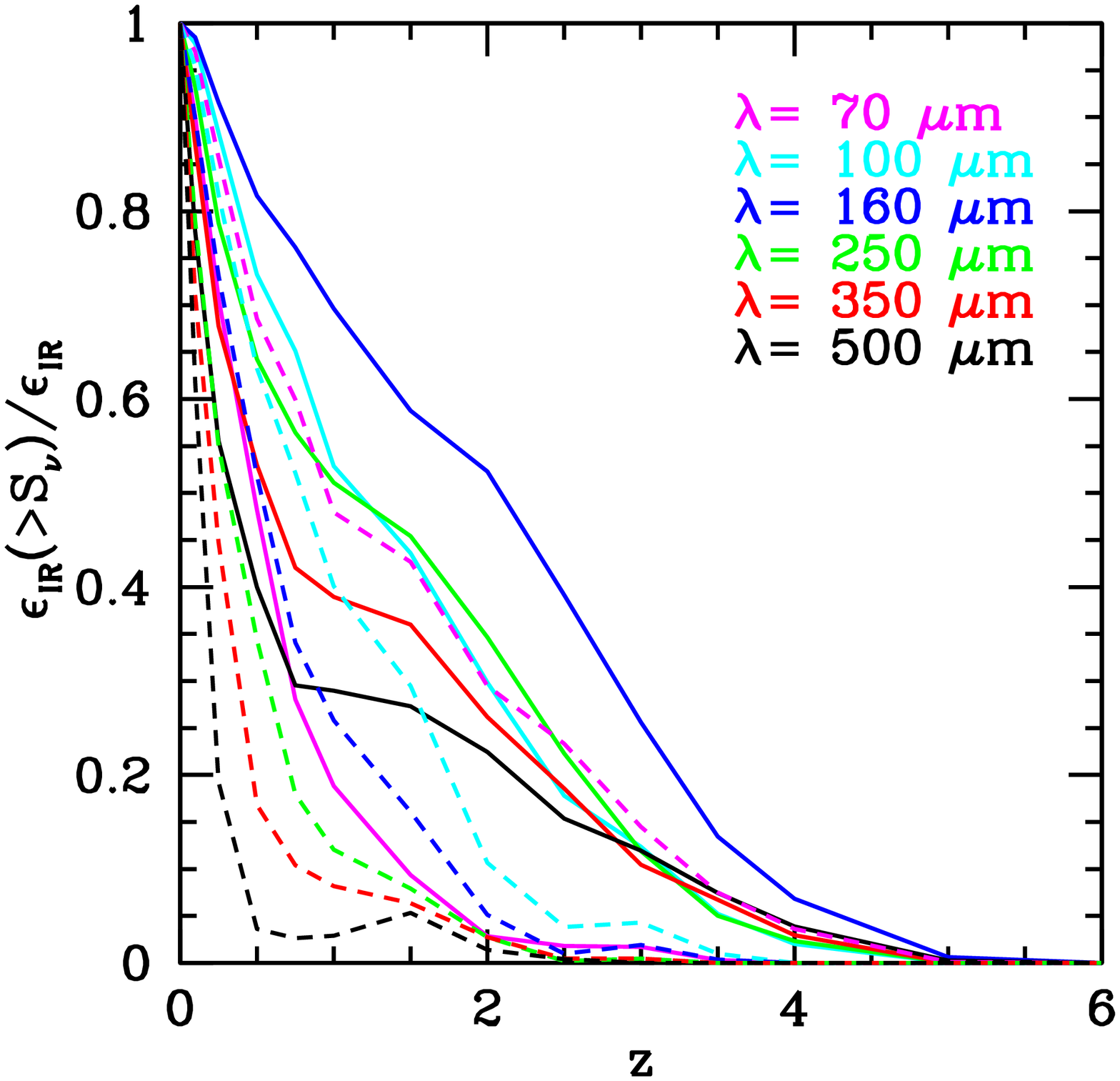}

\includegraphics[width=7cm]{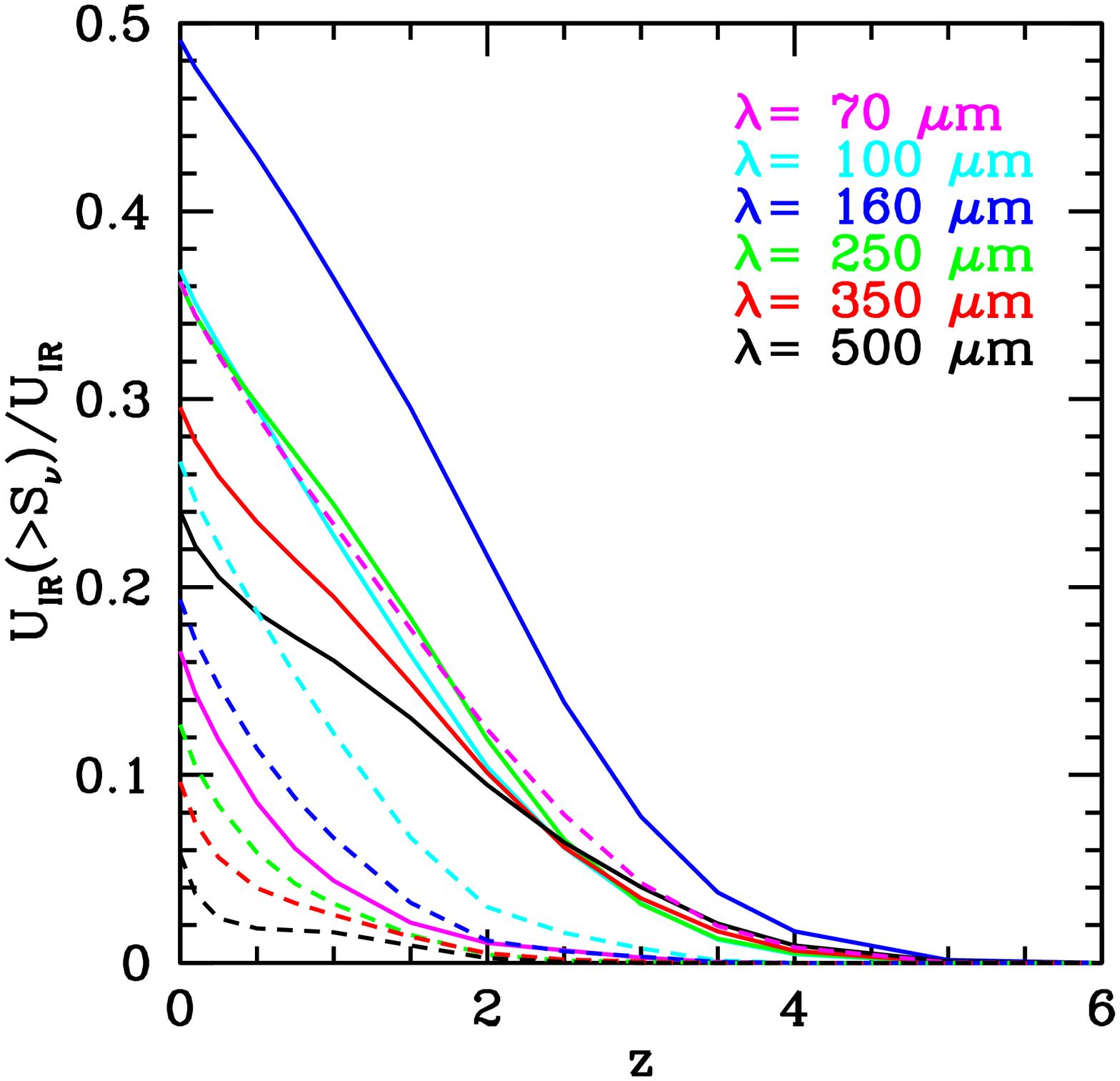}

\end{center}

\caption{(a) Fraction of IR luminosity density from
  galaxies brighter than expected \HERSCHEL\ survey flux limits at different
  wavelengths. The solid lines correspond to the faintest nominal flux
  limit at each wavelength (ignoring confusion) from
  Table~\ref{tab:survey_params}, while the dashed lines correspond to
  the confusion limit at each wavelength from
  Table~\ref{tab:conf_limits}. (b) Similar to (a), but showing the
  fraction of the time integral of the IR luminosity density up to
  redshift $z$ which is in galaxies brighter than flux $S_{\nu}$ in a
  \HERSCHEL\ band.}

\label{fig:frac_rhoIR_Snu}
\end{figure}

We next consider the fraction of the total IR emissivity,
$\epsilon_{IR}$, which is
produced by galaxies brighter than total luminosity $L_{IR}$ for
different redshifts. This is obtained by integrating over the total IR
luminosity function, which was shown in Fig.~\ref{fig:lf-evoln}. The
results are shown in the top and bottom panels of
Fig.~\ref{fig:frac_rhoIR}. The top panel shows the fraction of the IR
emissivity in galaxies brighter than $L_{IR}$ as a function of
$L_{IR}$ for different redshifts, while the bottom panel shows the
same fraction as a function of redshift for a number of different
minimum luminosities. We see from the top panel that the value of
$L_{IR}$ above which 50\% of the IR emissivity is contributed
increases with increasing $z$, from $3\times 10^{10} h^{-2}\Lsol$ at
$z=0$ to $2\times10^{11} h^{-2}\Lsol$ at $z=3$, due to the general
brightening in the luminosity function with $z$ discussed in
\S\ref{sec:LF-evoln}. This evolutionary effect partly compensates for
the effect of increasing luminosity distance, which generally means
that the minimum $L_{IR}$ for which we can detect galaxies in surveys
increases with $z$.

The middle panel in Fig.~\ref{fig:frac_rhoIR} shows the fraction of
star formation occuring in galaxies brighter than $L_{IR}$, as a
function of $L_{IR}$ for different redshifts. We show this both for
the total SFR integrated over all stellar masses (solid lines) and for
the high-mass ($m>5\Msol$) SFR (dashed lines). As a consequence of the
bright end of the total IR LF being dominated by bursts with a
top-heavy IMF, galaxies brighter than a given $L_{IR}$ account for a
larger fraction of the high-mass SFR than of the total SFR. For
example, at $z=3$, galaxies with $L_{IR} > 2\times10^{11} h^{-2}\Lsol$
account for 50\% of the high-mass SFR but only 36\% of the total SFR.

The critical question is: what fraction of the total IR emissivity
will different \HERSCHEL\ surveys be able to resolve at different
redshifts? To answer this, we plot in the top panel of
Fig.~\ref{fig:frac_rhoIR_Snu} the fraction of the total IR emissivity
$\epsilon_{IR}$ produced by galaxies brighter than some flux $S_{\nu}$
as a function of redshift. Each of the six \HERSCHEL\ imaging bands is
plotted in a different colour, and for each band we show the result
for two different flux limits: the solid line shows the result for the
faintest planned flux limit at that wavelength ignoring confusion,
taken from the list of surveys in Table~\ref{tab:survey_params}, while
the dashed line shows the result at the estimated confusion limit,
taken from Table~\ref{tab:conf_limits}. The lower panel of
Fig.~\ref{fig:frac_rhoIR_Snu} instead shows the fraction of the
time-integrated emissivity $U_{IR}(z) = \int_0^{t(z)} \epsilon_{IR}\,
dt$ which is resolved at a given flux limit.  We see from this figure
that if source confusion is ignored (or can be circumvented), then the
GOODS-Herschel Ultra-deep survey at 160$\mum$ (with a planned flux
limit of 0.9$\mJy$) should resolve the largest fraction of the total
IR emissivity into sources at all redshifts. Integrated over all
redshifts, a survey at this flux limit would resolve 49\% of all IR
emission into sources. On the other hand, if confusion sets a hard
limit, then the GOODS-Herschel Ultra-deep survey at 100$\mum$ (with a
confusion limit of 1.6$\mJy$) should instead resolve the largest
fraction of the IR emissivity at all redshifts, and a total fraction
of 27\% when integrated over all redshifts. (In principle, a
confusion-limited survey at 70$\mum$ would resolve a larger fraction,
but no such survey is planned). Note that the question we have
asked (and answered) here is different from asking what fraction of
the present-day cosmic IR background (CIB) is resolved into sources at
different fluxes and wavelengths. The energy density in the CIB
at $z=0$ is given by the integral $\int_0^{t_0} \epsilon_{IR}/(1+z) \,
dt$, which differs from the time-integrated emissivity $U_{IR}(z=0)$
by a factor $1/(1+z)$ in the integrand due to redshifting of the
photon energies.

In summary, the planned surveys with \HERSCHEL\ should be able to
resolve into sources around 30-50\% of the total IR dust emission, and thus a
similar fraction of the dust-obscured massive star formation, over the
history of the Universe.


\begin{figure}
\begin{center}
\includegraphics[width=7cm]{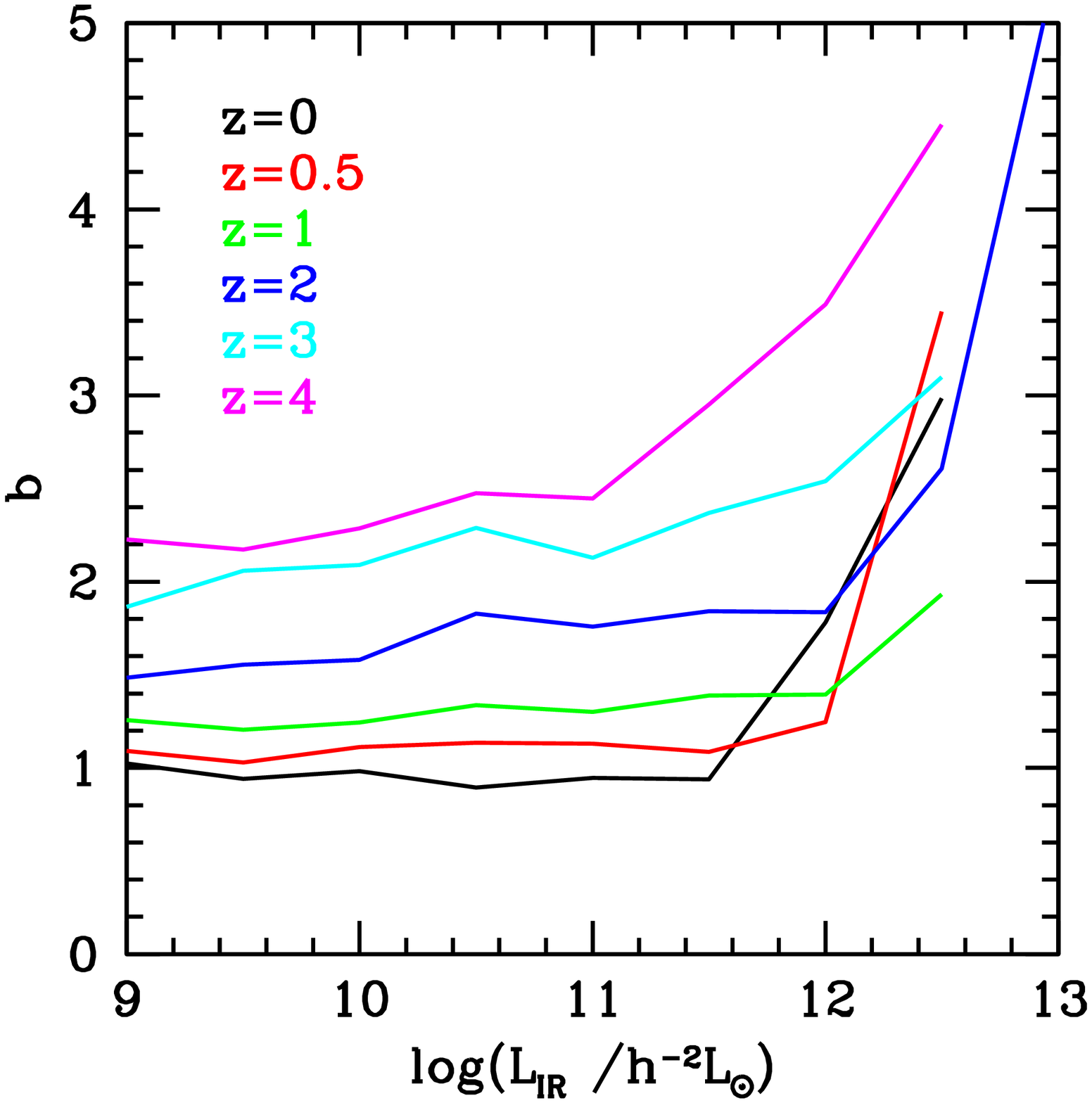}
\includegraphics[width=7cm]{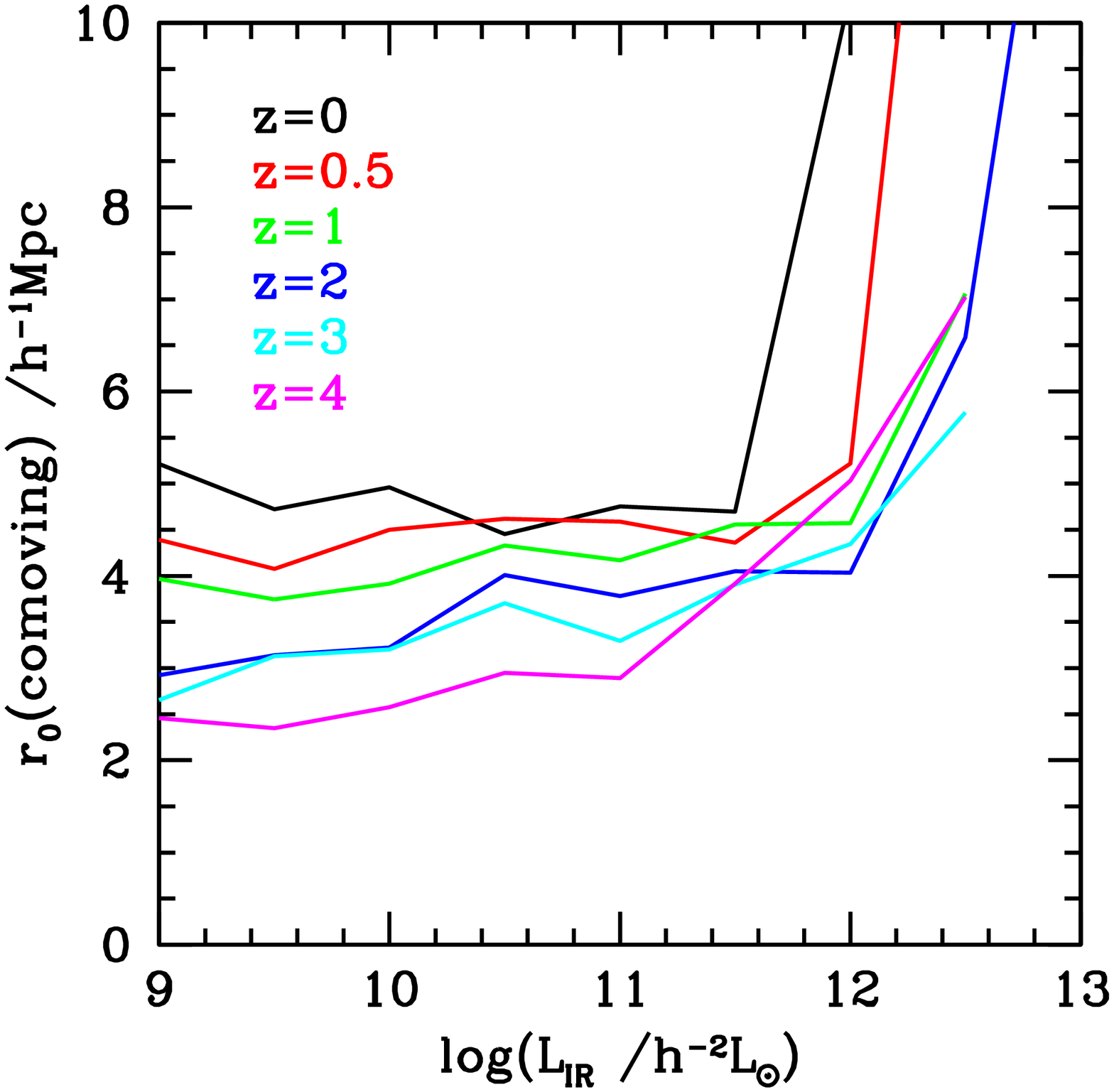}
\end{center}

\caption{(a) Mean clustering bias $b$ as a function of $L_{IR}$ for
  galaxies at different redshifts, as indicated in the key. (b)
  Clustering length $r_0$ obtained from the bias and power spectrum as
  function of $L_{IR}$ for different redshifts.  }

\label{fig:clustering}
\end{figure}

\section{Clustering of galaxies in the far-IR}
\label{sec:clustering}

The final topic we consider is the predicted clustering of galaxies
detected in the far-IR by \HERSCHEL. In the \GALFORM\ model, the
clustering of galaxies is determined (on large scales) by that of
their host dark matter halos, and (on small scales) by the spatial
distribution of galaxies within dark matter halos.  Observational
measurements of clustering for galaxies of different classes and
luminosities can thus provide robust information about the masses of
the dark halos hosting these galaxies. Such information is difficult
to obtain in other ways, particularly for high redshift galaxies (for
example, measurements of galaxy circular velocities do not probe the
radii where most of the halo mass is, and so only weakly constrain
halo masses). Clustering measurements can thus provide an essential
test of the link between mass (mostly in dark matter) and light, which
provides a fundamental test of the galaxy formation model.

We limit ourselves here to a few predictions for large-scale
clustering, and defer a more detailed analysis of clustering
properties to a future paper. On large scales, where the density
fluctuations for the dark matter are still in the linear regime, the
clustering of galaxies and halos can be described by a linear bias
factor $b$, such that $\xi(r) = b^2 \xi_{dm}(r)$, where $\xi(r)$ is
the 2-point corelation function for the galaxies or halos, and
$\xi_{dm}(r)$ is that for the dark matter. We can calculate the bias,
$b_{halo}(M,z)$, for halos as a function of halo mass and redshift
using the analytical approximation of \citet{Sheth01}. We can then
calculate the bias, $b_{gal}(L_{IR},z)$, for galaxies of a specified
IR luminosity as a mean of the halo bias, weighted by the distribution
of host halo masses for galaxies of this luminosity and redshift. The
latter information is provided by the semi-analytical model. We show
the results of this calculation in the top panel of
Fig.~\ref{fig:clustering}. We see from this that the clustering bias
is predicted to depend only weakly on total IR luminosity over the
range $10^9 \lsim L_{IR} \lsim 10^{11.5} h^{-2}\Lsol$, but to increase
steeply at the highest luminosities. The weak dependence on IR
luminosity stems from the weak correlation of IR luminosity with host
halo mass, which was seen in Fig.~\ref{fig:Mhalo_flux}. The bias at a
given IR luminosity is seen to increase gradually with redshift, from
$b\sim 1$ at $z=0$, to $b\sim 2$ at $z=3$, for the luminosity range
where $b$ is roughly constant.

By combining our analytical estimate of the bias with an estimate of
the correlation function, $\xi_{dm}(r,z)$, of the dark matter, we can
calculate the correlation function of the galaxies,
$\xi_{gal}(r,L_{IR},z) = b_{gal}^2 (L_{IR},z) \xi_{dm}(r,z)$. This
estimate will be valid on scales large enough that the linear bias
approximation is valid. We calculate the dark matter correlation
function using the approximate analytical model of \citet{Smith03},
which allows for non-linear effects in the dark matter evolution. From
this, we can calculate the correlation length, $r_0$, which we define
by $\xi(r_0)=1$, and which provides a simple, directly measureable,
characterization of how strongly a particular population of galaxies
is clustered. This approach to calculating $r_0$ is valid provided $b
\geq 1$, since then $\xi_{dm}(r_0) \leq 1$, and the dark matter is
still approximately in the linear regime on the scale $r_0$. We show
our predictions for $r_0$ in the lower panel of
Fig.~\ref{fig:clustering}. We see from this that, as for the bias,
$r_0$ is almost independent of luminosity over the range $10^9 \lsim
L_{IR} \lsim 10^{11.5} h^{-2}\Lsol$, but increases steeply for $L_{IR}
\gsim 10^{11.5} h^{-2}\Lsol$. At a given luminosity, $r_0$ decreases
with increasing redshift, but typically only gradually.  For $10^9
\lsim L_{IR} \lsim 10^{11.5} h^{-2}\Lsol$, $r_0$ decreases from $r_0
\sim 5 h^{-1} \Mpc$ at $z=0$ to $r_0 \sim 3 h^{-1} \Mpc$ at $z=3$.

\citet{Viero09} have estimated the clustering of
galaxies in the SPIRE bands from the BLAST observations by measuring
the angular power spectra of the total intensity maps. From this,
they estimate a bias $b\sim 4$ for the galaxies responsible for the
far-IR background, larger than the values typical in our
model. However, their estimate of the bias relies on a model for the
source redshift distribution, and assumes that the bias is independent
of both redshift and IR luminosity. The latter assumption seems
unrealistic in the light of our own predictions for the bias shown in
Fig.~\ref{fig:clustering}(a). We therefore do not make a more detailed
comparison with their results.

\begin{figure*}
\begin{center}

\begin{minipage}{7cm}
\includegraphics[width=7cm]{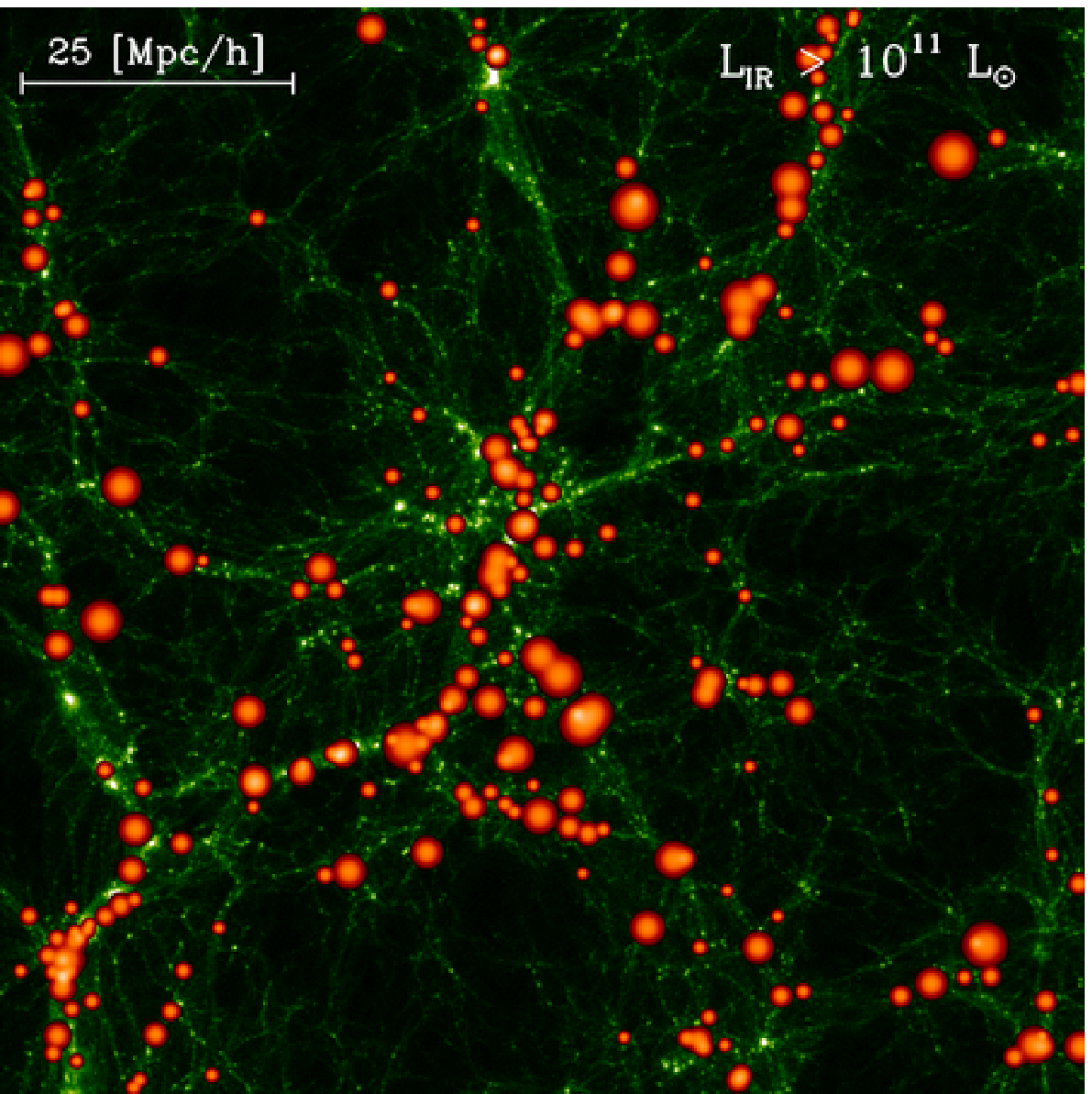}
\end{minipage}
\hspace{1cm}
\begin{minipage}{7cm}
\includegraphics[width=7cm]{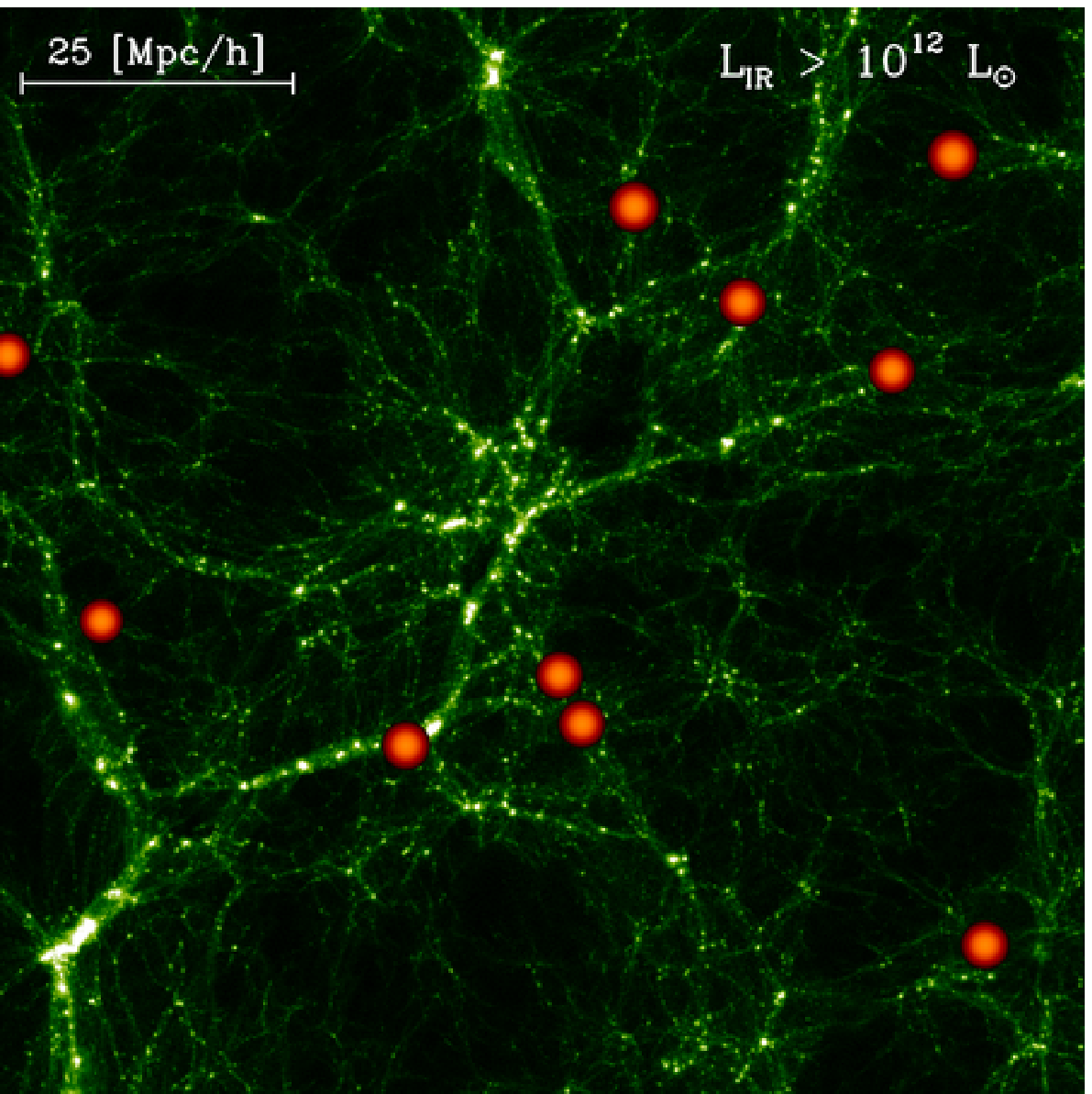}
\end{minipage}

\begin{minipage}{7cm}
\includegraphics[width=7cm]{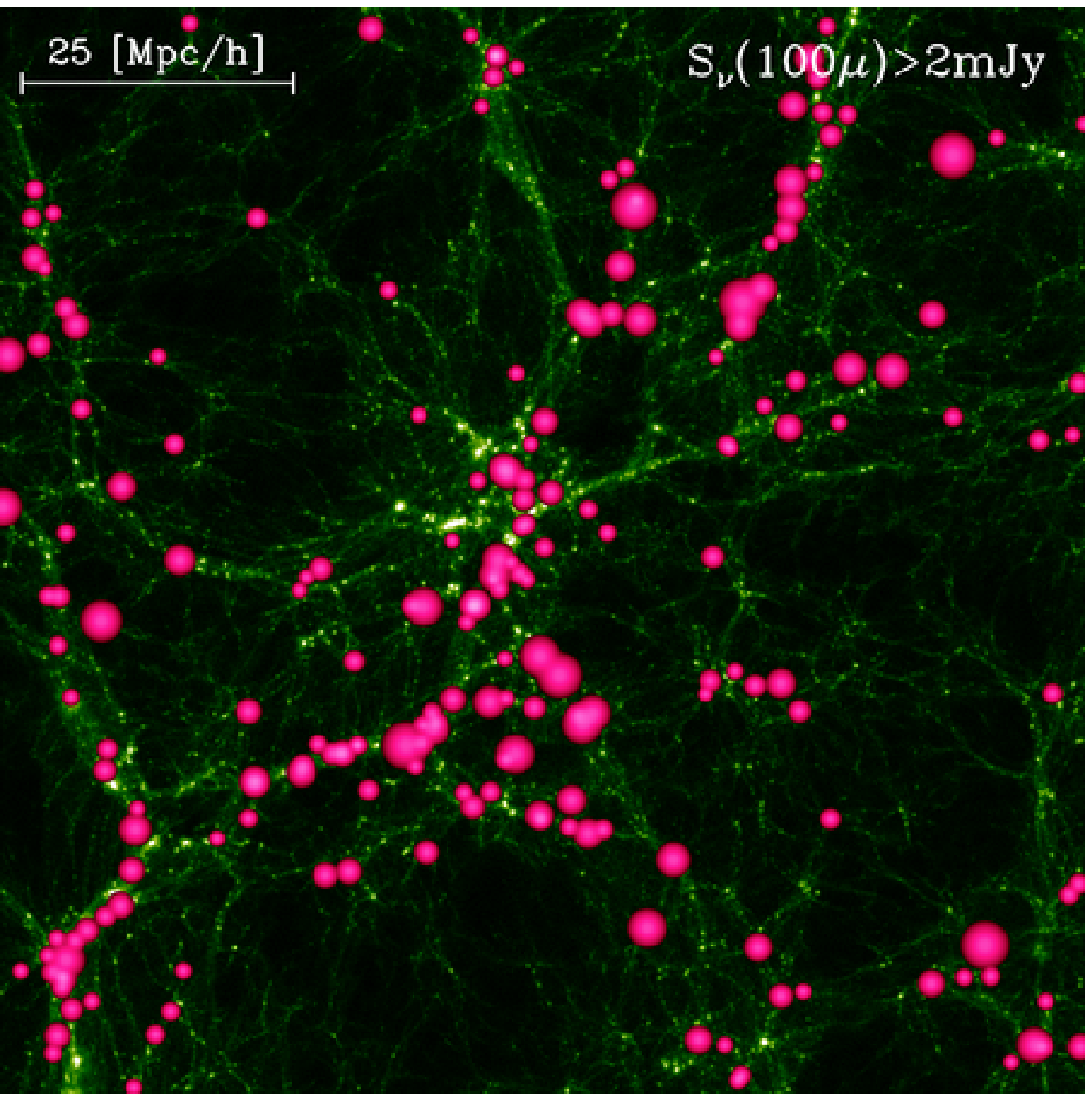}
\end{minipage}
\hspace{1cm}
\begin{minipage}{7cm}
\includegraphics[width=7cm]{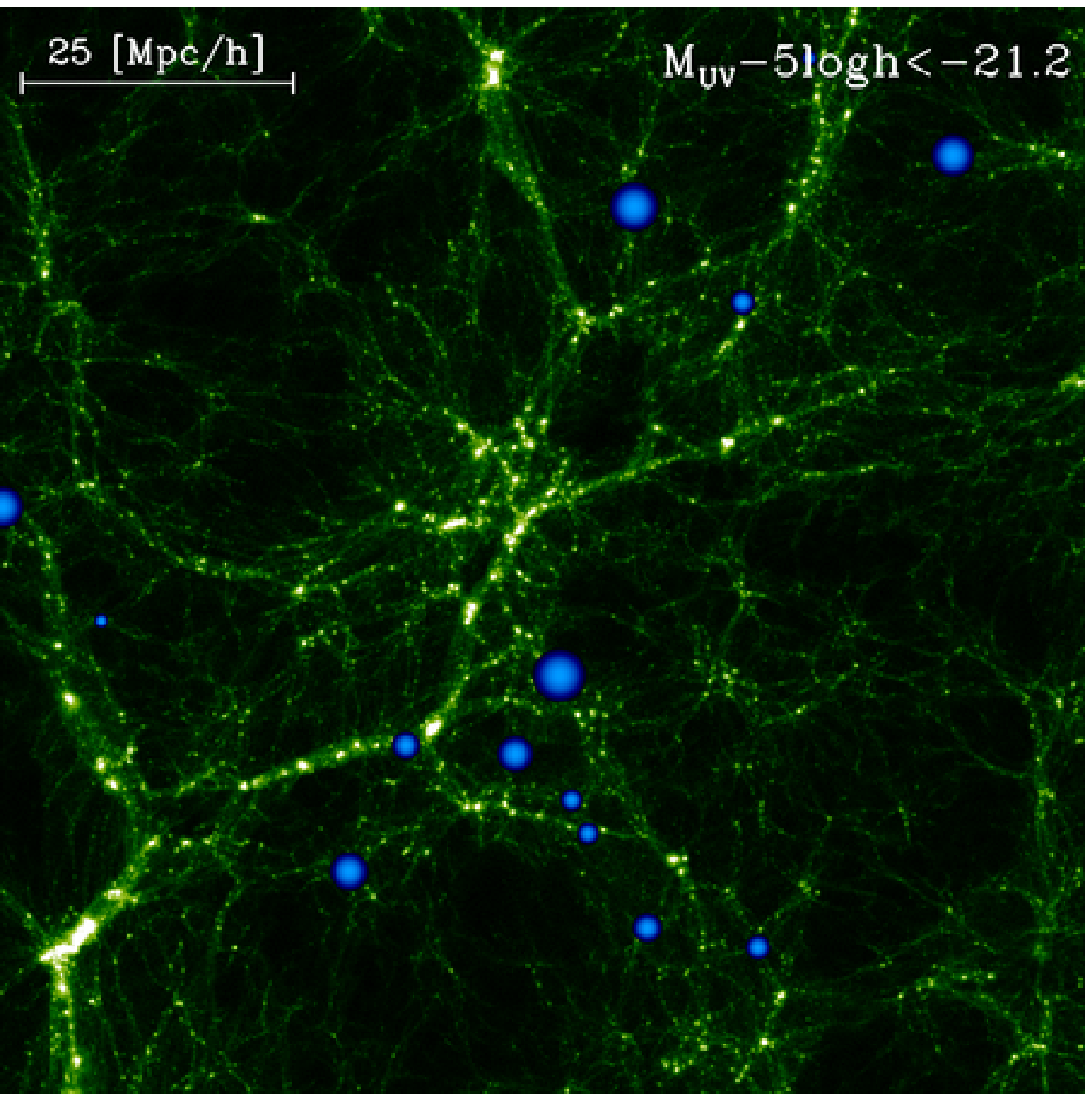}
\end{minipage}

\end{center}

\caption{Images of a simulated slice of the universe 100$\hMpc$ wide
  and 10$\hMpc$ thick at $z=1$. Each panel shows the same slice, with
  the dark matter density plotted in green, and with the galaxies
  plotted as coloured blobs, the blob size increasing with luminosity
  or flux. Each panel shows galaxies selected at a different
  wavelength and/or lumninosity/flux. (a) $L_{IR}>10^{11} \hLsol$. (b)
  $L_{IR}>10^{12} \hLsol$. (c) $S_{\nu}(100\mum)>2\mJy$. (d)
  $M_{UV}-5logh<-21.2$. }

\label{fig:slice}
\end{figure*}

Finally, in Fig.~\ref{fig:slice}, we show simulated images of a slice
through the universe at $z=1$, 100$\hMpc$ wide by 10$\hMpc$
thick. These have been obtained by combining our semi-analytical model
with the Millennium dark matter simulation \citep{Springel05}, using
the same method as in \citet{Orsi08}. In these images, the dark matter
distribution is shown in green, while the positions and luminosities
of the galaxies are shown as coloured blobs. For making these images,
we have calculated the far-IR luminosities of galaxies using the
simplified dust emission model described in \citet{Gonzalez09b}, since
running the \GRASIL\ dust code separately on each galaxy in the
Millennium simulation would not have been feasible computationally. In
a future paper, we will present images calculated using a more
accurate approximation to \GRASIL\ based on Artificial Neural Nets
\citep{Almeida09}. The top two panels show galaxies selected based on
their total IR luminosities, brighter than $10^{11} h^{-2}\Lsol$ and
$10^{12} h^{-2}\Lsol$ respectively in the left and right panels. The
lower left panel shows galaxies selected to have 100$\mum$ fluxes
brighter than 2$\mJy$, which is similar to the planned flux limit in
the PEP GOODS-S survey. Finally, the lower right panel shows galaxies
selected according to their dust-extincted rest-frame far-UV
luminosities, with $M_{AB}(1500\AA) -5\log h < -21.2$. This absolute
magnitude limit has been chosen because it corresponds to about the
same SFR in a completely unextincted galaxy as does $L_{IR}=10^{11}
h^{-2}\Lsol$ in a completely extincted galaxy. Comparing the top left
and lower right panels illustrates how incomplete surveys for
star-forming galaxies can be if they use only observations in the
rest-frame far-UV and ignore the far-IR.

\section{Summary and Conclusions}
\label{sec:conc}

We have used a detailed model of hierarchical galaxy formation and of
the reprocessing of starlight by dust to make predictions for the
evolution of the galaxy population at the far-infrared wavelengths
(60-670$\mum$) which will be probed by observations with the
\HERSCHEL\ Space Observatory. We calculated galaxy formation in the
framework of the $\Lambda$CDM model using the \GALFORM\ semi-analytical
model, which includes physical treatments of the hierarchical assembly
of dark matter halos, shock-heating and cooling of gas, star
formation, feedback from supernova explosions and photo-ionization of
the IGM, galaxy mergers and chemical enrichment. We computed the IR
luminosities and SEDs of galaxies using the \GRASIL\ multi-wavelength
spectrophotometric model, which computes the luminosities of the
stellar populations in galaxies, and then the reprocessing of this
radiation by dust, including radiative transfer through a two-phase
dust medium, and a self-consistent calculation of the distribution of
grain temperatures in each galaxy based on a local balance between
heating and cooling. 

Our galaxy formation model incorporates two different IMFs: quiescent
star formation in galaxy disks occurs with a normal solar
neighbourhood IMF, but star formation in starbursts triggered by
galaxy mergers happens with a top-heavy $x=0$ IMF. In a previous paper
\citep{Baugh05}, we found that the top-heavy IMF in bursts was
required in order that the model reproduces the observed number counts
of the faint sub-mm galaxies detected at 850 $\mum$, which are
typically ultra-luminous starbursts at $z \sim 2$, with total IR
luminosities $L_{IR} \sim 10^{12} - 10^{13} \Lsol$. We subsequently
found that the same model also reproduces the evolution of the galaxy
population at mid-infrared wavelengths found by \SPITZER\
\citep{Lacey08}. We have used the same model, with identical
parameters, to make predictions for \HERSCHEL\ in the present paper.

We began (\S\ref{sec:LF-evoln}) by showing the predicted evolution of
the galaxy luminosity function at far-IR wavelengths. This brightens
by a factor $\sim 10$ going back from $z=0$ to $z\sim 3$, reflecting
both the evolution of star formation rates in galaxies and the
increasing importance of the top-heavy burst mode with increasing
redshift. We next (\S\ref{sec:counts_redshifts}) presented predictions
for galaxy number counts as functions of flux in the \HERSCHEL\ PACS
and SPIRE imaging bands (covering the wavelength range
70-500$\mum$). We calculated the confusion limits for the \HERSCHEL\
bands, and found that source confusion is likely to be a serious
problem for all of the deepest cosmological surveys planned (PEP,
GOODS-Herschel and HERMES) (except at 70$\mum$). The number of faint
galaxies which can be resolved in these surveys will depend
dramatically on whether or not the confusion limit can be
circumvented, e.g. by using multi-wavelength data. We also
investigated the predicted redshift distributions in these deep
cosmological surveys and in the wide-area ATLAS survey. We found that
the deep surveys should reach median redshifts $\sim 1-1.8$, depending
on the wavelength and on whether it is possible in practice to probe
sources fainter than the confusion limit. The highest median redshift
(1.4-1.8, depending on confusion) should be attained in the HERMES
survey at 500$\mum$. For the ATLAS survey, the median redshift should
be 0.2-0.4, with the highest value at 250$\mum$. At the faintest
fluxes and highest redshifts, the galaxy source population is
predicted to be dominated by starbursts.

Following on from the predictions for source counts and redshifts, in
\S\ref{sec:properties} we showed what the model predicts for some of
the basic physical properties of galaxies detected in different
\HERSCHEL\ bands as functions of flux and redshift. As expected, at
each redshift there are nearly linear corelations between the fluxes
in different bands and the total IR luminosity $L_{IR}$ (integrated
over the wavelength range 8-1000$\mum$). On the other hand, the
relation between the total star formation rate SFR and flux shows more
non-linearity, principally due to the different IMFs assumed in
starburst galaxies (which dominate at high luminosity) and quiescent
galaxies (which dominate at low luminosity). The deepest surveys
should resolve galaxies with $L_{IR} \gsim 10^{11} \hLsol$ at $z=1$
and $L_{IR} \gsim 10^{12} \hLsol$ at $z=2$, corresponding to SFRs
$\gsim 10-100\hMsol\yr^{-1}$. The stellar and dark halo masses of
\HERSCHEL\ galaxies show much weaker correlations with flux and
redshift, and with much more scatter, in large part because of the
role of starbursts triggered by galaxy mergers. In the deep
surveys at $z=1-2$, the typical \HERSCHEL-detected galaxy should have
a stellar mass $\sim 10^{10} \hMsol$ and a halo mass $\sim 10^{12}
\hMsol$, and a typical gas mass $\sim 10^{10} \hMsol$. Finally, the
morphologies of the galaxies detected in the \HERSCHEL\ cosmological
surveys should be quite mixed, rannging from highly bulge-dominated to
highly disk dominated systems, reflecting our assumption that
starbursts can be triggered by both major and minor galaxy mergers.

Since our model is a multi-wavelength model, we used it in
\S\ref{sec:multi-wavelength} to predict what should be the fluxes at
other wavelengths (ranging from the far-UV to the radio) of galaxies
detected in \HERSCHEL\ surveys. Follow-up data at other wavelengths
will be essential both to determine redshifts of \HERSCHEL\ sources
and to investigate their physical properties. Specifically, we
presented predictions for the \GALEX\ NUV band, the SDSS $r$ band, the
\SPITZER\ 3.6 and 24$\mum$ bands, the \SCUBA\ 850$\mum$ band and
finally the 1.4GHz radio flux.

One of the primary goals of \HERSCHEL\ will be to unveil the history
of dust-obscured star formation in the Universe. Therefore, in
\S\ref{sec:sfrhist} we presented the predictions of our model for the
cosmic star formation history and for the evolution of the cosmic
emissivity of galaxies in the UV and the mid/far-IR. Our model
predicts that, over the history of the Universe, about 90\% of the UV
radiation from massive young stars has been reprocessed by dust into
the mid/far-IR wavelength range. We used our model to investigate what
fraction of the total energy emitted by dust heated by stars over the
history of the Universe should be resolved into galaxies by different
planned \HERSCHEL\ surveys. We find that the fraction of the total IR
emission resolved in this way should be $\sim 50\%$ if individual
sources can be resolved all the way down to the nominal survey flux
limits set by signal-to-noise, but only $\sim 30\%$ if source
confusion sets a hard limit. Of the currently planned surveys, those
at 100 or 160$\mum$ should be able to resolve the largest fraction of
the time-integrated IR emission. This then implies that \HERSCHEL\
should resolve a similar fraction of the massive star formation over
the history of the Universe (roughly $m \gsim 5\Msol$, since these
stars are responsible for most of the UV heating of dust
grains). Lower-mass stars make only a small contribution to powering
the mid/far-IR emission from dust, so determining the total SFRs from
\HERSCHEL\ observations relies on extrapolating to lower masses ($m
\lsim 5\Msol$) based on an assumed IMF. If, as assumed in our model,
the IMF is different in different types of galaxy (quiescent {\em vs}
starburst), then estimating total SFRs from IR data becomes much more
complicated and uncertain, but estimates of the SFRs in high-mass
stars should be much more robust.

Finally, in \S\ref{sec:clustering}, we briefly investigated the
predicted clustering of \HERSCHEL\ galaxies. We found that the typical
galaxies in \HERSCHEL\ cosmological surveys should have a modest
clustering bias ($b \sim 1-2$) relative to the dark matter, with
correlation lengths $r_0 \sim 3-5 \hMpc$ (in comoving units), except
at the highest luminosities ($L_{IR} \gsim 10^{12}
\hLsol$). Measurements of the clustering in \HERSCHEL\ surveys will
provide an essential test of the relation between galactic star
formation rates and halo masses predicted by our galaxy formation
model.

\section*{Acknowledgements} 
We thank the referee Steve Eales for a helpful and positive
report. This work was supported in part by the Science and Technology
Facilities Council rolling grant to the ICC. CSF acknowledges a Royal
Society Wolfson Research Grant Award, and CMB a Royal Society
University Research Fellowship. AO is supported by a Gemini Fellowship
from the STFC. AJB acknowledges the support of the Gordon and Betty
Moore Foundation.

\bibliographystyle{mn2e}
\bibliography{paper}

\end{document}